\documentclass[twocolumn,
 amssymb, amsmath,%
 aip,jcp,
groupedaddress,10pt,%
]{revtex4-1}

\usepackage{docs}%
\usepackage{amsmath,bm}%
\usepackage{graphicx}
\usepackage{subcaption}
\usepackage{braket}
\usepackage{color,soul}%
\usepackage[framemethod=tikz]{mdframed}

\expandafter\ifx\csname package@font\endcsname\relax\else
 \expandafter\expandafter
 \expandafter\usepackage
 \expandafter\expandafter
 \expandafter{\csname package@font\endcsname}%
\fi
\newcommand{\bea}{\begin{eqnarray}}
\newcommand{\eea}{\end{eqnarray}}

\begin{document}
\title{Generalized input-output method: A new route to quantum transport junctions}


\author{Junjie Liu}
 \address{Department of Chemistry and Centre for Quantum Information and Quantum Control,
 University of Toronto, 80 Saint George St., Toronto, Ontario, M5S 3H6, Canada}
\author{Dvira Segal}
\address{Department of Chemistry and Centre for Quantum Information and Quantum Control,
University of Toronto, 80 Saint George St., Toronto, Ontario, M5S 3H6, Canada}
\address{Department of Physics, 60 Saint George St., University of Toronto, Toronto, Ontario, Canada M5S 1A7}

%
%
\begin{abstract}
The interaction of electrons with atomic motion critically influences
charge transport properties in molecular conducting junctions and quantum dot systems, 
and it is responsible for a plethora of transport phenomena.
%
Nevertheless, theoretical tools are still limited to
treat simple model junctions in specific parameter regimes. 
%
In this work, we put forward a generalized input-output method (GIOM)
for studying charge transport in molecular junctions accounting for strong electron-vibration interactions
and including electronic and phononic environments.
The method radically expands the scope of the input-output theory, 
which was originally put forward to treat quantum optic problems.
Based on the GIOM we derive a Langevin-type equation of motion for molecular operators, 
which posses a great generality and accuracy,
and permits the derivation of a stationary charge current expression involving only
two types of transfer rates.
Furthermore, we devise the so-called ``Polaron Transport in Electronic Resonance" (PoTER) approximation,
which allows to feasibly simulate electron dynamics in generic tight-binding models with strong electron-vibration interactions. 
To illustrate the breadth of applications of the GIOM-PoTER technique,
we analyze prototype molecular junction models with primary and secondary vibrational modes.
For short chains, the charge current 
reduces to known limits and reasonably agrees with exact numerical simulations (when available). 
For extended junctions the current displays a turnover from phonon-assisted to phonon-suppressed
transport. Nevertheless, the onset of ohmic behavior requires extensions beyond the PoTER approximation.
%
As an additional application, we consider a cavity-coupled molecule junction. Here we identify
a cavity-induced suppression of charge current in the single-site case,
and observe signatures of polariton formation in the current-voltage characteristics in the strong 
light-matter coupling regime.
A critical understanding gained from the GIOM-PoTER scheme is that
the single-site vibrationally-coupled model is deceptively simpler, and amenable to approximations than
multi-site models. Therefore, benchmarking of methods should not be concluded with the single-site case.
%
The work manifests that the input-output framework, which is normally employed in quantum optics,
can serve as a powerful and feasible tool in the realm of electron transport junctions.
\end{abstract}

\date{\today}


\maketitle
\section{Introduction}
\label{intro}

Single-molecule devices offer a rich and versatile platform for exploring the 
fundamentals of charge and energy transport at the nanoscale
and for making progress in electronic, photovoltaic, and thermoelectric systems \cite{Cuevas.10.NULL}.
A confounding aspect of electronic components fabricated from molecular building blocks is that they operate based on
quantum mechanical principles, yet the surrounding environments
in the form of electrical contacts, internal nuclear motion, solvent, and 
electromagnetic fields fundamentally impact and alter the charge transport behavior.
These environments do not fully erase quantum signatures in the device,  
but in fact allow for the realization of a plethora of compound many-body 
quantum transport effects 
that build upon hybrid quasi-particles, polarons, polaritons, plasmons.

Efforts to utilize molecules as active electronic elements 
\cite{Aviram.74.CPL} have converged into the field of organic and molecular electronics \cite{Cuevas.10.NULL}. 
Advanced experimental techniques 
\cite{Wold.01.JACS,Holmlin.01.JACS,Zhitenev.02.PRL,Agrait.03.PR,Tao.06.NN,Venkataraman.06.N,Ward.08.JP,McCreery.09.AM,Poulsen.09.NN,Tuccitto.09.NM,Lafferentz.09.S,Aradhya.13.NN,Xiang.16.CR}
 have led to the discovery of a variety of intriguing many-body phenomena in molecular junctions (MJs) 
in a metal-molecule-metal motif \cite{Akkerman.08.JP,Su.16.NRM,Cui.17.JCP,Thoss.18.JCP}, 
including length and temperature-dependent charge transfer 
\cite{Selzer.04.JACS,Selzer.04.NN,Weiss.05.JACS,Selzer.06.ARPC,Poot.06.NL,Goldsmith.08.JPCA,Choi.S.08,Lu.09.ACSN,Tuccitto.09.NM,Lafferentz.09.S,Luo.10.JACS,Choi.10.JACS,Hines.10.JACS,Luo.11.CM,Sedghi.11.NN,Li.12.NL,Taherinia.16.ACSN,Thomas.19.NC}, 
quantum interference effects 
\cite{Solomon.10.NC,Guedon.12.NN,Vazquez.12.NN,Aradhya.13.NN,Garner.18.N,Evers.19.NULL},
 molecular thermoelectricity \cite{Reddy.07.S,Malen.09.NL,Zimbovskaya.16.JCP,Garcia.16.CSR}, 
giant magnetoresistance \cite{Schmaus.11.NN}, Kondo resonance \cite{Liang.02.N,Scott.10.ACSN}, 
chirality induced spin selectivity \cite{Gohler.S.11,Naaman.15.ARPC,Naaman.19.NCR} 
and Franck-Condon blockade (FCB) directed by vibrational effects \cite{Koch.05.PRL,Koch.06.PRB,Leturcq.09.NP,Burzuri.14.NL,Lau.16.NL}, to name just a few.

The potential to rationally design molecular electronic devices hinges on our 
understanding of the underlying transport phenomena \cite{Thoss.18.JCP,Evers.19.NULL}. To this end, 
simplified theoretical models capable of pinpointing fundamental mechanisms are an indispensable tool. 
However, faithful modeling inevitably needs to take into account many-body interactions,
specifically the coupling between electrons and the nuclei's motion. 
Moreover, the application of voltage bias on the contact electrodes necessitates a 
description of the junction in the out-of-equilibrium regime. 

Numerous approaches have been put forward to address 
vibrationally-coupled electron transport (VCET) processes in MJs, and similarly quantum dot systems. 
Partial list of approximate analytic methods include the inelastic scattering theory 
\cite{Ness.01.PRB,Cizek.04.PRB,Toroker.07.JCP,Benesch.08.JPCC,Zimbovskaya.09.JCP,Jorn.09.JCP}, 
which only accounts for coherent scattering events, 
mixed quantum-classical approaches \cite{Koch.06.PRBa,Shen.07.PRB,Dzhioev.11.JCP,Li.14.JCP} 
where the vibrations are treated in a classical-like fashion, 
quantum master equation (QME) techniques 
\cite{Segal.00.JPCB,May.02.PRB,Mitra.04.PRB,Pedersen.05.PRB,Harbola.06.PRB,Donarini.06.PRL,Leijnse.08.PRB,Timm.08.PRB,Esposito.09.PRB,Volkovich.11.PCCP,Hartle.11.PRB,Sowa.17.PCCP},
which is often limited to weak molecule-lead couplings and become inadequate to describe off-resonant tunneling, 
and the nonequilibrium Green's function method (NEGF) 
\cite{Flensberg.03.PRB,Galperin.04.JCP,Galperin.06.PRBa,Galperin.06.PRB,Ryndyk.06.PRB,Frederiksen.07.PRB,Entin.08.PRB,Hartle.08.PRB,Hartle.09.PRL,Erpenbeck.15.PRB,Cabra.18.JCP}.
Unambiguously, the NEGF is the state-of-the-art among these tools. 
However, its complicated structure limits its applicability to simple systems. 
Even so, standard formulations of the NEGF method 
can not account for strong electron-vibration interactions due to the cumulant expansion 
employed \cite{Galperin.07.JP}.  

To provide benchmark calculations for analytic and perturbative studies, 
various numerically-exact methods have been developed, among them
the hierarchical quantum master equation (HEOM) 
\cite{Jin.08.JCP,Li.12.PRL,Jiang.12.PRB,Hartle.13.PRBa,Hartle.14.PRB,Hartle.15.PRB,Schinabeck.16.PRB,Schinabeck.18.PRB,Erpenbeck.18.JCP}, 
multiconfigurational wave-function methods \cite{Wilner.14.PRB,Wang.14.JCP},  path integral techniques based on
Monte Carlo sampling \cite{Werner.09.PRB,Gull.11.RMP,Cohen.13.PRB,Hartle.15.PRB} and the influence functional formalism
\cite{Muhlbacher.08.PRL,Weiss.08.PRB,Segal.10.PRB}. 
%
Complementing model system calculations, 
first principle density functional theory (DFT) simulations were integrated within the NEGF formalism
\cite{Kornilovitch.02.PRB,Verdozzi.06.PRL,Frederiksen.07.PRB,Arnold.07.JCP,Thoss.18.JCP} 
to include structural details of the junction and provide much insight into the transport process
\cite{Thoss.18.JCP,Evers.19.NULL}.

To make progress in organic and single-molecule electronics,
it is imperative to develop a computationally-feasible technique 
that treats (possibly strong) vibrational effects in electronic conduction and can handle extended models.
In this study, we focus on electron transport in molecular transport junctions
and introduce an alternative route to common QME and NEGF frameworks.
Our approach builds upon the quantum optical input-output method, which was previously advanced mainly for optical 
cavities \cite{Collett.84.PRA,Gardiner.85.PRA,Search.02.PRA,Gardiner.04.NULL,Ciuti.06.PRA,Genes.08.PRA,Clerk.10.RMP,Rabl.11.PRL,Nunnenkamp.11.PRL,Reitz.19.PRL,Kiilerich.19.PRL}.
The framework is termed the `generalized input-output method' (GIOM),  and it
possesses a simple structure---and a wide domain of applicability. 


The GIOM relies on the definitions of generalized input and output fields for the environments,
which circumvent the state-independent coupling approximation 
adopted in the quantum optical input-output theory \cite{Collett.84.PRA,Gardiner.85.PRA,Gardiner.04.NULL}.
Furthermore, the standard input-output theory is generalized here to include the coupling of system 
(molecular) operators to different (bosonic, fermionic) environmental degrees of freedom.
Altogether, these extensions result in a Langevin-type equation of motion (EOM) for molecular operators that 
takes into account strong couplings of the molecule to the metal electrodes as well as to vibrational modes, 
which comprise primary and secondary vibrations.

\begin{figure}[tbh!]
 \centering
\includegraphics[width=1\columnwidth] {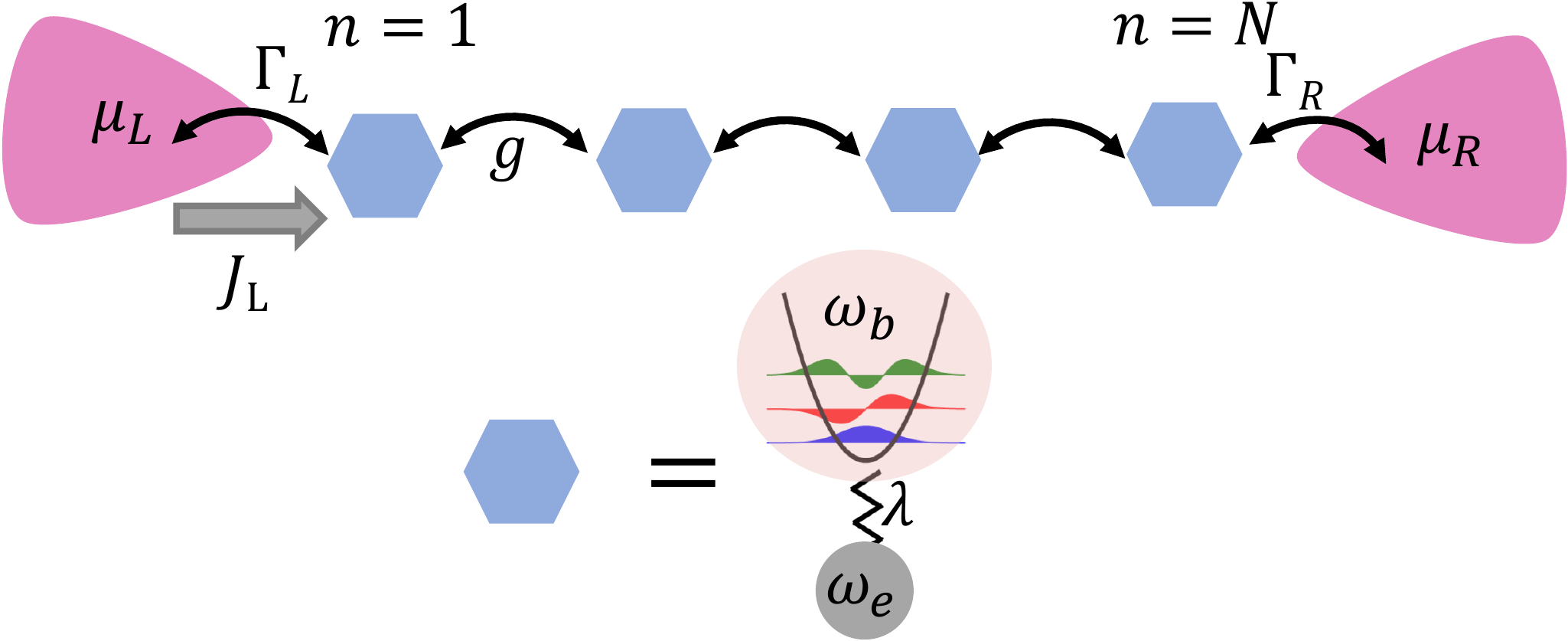} 
\caption{Scheme of a metal-molecule-metal junction with $N$ repeating molecular units.
As a minimal model, each monomer (represented by a hexagon) includes a single electronic level of energy 
$\omega_e$. The charge on each site is coupled to a local, `primary'
vibrational mode of frequency $\omega_b$, that is furthermore equilibrated by a thermal bath with `secondary' modes,
represented by the shaded region surrounding the harmonic oscillator.
Between monomers, $g$ quantifies the charge transfer transition amplitude.
$\Gamma_{L,R}$ is the hybridization energy of the first and last sites to the left and right metal electrodes, respectively.
Setting $\mu_L>\mu_R$, the  main observable of interest is the steady state charge current
evaluated here at the left metal-molecule interface, $J_L$. 
}
\label{fig:fig0}
\end{figure}

The appealing features of the GIOM are apparent in a generic tight-binding model describing VCET in MJs,
schematically represented in Fig. \ref{fig:fig0}.
In this case, the method provides a formally-exact solution, which is amenable to approximations and simulations.
Remarkably, a closed-form expression for the charge current can be achieved,
involving only two types of generalized transfer rates.
We devise the  ``Polaron Transport in Electronic Resonance" (PoTER) approximation,
and readily evaluate the charge current,
which is in fact exact in the coherent limit, regardless of molecular complexities. 
%
Given this simplicity, the GIOM-PoTER framework can be readily implemented in large systems, 
thereby holding promise to bridge microscopic-oriented modelings and effective phenomenological descriptions. 

To benchmark the GIOM-PoTER, we focus on VCET in single-site and two-site models, which see abundant applications in molecular transport junctions and quantum dot physics.
We simplify the charge current expression and show analytically that it recovers
the coherent and incoherent results as special limits. 
Further numerical calculations over a wide domain of the parameter space demonstrate the capabilities of the GIOM-PoTER 
in capturing essential features of VCET in those prototype models, covering weak-to-strong coupling effects to the metal electrodes and to the primary vibrations.
 
Strong light-matter interaction is now achievable 
at the single molecule level \cite{Chikkaraddy.16.N,Liu.17.PRL}.  
As the GIOM can naturally treat quantum optical setups, we consider cavity-coupled single-site MJ
in which we identify a cavity-induced suppression of charge current. This should be contrasted to
the enhancement observed in extended systems \cite{Orgiu.15.NM,Hagenmuller.17.PRL,Hagenmuller.18.PRB}.  
From the current-voltage characteristics, we further observe signatures of polariton formation 
in the strong light-matter coupling regime, suggesting that cavity-coupled MJs can 
provide a rich platform for studying nonequilibrium polaritonics.

The GIOM-PoTER favorably scales with system size, and we can
readily perform simulations for extended models with many sites.
Strong signatures of VCET effects are observed in the crossover behavior of the current 
with the electron-vibration coupling strength, transition that is missing in single-site MJs. 
Nevertheless, the sequential transport limit is not developed in uniform chains, within PoTER,
as the obtained distance dependence is missing the ohmic component, presumably due to the approximate nature of 
the method. This calls for additional research into first-principle modelling of 
ohmic conduction, and further
developments of the GIOM and other computational tools for extended models.

The paper is organized as follows. 
We present the GIOM for generic VCET models in Section \ref{sec:1}. 
In Section \ref{sec:2}, we focus on a tight-binding model. We
devise the PoTER approximation scheme, and derive
a general and computationally-feasible charge current expression in the steady state limit. 
In Section \ref{sec:3}, we benchmark our method on single-site and two-site MJs, 
and further present simulations on extended systems.
We summarize our findings and discuss future directions in section \ref{sec:4}.

\section{Generalized input-output method}
\label{sec:1}

\subsection{Quantum transport model}

We start by defining a general model for describing VCET in MJs. 
The total Hamiltonian contains three different parts,
\begin{equation}\label{eq:totalH}
H~=~H_M+H_{E}+H_{I}.
\end{equation}
First, the molecular part $H_M$ accounts for a collection of electronic states with on-site energies $\{\omega_{e,n}\}$ and 
fermionic annihilation operators $\{d_n\}$, 
local molecular vibrations with frequencies $\{\omega_{b,n}\}$ and bosonic annihilation operators $\{b_n\}$, 
and electron-vibration couplings as measured by  dimensionless coupling strengths $\{\lambda_n\}$ 
(setting $\hbar=1$, $e=1$, $k_B=1$ and the Fermi energy $\epsilon_F=0$ hereafter),
\begin{eqnarray}
H_M &=& H_e(\{\omega_{e,n},d_n\})+\sum_{n}\omega_{b,n}b_n^{\dagger}b_n\nonumber\\
&&+\sum_n\lambda_n\omega_{b,n}(b_n^{\dagger}+b_n)d_n^{\dagger}d_n.
\label{eq:HM}
\end{eqnarray}
Here, we consider the coupling of each electronic site to a single local vibrational mode.
%
However, generalizations to include additional vibrational modes per site, even yet 
creating a ``bath" of primary modes is straightforward within our scheme. 
Noting that the electronic part $H_e$ may also depend on additional internal parameters, for the moment, however, 
its detailed form is not relevant and we do not specify it. 

The environmental part $H_E$ contains two metallic leads ($L$ and $R$) and
thermalized phonon baths that are associated with the local nuclear motions of molecules.  These secondary modes 
allow vibrational relaxation of the primary modes, 
\begin{equation}
H_E~=~\sum_{k,v=L,R}\epsilon_{kv}c_{kv}^{\dagger}c_{kv}+\sum_{n,j}\omega_{n,j}r_{n,j}^{\dagger}r_{n,j}.
\end{equation}
Here $c_{kv}$ annihilates an electron in lead $v$ with momentum $k$. 
The thermal phonon baths are represented by collections of harmonic oscillators with annihilation operators $r_{n,j}$ and 
frequencies $\omega_{n,j}$. 
The third part, $H_{I}$, stands for the interaction between the molecule and its electronic and vibrational environments,
\begin{equation}
H_{I}~=~\sum_{k,v=L,R}t_{kv}(c_{kv}^{\dagger}d_{\sigma}+d_{\sigma}^{\dagger}c_{kv})+\sum_{n,j}\gamma_{n,j}(r_{n,j}^{\dagger}b_n+b_n^{\dagger}r_{n,j}),
\label{eq:HI}
\end{equation}
here, $\sigma=1(N)$ for $v=L (R)$. We refer to $d_{\sigma}$ as `boundary operators' since 
they are associated with terminal electronic sites of the molecule. 
We assume that the interaction between the primary vibrational modes and the secondary-thermalized modes is rather weak 
such that the rotating wave approximation is justified. 
The influence of each thermal bath, acting on sub-unit $n$, is characterized by the spectral density 
$I_n(\omega)=\pi\sum_j\gamma_{n,j}^2\delta(\omega-\omega_{n,j})$. 
Similarly, we introduce spectral densities for the metal leads as 
$\Gamma_v(\epsilon)=\pi\sum_kt_{kv}^2\delta(\epsilon-\epsilon_{kv})$.

To handle potentially strong electron-vibration couplings,
we perform the small polaron transformation with the 
unitary operator $G\equiv\prod_n(\mathcal{D}_{n,b})^{d_n^{\dagger}d_n}$ 
and displacement operators 
\bea
\mathcal{D}_{n,b}\equiv\exp[\lambda_{n}(b_{n}^{\dagger}-b_{n})].
\label{eq:dispb}
\eea 
We neglect the effect of this transformation on the coupling of primary modes to the thermal bath \cite{Galperin.06.PRBa}. 
In other words, in the polaron-transformed Hamiltonian we ignore
the effective weak interaction that forms between charge carriers and the secondary phonons.
%
The transformed Hamiltonian then reads 
\bea
\tilde{H}=GHG^{\dagger}=\tilde{H}_M+H_E+\tilde{H}_{I},
\label{eq:polaron}
\eea
with
\bea
\tilde{H}_M &=& \tilde{H}_e(\{\tilde{\omega}_{e,n},\tilde{d}_n\})+\sum_{n}\omega_{b,n}b_{n}^{\dagger}b_{n},
\nonumber\\
\tilde{H}_{I} &=& \sum_{kv}t_{kv}(c_{kv}^{\dagger}\tilde{d}_{\sigma}+\tilde{d}_{\sigma}^{\dagger}c_{kv})+\sum_{n,j}\gamma_{n,j}(r_{n,j}^{\dagger}b_n+b_n^{\dagger}r_{n,j}).
\nonumber\\
\label{eq:HP}
\end{eqnarray}
As can be seen, the transformation amounts to the renormalization of on-site energies, 
$\omega_{e,n}\to\tilde{\omega}_{e,n}=\omega_{e,n}-\lambda_{n}^2\omega_{b,n}$, and the dressing of the
tunneling transition frequency;
in the above expressions, we introduce the polaron operator as 
\bea
\tilde{d}_{n}\equiv \mathcal{D}_{n,b}^{\dagger}d_n.
\eea 

Before presenting the GIOM we point out that while we discuss the model, the theoretical framework 
and our results in the context of molecular transport junctions, 
this work can be immediately applied to investigate related problems such as:
(i) Charge transport in quantum dot setups defined e.g. within nanotubes, with electrons coupled to 
different phonon modes of the nanotube \cite{Leturcq.09.NP,Sowa.17.PRB}. 
(ii) Quantum optics scenarios with quantum dots defined within heterostructures coupled to cavity modes and phonons of 
the substrate \cite{Petta15}.

\subsection{Input-output equations of motion}

To develop an input-output theory for the system, 
we first write down Heisenberg EOMs for the annihilation operators of the environments (electronic and secondary phonon modes)
in the polaron frame,
\begin{subequations}
\begin{align}
\dot{r}_{n,j} &= -i\omega_{n,j}r_{n,j}-i\gamma_{n,j}b_n,\label{eq:eom1}\\
\dot{c}_{kv} &= -i\epsilon_{kv}c_{kv}-it_{kv}\tilde{d}_{\sigma},
\label{eq:eom3}
\end{align}
\end{subequations}
where we have introduced the notation $\dot{A}\equiv dA/dt$. 
From the above EOMs, we get the following formal solutions,
\begin{subequations}
\begin{align}
r_{n,j}(t) &= e^{-i\omega_{n,j}(t-t_0)}r_{n,j}(t_0)-i\gamma_{n,j}\int_{t_0}^t\,e^{-i\omega_{n,j}(t-\tau)}b_n(\tau)d\tau,\label{eq:s1}\\
c_{kv}(t) &= e^{-i\epsilon_{kv}(t-t_0)}c_{kv}(t_0)-it_{kv}\int_{t_0}^t\,e^{-i\epsilon_{kv}(t-\tau)}\tilde{d}_{\sigma}(\tau)d\tau.
\label{eq:s3}
\end{align}
\end{subequations}
Here $t_0$ is the initial time at which the dynamical evolution begins. 
As the coupling of primary modes to the phonon thermal bath is weak, 
we approximate $b_n(\tau)$ by $b_n(t)e^{i\omega_{b,n}(t-\tau)}$ in Eq. (\ref{eq:s1}), yielding 
\bea
&&\sum_j\gamma_{n,j}r_{n,j}(t)
\nonumber\\
&&=\sqrt{2\pi}b_{in}^n(t)
-ib_n(t)\sum_j\gamma_{n,j}^2\int_{t_0}^te^{-i(\omega_{n,j}-\omega_{b,n})(t-\tau)}d\tau,
\nonumber\\
\label{eq:10}
\eea 
where we have defined the  input field 
\bea
b_{in}^n(t)\equiv \frac{1}{\sqrt{2\pi}} \sum_j\gamma_{n,j}e^{-i\omega_{n,j}(t-t_0)}r_{n,j}(t_0).
\label{eq:inputfield}
\eea
Assuming that $I_n(\omega)$ is about a constant at the vicinity of $\omega_{b,n}$,
we find that $\sum_j\gamma_{n,j}^2e^{-i(\omega_{n,j}-\omega_{b,n})t}$ is nonzero only around $t=0$.
We then extend the lower limit of integration on the right-hand-side of Eq. (\ref{eq:10}) to $-\infty$ and proceed  as 
$\sum_j\gamma_{n,j}^2\int_{0}^{\infty}e^{-i(\omega_{n,j}-\omega_{b,n})\tau}d\tau\approx I_n(\omega_{b,n})\equiv\nu_n$; 
we have neglected the Cauchy principal value, which just stands for a minute frequency renormalization. 
Overall, we get 
\begin{equation}\label{eq:finals_1}
\sum_j\gamma_{n,j}r_{n,j}(t)~=~\sqrt{2\pi}b_{in}^n(t)-i\nu_n b_n(t).
\end{equation}
Here, $\nu_n$ is the damping rate (energy over $\hbar$) on site $n$ of primary modes of frequency $\omega_{b,n}$
to the associated thermal bath.
Contrasting this derivation to steps in the quantum optical input-output theory 
\cite{Collett.84.PRA,Gardiner.85.PRA,Gardiner.04.NULL}, 
we note that here we define the input field from the summation $\sum_j\gamma_{n,j}r_{n,j}$, 
instead of $\sum_jr_{n,j}$. 
By doing so, we circumvent the state-independent coupling approximation adopted in the quantum optical input-output theory 
\cite{Collett.84.PRA,Gardiner.85.PRA,Gardiner.04.NULL}, which assumes that the coupling coefficients $\gamma_{n,j}=\gamma_n$ are 
independent of the state index $j$. 

Proceeding with the metallic leads, we consider the wide-band limit \cite{Wingreen.89.PRB} 
such that we can exactly turn Eq. (\ref{eq:s3}) into
\begin{equation}
\label{eq:finals_3}
\sum_kt_{kv}c_{kv}(t)~=~ \sqrt{2\pi}d_{in}^v(t)-i\Gamma_v\tilde{d}_{\sigma}(t)
\end{equation} 
without compromising the magnitude of hybridization energy $\Gamma_v$. 
The input fields from the metal leads are defined as 
\bea
d_{in}^v(t)\equiv \frac{1}{\sqrt{2\pi}} \sum_kt_{kv}e^{-i\epsilon_{kv}(t-t_0)}c_{kv}(t_0).
\label{eq:14}
\eea 
%

The definitions of input fields in terms of environment operators at the initial time
ensure that they can be specified as initial conditions. 
Here, we prepare the initial state of the composite system to be such that, 
at $t=t_0$, the molecule and its environments factorize. 
Specifically, we assume that the metal leads and the vibrational thermal baths are initially in their thermal 
equilibrium states characterized by the Dirac-Fermi distribution function
$n_F^v(\epsilon)=\{\exp[(\epsilon-\mu_v)/T]+1\}^{-1}$ with 
$\mu_v$ the chemical potentials and $T$ the temperature, 
and the Bose-Einstein distribution function $n_B(\omega)=[\exp(\omega/T)-1]^{-1}$, respectively.
The resulting anticommutation/commutation relations, 
correlation functions for input fields, which define their statistics, 
as well as input-output relations that connect input and output fields can be found in Appendix \ref{a:1}. 

We now write down the Heisenberg EOM for an arbitrary molecular operator $\mathcal{O}$,
\begin{eqnarray}
\dot{\mathcal{O}} &=& i[\tilde{H}_M,\mathcal{O}]-i\sum_{kv}t_{kv}\left\{[\mathcal{O},c_{kv}^{\dagger}\tilde{d}_{\sigma}]+[\mathcal{O},\tilde{d}_{\sigma}^{\dagger}c_{kv}]\right\}\nonumber\\
&&-i\sum_{n,j}\gamma_{n,j}\left\{[\mathcal{O},r_{n,j}^{\dagger} b_n]+[\mathcal{O},b_n^{\dagger}r_{n,j}]\right\}.
\end{eqnarray}
As the molecular system contains both fermionic and bosonic operators, we should treat them separately. 
To this end, we redefine quantum commutator and anti-commutator as 
$[A,B]_-\equiv[A,B]$ and $\{A,B\}\equiv[A,B]_+$, respectively.
The EOM for $\mathcal{O}$ can be expressed as
\begin{eqnarray}
\dot{\mathcal{O}} &=& i[\tilde{H}_M,\mathcal{O}]_{-}-i\sum_{kv}t_{kv}\left\{\mp c_{kv}^{\dagger}[\mathcal{O},\tilde{d}_{\sigma}]_{\pm}+[\mathcal{O},\tilde{d}_{\sigma}^{\dagger}]_{\pm}c_{kv}\right\}\nonumber\\
&&-i\sum_{n,j}\gamma_{n,j}\left\{r_{n,j}^{\dagger}[\mathcal{O},b_n]_-+[\mathcal{O},b_n^{\dagger}]_-r_{n,j}\right\}.
\end{eqnarray}
Here,  the top signs apply if $\mathcal{O}$ is a fermionic operator; the bottom signs apply if 
$\mathcal{O}$ is bosonic. 
Making use of Eqs. (\ref{eq:finals_1}) and (\ref{eq:finals_3}), we obtain the EOM
\begin{equation}
\label{eq:eom_o}
\dot{\mathcal{O}}~=~ i[\tilde{H}_M,\mathcal{O}]_{-}-i\sum_v\mathbb{L}_{\pm}^v-i\sum_n\mathbb{X}_n,
\end{equation}
where 
\bea
\mathbb{L}_{\pm}^v&\equiv&\mp\left(i\Gamma_v \tilde{d}_{\sigma}^{\dagger}+\sqrt{2\pi}d_{in}^{v,\dagger}\right)[\mathcal{O},\tilde{d}_{\sigma}]_{\pm}
\nonumber\\
&+&[\mathcal{O},\tilde{d}_{\sigma}^{\dagger}]_{\pm}\left(-i\Gamma_v\tilde{d}_{\sigma}+\sqrt{2\pi}d_{in}^v\right), 
\nonumber\\
\mathbb{X}_n&\equiv &\left(i\nu_n b_{n}^{\dagger}+\sqrt{2\pi}b_{in}^{n,\dagger}\right)[\mathcal{O},b_n]_{-}
\nonumber\\
&+&[\mathcal{O},b_{n}^{\dagger}]_{-}\left(-i\nu_nb_{n}+\sqrt{2\pi}b_{in}^n\right).
\eea
This Langevin-type EOM constitutes the first main result of this study.
Before proceeding to derive it for tight-binding models, 
there are several features of Eq. (\ref{eq:eom_o}) that are worth mentioning. 
First, it does not rely on molecular details and thus possesses a great generality. 
Second, the electronic part is treated exactly in the wide band limit.
Third, there is no corresponding Lindblad master equation for Eq. (\ref{eq:eom_o}) as it can treat possible strong hybridization energy, in direct contrast to standard quantum optical input-output theory \cite{Gardiner.85.PRA}.
Moreover, the equation conserves the overall charge.

The main observable of interest in the steady state limit is the charge current across the system.
Introducing the charge occupation number operator of the left lead (source), 
$n_L\equiv\sum_{k}c_{kL}^{\dagger}c_{kL}$, the charge current out of the $L$ metal is formally given by
\bea
J_L=-\frac{d}{dt}\langle n_L \rangle=i\sum_{k}t_{kL}\langle(c_{kL}^{\dagger}\tilde{d}_{1}-\tilde{d}_{1}^{\dagger}c_{kL})\rangle,
\eea 
with the average performed over the initial factorized state of the composite system.
In the language of the input field, using Eq. (\ref{eq:finals_3}), we get
%
\begin{equation}\label{eq:js}
J_L~=~2\left(\sqrt{2\pi}\mathrm{Im}\langle \tilde{d}_{1}^{\dagger}d_{in}^L\rangle-\Gamma_L\langle d_{1}^{\dagger}d_{1}\rangle\right).
\end{equation}
Here, ``Im" refers to an imaginary part. 

In a complete analogy to the input fields, 
one can define output fields by solving Eqs. (\ref{eq:eom1})-(\ref{eq:eom3}) for $t_1>t$, 
rather than from the initial condition $t_0$ (see Appendix  \ref{a:1}). 
This allows the derivation of a Langevin-type equation, which is parallel to 
Eq. (\ref{eq:eom_o}), but given in terms of the output fields. However, due to the existence of input-output relations as given in Appendix  \ref{a:1}, it is sufficient to work with the input fields.

Before proceeding, we recall other EOM methods that has been developed to treat transport problems,
such as the Heisenberg EOM approach 
\cite{Topp.15.EPL,Jussiau.19.PRB} and methods written in the form of the Langevin equation 
\cite{Shen.07.PRB,Yang.15.PRB}. 
However, so far, the Heisenberg EOM approach has been only applied to 
simple noninteracting electronic systems, since it requires the inverse Laplace transform  in order
to calculate the dynamics of observables, 
typically a tedious and prohibitive task in molecular systems. 
Langevin equation techniques discussed in the literature for quantum transport are formulated 
for density matrix elements within the scope of the QME \cite{Shen.07.PRB,Yang.15.PRB}. 
In contrast,  the GIOM formulates the dynamics at the level of operators, rather than states, and it is nonperturbative.



\section{GIOM for tight-binding models}
\label{sec:2}

We now apply the GIOM Langevin-type equation (\ref{eq:eom_o}) to a generic tight binding model
with electron-vibration couplings, culminating with a closed-form expression for the stationary charge current.
The section includes two powerful results of theoretical and computational importance:

(i) From the theoretical side, we arrive at a formally exact EOM for molecular electronic operators 
in this open, many-body system, Eq. (\ref{eq:m1}), and write down its solution, Eq. (\ref{eq:exact}).
This equation describes exact electron dynamics in a molecular junction, barring two elements from our modeling:
The effective electron-secondary phonon bath coupling was omitted from the polaronic Hamiltonian,
and the wide band limit for the reservoirs was enforced.

(ii) To allow feasible calculations, 
we further devise an approximate solution to Eq. (\ref{eq:m1}),  that is Eq. (\ref{eq:ddd}). 
This solution neglects some aspects of electron-vibration interactions (as we describe in this section),
and we refer to it as the Polaron Transport in Electronic Resonance approximation.
As evident from its title, this PoTER approximation describes 
the transport of polarons through electronic resonances,
eigenstates of electronic Hamiltonian $H_e$ that are broadened by their hybridization to the metal reservoirs. 
Notably, in the context of the GIOM the electronic current 
is given in terms of only two types of rates, and the PoTER approximation 
allows for an economical simulation of these rates, and therefore the 
charge current, for a broad range of parameters. 

\subsection{Exact GIOM equations of motion} 
\label{subsec:21}

We specify the model Hamiltonian of Eqs. (\ref{eq:polaron})-(\ref{eq:HP}), and apply the GIOM.
The model includes a tight-binding chain with vibrational coupling, see Fig. \ref{fig:fig0}.
We assign a single electronic state to each repeat unit of a molecular wire. 
For the electronic part, we therefore have
\begin{equation}
\tilde{H}_e~=~\sum_{n=1}^N\tilde{\omega}_{e,n}d_n^{\dagger}d_n+\sum_{n=1}^{N-1}g_n(\tilde{d}_n^{\dagger}\tilde{d}_{n+1}+\tilde{d}_{n+1}^{\dagger}\tilde{d}_n),
\end{equation}
with $g_n$ as the hopping element between sites $n$ and $n+1$;
$N$ is the total number of electronic states. 
The single-site case, $N=1$, corresponds to the eminent single-impurity Anderson-Holstein model.
Multisite extensions involve two electronic sites, and beyond.

Inserting the above form into Eq. (\ref{eq:eom_o}), we obtain the following coupled EOMs for the bare 
electronic operators $d_{n}$,
\begin{widetext}
\begin{eqnarray}\label{eq:eom_d1}
&&\dot{d}_{1} 
~=~ -\left(\Gamma_L+i\tilde{\omega}_{e,1}\right)d_{1}-ig_1\mathcal{D}_{1,b}\mathcal{D}_{2,b}^{\dagger}d_2-i\sqrt{2\pi}\mathcal{D}_{1,b}d_{in}^L,\nonumber\\
&&\dot{d}_{n\neq1,N} ~=~ -i\tilde{\omega}_{e,n}d_n-ig_{n-1}\mathcal{D}_{n,b}\mathcal{D}_{n-1,b}^{\dagger}d_{n-1}-ig_n\mathcal{D}_{n,b}\mathcal{D}_{n+1,b}^{\dagger}d_{n+1},\nonumber\\
&&\dot{d}_{N} 
~=~ -\left(\Gamma_R+i\tilde{\omega}_{e,N}\right)d_{N}-ig_{N-1}\mathcal{D}_{N,b}\mathcal{D}_{N-1,b}^{\dagger}d_{N-1}-i\sqrt{2\pi}\mathcal{D}_{N,b}d_{in}^R.
\end{eqnarray}
\end{widetext}
Notably, EOMs for boundary operators $d_{1,N}$ 
naturally incorporate level broadening due to molecule-lead coupling. 
By introducing column vectors $\boldsymbol{d}=(d_1,d_2,\cdots,d_N)^T$, 
$\boldsymbol{\tilde d}_{in}=(\mathcal{D}_{1,b}d_{in}^L,0,\cdots,0,\mathcal{D}_{N,b}d_{in}^R)^T$, we recast Eq. (\ref{eq:eom_d1}) into a matrix form,
\begin{equation}
\label{eq:m1}
\boldsymbol{\dot{d}}~=~-\boldsymbol{M}\cdot\boldsymbol{d}-i\sqrt{2\pi}\boldsymbol{\tilde d}_{in}.
\end{equation}
Note that $\boldsymbol{\tilde d}_{in}$ includes the dressing of input fields by displacement operators.
The drift matrix $\boldsymbol{M}$ is of a tridiagonal structure with elements 
$[\boldsymbol{M}]_{nm}=W_{nm}\mathcal{D}_{n,b}\mathcal{D}_{m,b}^{\dagger}$. Here, 
$W_{nn}=\Gamma_L\delta_{n1}+\Gamma_R\delta_{nN}+i\tilde{\omega}_{e,n}$ and $W_{n,n+1}=W_{n+1,n}=ig_n$;
$\delta_{kp}$ is the Kronecker delta function. 

We now formally introduce the matrix $\boldsymbol{R}$ whose rows are made of the left eigenvectors of the drift matrix $\boldsymbol{M}$, 
\bea\label{eq:eigen_M}
\boldsymbol{R}\cdot\boldsymbol{M}=\boldsymbol{\Lambda}_M\cdot\boldsymbol{R}.
\eea
Here $\boldsymbol{\Lambda}_M=\mathrm{diag}(\Lambda_1,\cdots,\Lambda_N)$ is a diagonal matrix with eigenvalues $\{\Lambda_n\}$ representing electronic resonances. 
The fact that electronic resonances  
are characterized by complex {\it c}-numbers implies the form for matrix elements in the site basis, 
$[\boldsymbol{R}]_{nm}=\tilde{s}_{nm}\mathcal{D}_{n,b}\mathcal{D}_{m,b}^{\dagger}$ with $\tilde{s}_{nm}$ complex {\it c}-numbers according to Eq. (\ref{eq:eigen_M}). 

We now make an important observation: The determinant of $\boldsymbol{M}-\Lambda \boldsymbol{I}$ is equivalent to that of another
tridiagonal matrix, ${\boldsymbol{M_0}}-\Lambda\boldsymbol{I}$, having the same $\Lambda$ and $[{\boldsymbol{M_0}}]_{nm}=W_{nm}$.
This can be proved
by noting that the determinant of a tridiagonal matrix can be easily evaluated through the continuant of its elements. 
Defining $f_N=\mathrm{det}[\boldsymbol{M}_{N\times N}-\Lambda\boldsymbol{I}_{N\times N}]$, the sequence $\{f_N\}$ is called the continuant and satisfies the recurrence relation $f_N=(W_{NN}-\Lambda)f_{N-1}+g_{N-1}^2f_{N-2}$ with the boundary conditions $f_1=\Gamma_L+i\tilde{\omega}_{e,1}-\Lambda$, $f_0=1$ and $f_{-1}=0$. 
As can be seen, the sequence is fully equivalent to that of ${\boldsymbol{M_0}}-\Lambda\boldsymbol{I}$. 
Therefore, the task of determining $\boldsymbol{R}$ and $\boldsymbol{\Lambda}_M$ corresponds to 
the diagonalization problem of ${\boldsymbol{R_0}}\cdot{\boldsymbol{M_0}}=\boldsymbol{\Lambda}_M\cdot{\boldsymbol{R_0}}$
 with $[{\boldsymbol{R_0}}]_{nm}=\tilde{s}_{nm}$, thereby avoiding the displacement operators that appear in original matrices. 
In fact, this correspondence holds as well for one-dimensional models with long-range interactions, noting that 
displacement operators involved in $[\boldsymbol{M}]_{nm}$ and $[\boldsymbol{M}]_{mn}$ are always complex conjugate.

For small $N$, this diagonalization permits an analytical treatment. For larger $N$, 
we resort to a numerical diagonalization of the tridiagonal matrix ${\boldsymbol{M_0}}$.
We reiterate that the diagonal matrix $\boldsymbol{\Lambda}_M$ involves the set of electronic resonances
of the system ({\it c}-numbers). These broadened energy levels describe the electronic states of a molecule hybridized 
to metal electrodes.
Equipped with the knowledge that the diagonal matrix $\boldsymbol{\Lambda}_M$ does not depend on nuclear coordinates
we rewrite Eq. (\ref{eq:m1}), 
\begin{equation}\label{eq:m2}
\boldsymbol{R}\cdot\boldsymbol{\dot{d}}~=~-\boldsymbol{\Lambda}_M\cdot\boldsymbol{R}\cdot\boldsymbol{d}-i\sqrt{2\pi}\boldsymbol{R}\cdot\boldsymbol{\tilde d}_{in}.
\end{equation}
%
To proceed, we further reorganize (\ref{eq:m2}),
\begin{equation}
\label{eq:mm2}
\frac{d}{dt}(\boldsymbol{R}\cdot\boldsymbol{d})~=~-(\boldsymbol{\Lambda}_M-\boldsymbol{\dot{R}}\cdot\boldsymbol{R}^{-1})\cdot\boldsymbol{R}\cdot\boldsymbol{d}-i\sqrt{2\pi}\boldsymbol{R}\cdot\boldsymbol{\tilde d}_{in}.
\end{equation}
Here, $\boldsymbol{R}^{-1}$ denotes the inverse matrix of $\boldsymbol{R}$ with the matrix elements $[\boldsymbol{R}^{-1}]_{nm}=s_{nm}\mathcal{D}_{n,b}\mathcal{D}_{m,b}^{\dagger}$. 
Unlike $\boldsymbol{\Lambda}_M$, $\boldsymbol{\dot{R}}\cdot\boldsymbol{R}^{-1}$ depends on the nuclear coordinates, and it includes nonzero off-diagonal matrix elements in the site basis,
\begin{eqnarray}
[\boldsymbol{\dot{R}}\cdot\boldsymbol{R}^{-1}]_{nm} &=& 
\sum_k\tilde{s}_{nk}s_{km}(\lambda_n\dot{P}_n-\lambda_k\dot{P}_k)\mathcal{D}_{n,b}\mathcal{D}_{m,b}^{\dagger}\nonumber\\
&=& \delta_{nm}\lambda_n\dot{P}_n-\sum_k\tilde{s}_{nk}s_{km}\lambda_k\dot{P}_k\mathcal{D}_{n,b}\mathcal{D}_{m,b}^{\dagger}.
\nonumber\\
\end{eqnarray}
For simplicity, we introduce the notation $P_n\equiv b_n^{\dagger}-b_n$.
In obtaining the second line we used the fact that 
$\sum_k\tilde{s}_{nk}s_{km}\mathcal{D}_{n,b}\mathcal{D}_{m,b}^{\dagger}=\delta_{nm}$ since
 $\boldsymbol{R}\cdot\boldsymbol{R}^{-1}=\boldsymbol{I}$.

The exact dynamical evolution of electronic degrees of freedom is given by the formal solution of 
Eq. (\ref{eq:mm2}) in the steady state limit,
\begin{eqnarray}
\boldsymbol{d}(t) &=& -i\sqrt{2\pi}\int_{-\infty}^td\tau \boldsymbol{R}^{-1}(t)e^{-\boldsymbol{\Lambda}_M(t-\tau)}
\nonumber\\
&&\times e^{\int_{\tau}^t\boldsymbol{\dot{R}}(\tau')\boldsymbol{R}^{-1}(\tau')d\tau'}\boldsymbol{R}(\tau)\boldsymbol{\tilde d}_{in}(\tau).
\label{eq:exact}
\end{eqnarray}
The term resulting from the initial condition, $\boldsymbol{d}(t_0)$, is dropped
as it does not contribute to the ensemble average in the steady state limit (for the terms involved in charge current
Eq. (\ref{eq:js}), this is evident by noting the causality that $\boldsymbol{d}(t_0)$ does not correlate with the input
fields in the future). 

From the above formal solution we can find, for instance, the evolution of the operator $d_1$
\begin{eqnarray}
d_1(t) &=& -i\sqrt{2\pi}\sum_{nm}\int_{-\infty}^td\tau\Big[s_{1n}\tilde{s}_{m1}K_{nm}(t,\tau)d_{in}^L(\tau)
\nonumber\\
&&+s_{1n}\tilde{s}_{mN}K_{nm}(t,\tau)d_{in}^R(\tau)\Big],
\label{eq:exactd1}
\end{eqnarray}
where we have denoted by
\begin{eqnarray}
K_{nm}(t,\tau) &=& e^{-\Lambda_{n}(t-\tau)}\mathcal{D}_{1,b}(t)\mathcal{D}_{n,b}^{\dagger}(t)
\nonumber\\
&&\times\Big[\exp\Big(\int_{\tau}^t(\boldsymbol{\dot{R}}\boldsymbol{R}^{-1})_{\tau'}d\tau'\Big)\Big]_{nm}\mathcal{D}_{m,b}(\tau).
\nonumber\\
\label{eq:K}
\end{eqnarray}
This term incorporates all the vibrational effects.
Altogether, beginning with the exact EOM of the GIOM, Eq. (\ref{eq:m1}), we obtain the formally-exact
solution (\ref{eq:exact}) for electronic operators.
%
Interestingly, by inserting Eq. (\ref{eq:exactd1}) into Eq. (\ref{eq:js}) one immediately reveals that the charge current expression builds upon only two types of transfer processes involving one and two electronic resonances, which we will elaborate on later.
Nevertheless, since nuclear coordinates  appear in the combination 
$(\boldsymbol{\dot{R}}\boldsymbol{R}^{-1})_{\tau'}$, which evolves in time according to the full Hamiltonian,
the problem is formidable and it requires making approximations to become practical.

\subsection{PoTER approximation}

We now devise an approximation 
that allows us to reach a highly efficient computational scheme for solving the GIOM equations, (\ref{eq:m1}).
This so-called PoTER approximation includes two components. 
In this section, 
we elaborate on the first, critical part of the PoTER approximation: 
polarons are transmitted between the metals through purely electronic resonances.
In Section \ref{sec-current}, we perform the second part of the PoTER approximation: 
The vibrational degrees of freedom which form the 
polaron time-evolve without the back-action of electrons.

Under the first part of the PoTER approximation,
we simplify the kernel in Eq. (\ref{eq:K}) and replace
$\Big[\exp\Big(\int_{\tau}^t(\boldsymbol{\dot{R}}\boldsymbol{R}^{-1})_{\tau'}d\tau'\Big)\Big]_{nm}$ with $\delta_{nm}$.
This approximation amounts to 
(i) neglecting nonlocal vibration effects with $n\neq m$ in the dynamical evolution of electronic degrees of freedom,
and (ii) neglecting phonon corrections to the electronic resonances.  This reduces Eq. (\ref{eq:exact}) to
\begin{equation}
\label{eq:ddd}
\boldsymbol{d}(t)~\approx~-i\sqrt{2\pi}\int_{-\infty}^td\tau \boldsymbol{R}^{-1}(t)e^{-\boldsymbol{\Lambda}_M(t-\tau)}\boldsymbol{R}(\tau)\boldsymbol{\tilde d}_{in}(\tau).
\end{equation}
Specifically, Eq. (\ref{eq:exactd1}) becomes
\begin{eqnarray}\label{eq:d1t}
d_1(t) &\approx& -i\sqrt{2\pi}\sum_n\int_{-\infty}^td\tau\Bigg[s_{1n}\tilde{s}_{n1}{K}^{PoTER}_n(t,\tau)d_{in}^L(\tau)\nonumber\\
&&+s_{1n}\tilde{s}_{nN}{K}^{PoTER}_n(t,\tau)d_{in}^R(\tau)\Bigg],
\end{eqnarray}
with the PoTER kernel,
\bea
K^{PoTER}_n(t,\tau)=e^{-\Lambda_{n}(t-\tau)}\mathcal{D}_{1,b}(t)\mathcal{D}_{n,b}^{\dagger}(t)\mathcal{D}_{n,b}(\tau).
\eea
For simplicity, in what follows we replace the approximate symbol by an equality, since we consistently work 
 the PoTER approximation.
 
 Elaborating on this approximation: The term $(\boldsymbol{\dot{R}}\boldsymbol{R}^{-1})$ in Eq. (\ref{eq:exact})
involves time derivative of $\mathcal{D}_{n,b}\mathcal{D}_{m,b}^{\dagger}$. It is 
proportional to $\lambda_{n}\dot{P}_{n}-\lambda_{m}\dot{P}_{m}$,
or approximately to
$\lambda_{n}\omega_{b,n}({b}_{n}^{\dagger}+{b}_{n})-\lambda_{m}\omega_{b,m}({b}_{m}^{\dagger}+{b}_{m})$,  once we consider
free vibrations, $b_{n}^{\dagger}(t)=b_n^{\dagger}(0)e^{i\omega_{b,n}t}$. 
Therefore, the first part of the PoTER approximation amounts to assuming that 
nuclear displacements are uniform across the lattice. 
This assumption is expected to be reasonably valid
for a tight-binding model with identical (or similar) repeating units and when the dissipation to secondary modes is weak.
%

%

It is worth mentioning that: (i) The PoTER solution is exact in the coherent limit  ($\{\lambda_n\}\to0$) since then 
$\boldsymbol{R}$ is strictly time-independent. (ii) In Appendix \ref{a:4} we prove that the PoTER scheme does not impact the total charge conservation. (iii) The solution for $d_1^{\dagger}$ is just the Hermitian transpose of Eq. (\ref{eq:d1t}); the PoTER approximation
does not cripple this relation.
We show that by writing down the EOM for row vectors
 $\boldsymbol{d}^{\dagger}=(d_1^{\dagger},d_2^{\dagger},\cdots,d_N^{\dagger})$, $\boldsymbol{\tilde d}_{in}^{\dagger}=(d_{in}^{L,\dagger}\mathcal{D}_{1,b}^{\dagger},0,\cdots,0,d_{in}^{R,\dagger}\mathcal{D}_{N,b}^{\dagger})$,
\bea
\boldsymbol{\dot{d}}^{\dagger}=-\boldsymbol{d}^{\dagger}\cdot\boldsymbol{M}^{\dagger}+i\sqrt{2\pi}\boldsymbol{\tilde d}_{in}^{\dagger}.
\eea
We diagonalize $\boldsymbol{M}^{\dagger}$ in term of 
$\boldsymbol{M}^{\dagger}\cdot\boldsymbol{Q}=\boldsymbol{Q}\cdot\boldsymbol{\Lambda}_M^{\ast}$ with $\boldsymbol{Q}=\boldsymbol{R}^{\dagger}$ and find the PoTER solution 
$\boldsymbol{d}^{\dagger}(t)~=~i\sqrt{2\pi}\int_{-\infty}^td\tau \boldsymbol{\tilde d}_{in}^{\dagger}(\tau)\boldsymbol{Q}(\tau)e^{-\boldsymbol{\Lambda}_M^{\ast}(t-\tau)}\boldsymbol{Q}^{-1}(t)$, which is the Hermitian transpose of Eq. (\ref{eq:ddd}). 
Hence, we can obtain $d_1^{\dagger}$ directly from the Hermitian transpose of Eq. (\ref{eq:d1t}).

When do we expect the GIOM-PoTER treatment to be accurate? 
As mentioned above, the PoTER approximation does not affect the purely electronic problem,        
i.e. it exactly recovers the Landauer form with the correct transmission function. It is also expected to be accurate 
in the low temperature regime when nuclear motion is largely suppressed.
Moreover, for the single impurity problem {\it with} electron-vibration interaction
Eq. (\ref{eq:ddd}) is exact since $\boldsymbol{M}$ simply involves the electronic resonance
without vibrational corrections, which are fully delegated to the input fields.
Lastly, for short wires we expect corrections to Eq. (\ref{eq:ddd}) due to nonlocal 
vibrational effects and vibrational self energies to be minor. 
However, these terms could become important in long wires, particularly once energy dissipation is substantial.

So far, we discussed the first part of the PoTER approximation: polarons are transmitted through purely  electronic resonances.
The second part of the approximation is practiced in the next subsection, and
it concerns the time evolution of the vibrations forming the polaron:
We propagate the displacement operators $\mathcal{D}_{n,b}(t)$
while ignoring back-action from charge carriers. This approximation
allows us to prepare and evaluate the time correlation function of the polaron, a component in the expression of
the charge current.

\subsection{Charge current}
\label{sec-current}

In this section we derive a closed-form expression for the steady state charge current.
Our starting point is the definition, Eq. (\ref{eq:js}), with the average performed with respect to 
the initial total density matrix,
which is assumed to be in a factorized form, $\rho(t=0)=\rho_e\otimes \rho_b$.
Here $\rho_e$ is the initial state for the electronic degrees of freedom, factorized to the two baths
and the molecular electronic system, $\rho_e=\rho_L\otimes\rho_R\otimes\rho_S$. $\rho_b$ is the 
initial state for the  bosonic modes. It is factorized between the $N$ sites, and between the
the primary and secondary modes, the latter are assumed to be in a thermal state.  

Using the PoTER solution, Eq. (\ref{eq:d1t}), and the commutation relations of input fields from appendix \ref{a:1}, we 
get
\begin{equation}
\label{eq:dinL}
\sqrt{2\pi}\mathrm{Im}\langle \tilde{d}_{1}^{\dagger}d_{in}^L\rangle~=~\sum_n\mathrm{Re}[\Phi_n^L\chi_n^L].
\end{equation}
Here `` Re" refers to the real part. We have denoted by $\Phi_n^L=s_{1n}\tilde{s}_{n1}$ and defined the transfer rate
\begin{equation}
\label{eq:chinv}
\chi_n^v~=~2\Gamma_v\int\,\frac{d\epsilon}{2\pi}n_F^v(\epsilon)\int_{0}^{\infty}\,d\tau e^{i\epsilon\tau}e^{-\Lambda_{n}\tau}B_n(\tau),
\end{equation} 
with the vibrational correlation function 
$B_n(t-\tau)=\langle \mathcal{D}_{n,b}^{\dagger}(t)\mathcal{D}_{n,b}(\tau)\rangle$; the index $n$ recounts the electronic resonances. 
We refer to this rate as `first order' since it involves a single electronic resonance, $\Lambda_n$.
Clearly, $\chi_n^v$ describes a transfer event assisted by a single electronic resonance and dressed by local vibrational correlations. 

The derivation of  Eq. (\ref{eq:chinv}) is detailed in Appendix \ref{a:5}. Briefly, it involves
the second  part of the PoTER approximation: We
ignore the back-action of electrons on primary modes in the polaron frame and separately organize electronic and vibrational correlation functions.

We proceed and calculate
the vibrational correlation function using the decoupled EOMs for the primary modes, Eq. (\ref{eq:eom_o}), 
$\dot{b}_n\approx-(\nu_n+i\omega_{b,n})b_n-i\sqrt{2\pi}b_{in}^n$. We get
\begin{eqnarray}
\label{eq:corr_bb}
B_n(\tau) &=& \exp\Bigg[-\lambda_{n}^2\int d\omega\frac{I_n(\omega)}{\pi}\frac{1}{\nu_n^2+(\omega-\omega_{b,n})^2}\nonumber\\
&&\times\Big(\mathrm{coth}(\beta\omega/2)(1-\cos\omega\tau)+i\sin\omega\tau\Big)\Bigg].
\end{eqnarray}
Details can be found in Appendix \ref{a:2}.

The second contribution to the charge current involves the stationary charge occupation on the first site, 
\begin{equation}\label{eq:n11}
\langle d_1^{\dagger}d_1\rangle~=~\frac{1}{2}\sum_{nm,v}\frac{\Psi_{nm}^v}{\Gamma_v}\eta_{nm}^v.
\end{equation}
Here, we have denoted the coefficients 
$\Psi_{nm}^L=s_{1n}^{\ast}\tilde{s}_{n1}^{\ast}s_{1m}\tilde{s}_{m1}$, $\Psi_{nm}^R=s_{1n}^{\ast}\tilde{s}_{nN}^{\ast}s_{1m}\tilde{s}_{mN}$ and introduced the transfer rate (see Appendix \ref{a:5}), 
\begin{eqnarray}
\label{eq:eta_rate}
\eta_{nm}^v &=& 4\Gamma_v^2\int\,\frac{d\epsilon}{2\pi}n_F^v(\epsilon)\Big(\int_{0}^{\infty}d\tau e^{i\epsilon\tau}e^{-\Lambda_{n}\tau}B_n(\tau)\Big)^{\ast}\nonumber\\
&&\times\Big(\int_{0}^{\infty}d\tau e^{i\epsilon\tau}e^{-\Lambda_{m}\tau}B_m(\tau)\Big).
\end{eqnarray}
We refer to this rate as `second order' since it involves two resonances; diagonal terms 
are exceptions as they reduce to first order rates, $\eta_{nn}^v=\frac{2\Gamma_v}{\mathrm{Re}\Lambda_n}\mathrm{Re}[\chi_n^v]$ after performing a time integration. 
To evaluate the auto-correlation function of the polaron, 
we neglect the back-action of electrons on the vibrational dynamics.  

%

Inserting Eqs. (\ref{eq:dinL}) and (\ref{eq:n11}) into Eq. (\ref{eq:js}) 
we build the final expression for the charge current out of the left lead,
\begin{equation}\label{eq:jL}
J_L ~=~2\sum_n\mathrm{Re}[\Phi_n^L\chi_n^L]-\Gamma_L\sum_{nm,v}\frac{\Psi_{nm}^v}{\Gamma_v}\eta_{nm}^v.
\end{equation}
Similarly, the steady state charge current out of the right lead, $J_R$, 
can be expressed as $J_R=2\sum_n\mathrm{Re}[\Phi_n^R\chi_n^R]-\Gamma_R\sum_{nm,v}\frac{\tilde{\Psi}_{nm}^v}{\Gamma_v}\eta_{nm}^v$ with $\Phi_n^R=s_{Nn}\tilde{s}_{nN}$, $\tilde{\Psi}_{nm}^R=s_{Nn}^{\ast}\tilde{s}_{nN}^{\ast}s_{Nm}\tilde{s}_{mN}$ and $\tilde{\Psi}_{nm}^L=s_{Nn}^{\ast}\tilde{s}_{n1}^{\ast}s_{Nm}\tilde{s}_{m1}$.
These two analytic charge current expressions represent one of main results of this study. 
As the method maintains the charge conservation, it should satisfy $J_L=-J_R$ in the steady state limit.

Notably, the charge current involves only two types of rates as revealed by the GIOM, regardless of molecular complexities.
The PoTER approximation allows their facile evaluation:
The rates $\chi_n$ depend on individual eigenvalues $n$.
The rates $\eta_{nm}$ involve charge transfer jointly through two resonances. 

\subsection{GIOM-PoTER: Synopsis}

We summarize and enumerate the assumptions and approximations underlying the GIOM-PoTER scheme, 
as well as the technical advances.

The GIOM equations are derived under the following three assumptions:

{\bf a1.} The two types of environments, the metal electrodes and the secondary  phonon baths are treated in the
wide band limit. That is, the spectral density of the metal reservoirs is assumed to be about constant 
at the vicinity of molecular electronic levels. Similarly, the spectral function
of secondary phonon baths is assumed flat around frequencies of the primary modes.

{\bf a2.} The secondary phonon bath is coupled to the primary vibrational modes within a rotating-wave Hamiltonian.

{\bf a3.} In the polaron-transformed Hamiltonian we neglect the effective interaction  
between electronic degrees of freedom and the secondary bath modes.

These assumptions 
are justified based on weak coupling between primary and secondary modes.
Altogether, we regard the polaronic Hamiltonian (\ref{eq:HP}) as our starting point, and 
therefore refer to the resulting GIOM EOM as exact for this class of Hamiltonians.


The GIOM advances the quantum optical counterparts in four ways:
(i) The GIOM relies on the definition of generalized input and output fields for the environments, 
going beyond the state-independent 
coupling approximation \cite{Collett.84.PRA,Gardiner.85.PRA,Gardiner.04.NULL} 
frequently used in the quantum optical counterpart. This allows to faithfully model 
the coupling of molecular operators to metal electrodes and vibrational degrees of freedom.
(ii) The derived Langevin-type EOM Eq. (\ref{eq:eom_o}) describes
the coupling of molecular operators to different degrees of freedom, fermionic and bosonic, and
there is no corresponding Lindblad master equation for it. 
(iii)  Electron-vibration interactions are included nonperturbatively, and 
molecular electronic degrees of freedom are treated exactly in the wide band limit \cite{Wingreen.89.PRB}. 
(iv) For tight-binding models 
the GIOM equation has a compact-exact form  (\ref{eq:m1})
with the formally-exact steady state solution (\ref{eq:exact}).

To implement the GIOM efficiently, we furthermore make the PoTER approximation:

{\bf A1.} 
We disregard nonlocal phonon effects on charge transport, as well as corrections
to resonance energies due to nuclear effects. 
As a result, the vibrationally-dressed electrons cross the system through purely electronic resonances.

{\bf A2.} 
We time-evolve vibrations forming the polarons 
while ignoring back-action from the  electronic degrees of freedom.

Under the PoTER approximation {\bf A1}, Equation (\ref{eq:exact}) is replaced by (\ref{eq:ddd}), which is 
amenable to numerical simulations.  With the assistance of {\bf A2}, 
correlation functions of primary vibrations are feasibly calculated.
Specifically we gain the following advantages: (i) The charge current expression, which depends on two types of rates (even for extended systems),
is readily computed, with computational efforts scaling quadratically with system size,  $N^2$.
(ii) Analytic expressions can be obtained for short ($N=1,2$) junctions.
(iii) The GIOM-PoTER solution is exact in the coherent limit. Moreover, the first part of
the PoTER approximation, {\bf A1}, is redundant for the single-site case.
For relatively uniform structures, the approximation {\bf A1} is expected to be validated even in extended systems.

\begin{table}[tbh!]
\caption{Summary of parameters}
\centering
\begin{tabular}{ccc}
\hline\hline
$\omega_{e,n}$  & Molecular electronic energy on site $n$ & 0.1-2 eV\\ 
$\tilde\omega_{e,n}$ & Renormalized site energies,  $\omega_{e,n}-\lambda_n^2 \omega_{b,n}$ &\\
$g$  & intersite tunneling element & 0.01-0.5 eV\\ 
$\Gamma_{L,R}$ & Decay rate to the metals & 0.01-1 eV\\
$V$ & bias voltage &  0 -3 V\\ 
$\omega_{b,n}$  & Frequency of primary mode & 0.2 eV\\
$\lambda$  & Electron-vibration coupling  (dimensionless) & 0-3 \\ 
$\nu$  & Broadening of primary modes  & 0.005-0.01 eV\\ 
$T$ & Temperature of metals and vibrational baths & 300 K\\ 
$\omega_c$ & Cavity frequency & $\omega_e$ \\
$\Omega_R$ & Light-matter coupling energy &  0-0.8$\omega_c$ \\ 
$\kappa$ & Cavity loss rate & 0.05 eV\\
$N$ & Number of repeating units & 1-15\\
\hline\hline
\end{tabular}
\label{table1}
\end{table}

\section{Applications}
\label{sec:3}

The GIOM-PoTER treatment with the resulting charge current expression, Eq. (\ref{eq:jL}), is applicable to $N$-site tight-binding models. Here, we focus on two central models for MJs: the single-site and two-site MJs.  Results for longer chains are also described.
A central advantage of our technique is the feasibility of computations compared to other
transport methods. 
With existing benchmark results, we are able to fully elucidate the utility of the GIOM.
The list of physical parameters along with experimentally relevant values employed in this work are given in Table  1.

\subsection{single-site junction}
\label{ssec:1}

We first consider a monomeric MJ comprising a single electronic state.
We recall that the PoTER approximation is redundant in the single-site model,
and the only assumption involved is {\bf A2}.
The molecular electronic Hamiltonian is 
\begin{equation}
\tilde{H}_e~=~\tilde{\omega}_ed_1^{\dagger}d_1.
\end{equation}
The other terms in the Hamiltonian, describing the contributions of primary and secondary vibrational modes, 
as well as coupling of the molecule to metal electrodes 
are given by Eqs. (\ref{eq:HM})-(\ref{eq:HI}). 

One immediately resolves the resonance  energy,
$\Lambda_1=\Gamma+i\tilde{\omega}_e$, with $\Gamma=\Gamma_L+\Gamma_R$ 
and $s_{11}=1, \tilde{s}_{11}=1$. Hence, the nonvanishing coefficients of Eq. (\ref{eq:jL}) are $\Phi_1^L=1$ 
and $\Psi_{11}^{L,R}=1$, leading to the following simplified form of the charge current,
\begin{equation}
\label{eq:jl1}
J_L~=~2\frac{\Gamma_R\mathrm{Re}[\chi_1^L]-\Gamma_L\mathrm{Re}[\chi_1^R]}{\Gamma}.
\end{equation}
For convenience, we denote $\gamma_v=\mathrm{Re}[\chi_1^v]$ and introduce the notations
\bea\label{eq:rate_site1}
\gamma_v&=&
2\mathrm{Re}\Big[\Gamma_v\int\,\frac{d\epsilon}{2\pi}n_F^v(\epsilon)\int_{0}^{\infty}\,d\tau e^{i(\epsilon-\tilde \omega_e)\tau} e^{-\Gamma \tau}B(\tau)\Big],
\nonumber\\
\bar{\gamma}_v&=&2\mathrm{Re}\Big[\Gamma_v\int\,\frac{d\epsilon}{2\pi}(1-n_F^v(\epsilon))\int_{0}^{\infty}\,d\tau e^{-i(\epsilon-\tilde{\omega}_e)\tau}e^{-\Gamma\tau}B^{\ast}(\tau)\Big],
\nonumber\\
\eea
satisfying $\gamma_v+\bar{\gamma}_v=\Gamma_v$.
With that, one can recast Eq. (\ref{eq:jl1}) into the form 
\bea
J_L=2\frac{\bar{\gamma}_R\gamma_L-\bar{\gamma}_L\gamma_R}{\bar{\gamma}_L+\gamma_L+\bar{\gamma}_R+\gamma_R}.
\eea
This result was recently obtained using a generalized QME method  \cite{Sowa.18.JCP}
(noting the adopted definition of $\Gamma_v$ is half of theirs). 
Given the vast applications of the single-site Anderson model, achieving a closed-form expression for the
current, Eq. (\ref{eq:jl1}), is a significant achievement. Our derivation provides the foundation
of this approximate yet powerful result.

As noted in Ref. \cite{Sowa.18.JCP}, the charge current expression Eq. (\ref{eq:jl1}) 
is capable of bridging the coherent and incoherent descriptions. 
In the coherent limit of $\lambda=0$, we have 
\begin{equation}
\gamma_v~=~\int\,\frac{d\epsilon}{2\pi}n_F^v(\epsilon)\frac{2\Gamma_v\Gamma}{\Gamma^2+(\epsilon-\omega_e)^2},
\end{equation}
and the current in Eq. (\ref{eq:jl1}) reduces to
\begin{equation}
J_L^{LB}~=~\int\,\frac{d\epsilon}{2\pi}\frac{4\Gamma_L\Gamma_R}{\Gamma^2+(\epsilon-\omega_e)^2}[n_F^L(\epsilon)-n_F^R(\epsilon)].
\end{equation}
This is precisely the Landauer-B\"uttiker (LB) expression for a single noninteracting level \cite{Jauho.94.PRB,Datta.95.NULL}. 

In contrast, in the high temperature limit of $T\gg\Gamma_v$, one simplifies Eq. (\ref{eq:jl1})
by neglecting level broadening due to molecule-lead couplings.
Then, the rate $\gamma_v$ turns out to be the one obtained by the second order QME in the polaron frame \cite{Breuer.02.NULL}, 
and the corresponding charge current reads
\begin{equation}
J_L^{QME}~=~\int\,\frac{d\epsilon}{2\pi}\frac{4\Gamma_L\Gamma_R}{\Gamma_L+\Gamma_R}\tilde{B}(\epsilon)[n_F^L(\epsilon)-n_F^R(\epsilon)],
\end{equation}
where we have defined $\tilde{B}(\epsilon)=\mathrm{Re}[\int_0^{\infty}d\tau e^{i(\epsilon-\tilde{\omega}_e)\tau}B(\tau)]$. 
In the high bias limit of $V\to\infty$, 
 $J_L^{QME}=2\Gamma_L\Gamma_R/(\Gamma_L+\Gamma_R)$ which is expected as the influence of electron-vibration coupling is negligible in that limit.

\subsubsection{Electron-vibration effects}

In order to gain additional physical insight and evaluate the performance of the proposed GIOM-PoTER framework  
between the two limits, coherent and incoherent,
we utilize Eq. (\ref{eq:jl1}) to study the current-voltage characteristics while varying system parameters. 
For the sake of simplicity, in simulations we set $\Gamma_L=\Gamma_R=\Gamma/2$. 
A super-Ohmic spectrum $I(\omega)=\nu\frac{\omega^3}{\omega_{b,cut}^3}e^{-(\omega-\omega_{b,cut})/\omega_{b,cut}}$ with a cut-off frequency $\omega_{b,cut}=0.5$ eV is adopted for the thermal vibrational bath. 
We emphasize that the GIOM is not limited to a specific form of spectral density, and 
other functions such as the Debye-Drude form can be easily incorporated in the GIOM. 
However, the basic features of the current-voltage characteristics remain unaltered.

%
\begin{figure}[tbh!]
 \centering
\includegraphics[width=0.45\textwidth] {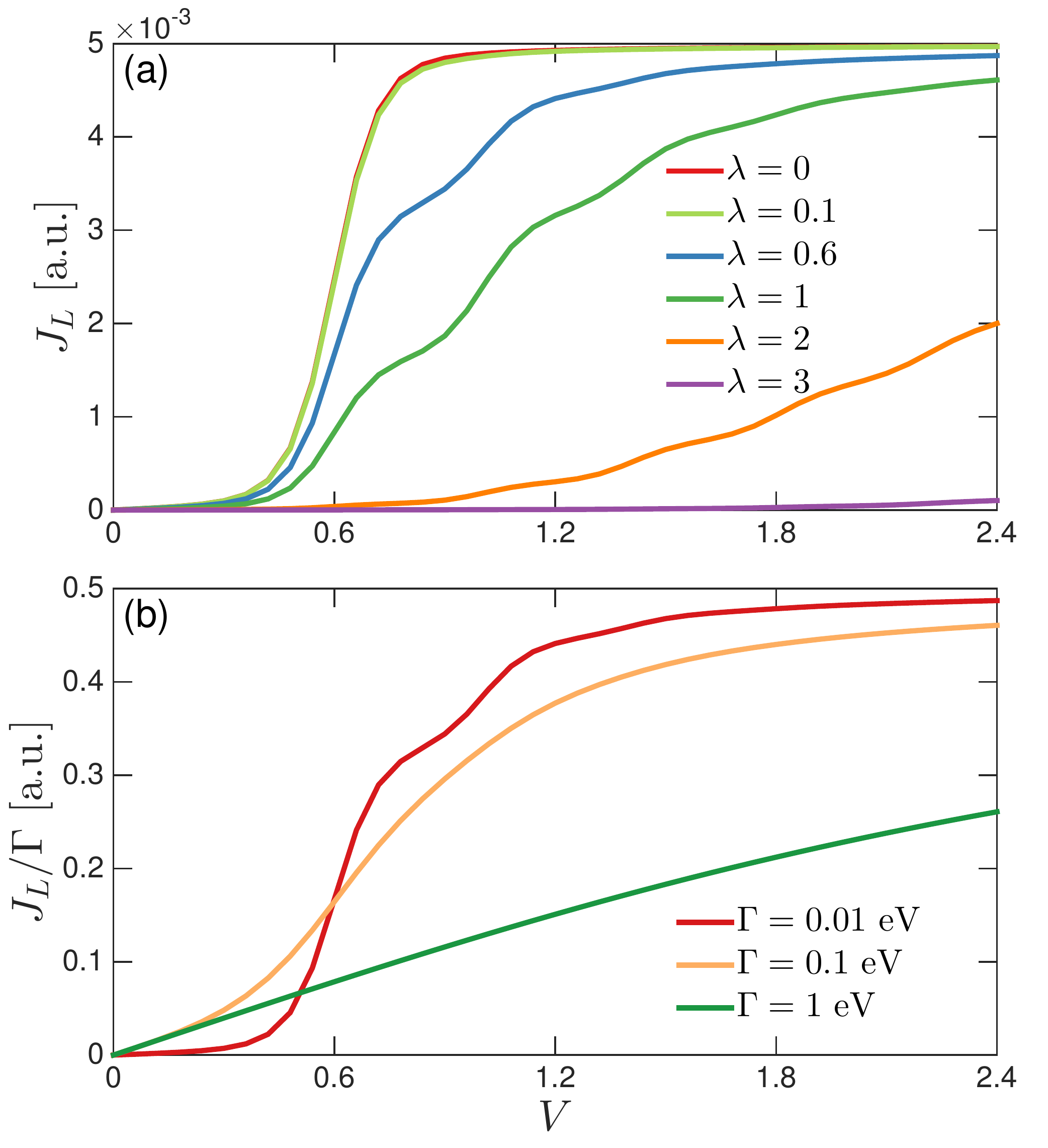} 
\caption{Current-voltage characteristics of a single-site MJ for
(a) different electron-vibration coupling strengths $\lambda$ with fixed $\Gamma=0.01$ eV, 
(b) different molecule-lead coupling strengths $\Gamma$ at fixed $\lambda=0.6$. 
Model parameters are $\mu_L=-\mu_R=V/2$, $\tilde{\omega}_e=0.3$ eV, $\omega_b=0.2$ eV, 
$\nu=0.005$ eV and $T=300$ K. Hereafter, ``a.u." denotes a short notation for ``arbitrary units".} 
\label{fig:fig1}
\end{figure}


A typical set of current-voltage curves is depicted in Fig. \ref{fig:fig1}. 
In panel (a), we vary the electron-vibration coupling strength $\lambda$. 
For vanishing and small ($\lambda=0.1$) electron-vibration couplings, 
the charge current is dominated by elastic processes. 
The step around $V=0.6$ marks the onset of resonant tunneling regime, noting that $\tilde{\omega}_e=0.3$ eV in simulations. 
Increasing $\lambda$ further, inelastic processes begin to play a role and the charge transfer 
is generally accompanied by excitation of molecular vibrations, leading to Franck-Condon step structures 
in the current-voltage curve. For high enough values of $\lambda$, 
the equilibrium coordinates in adjacent charge states are significantly shifted from each other, 
making diagonal Franck-Condon transitions between low-lying vibrational states exponentially suppressed. 
As a result, we observe the FCB \cite{Koch.05.PRL,Koch.06.PRB,Leturcq.09.NP,Burzuri.14.NL,Lau.16.NL} at low bias with
strong electron-vibration coupling, $\lambda=3$. 

In Fig. \ref{fig:fig1} (b), we further investigate the current-voltage characteristics while
 varying the molecule-lead hybridization $\Gamma$ with a fixed, moderate electron-vibration coupling $\lambda=0.6$. 
In the high bias limit, all curves approach the same value $\Gamma/2$ \cite{Gurvitz.96.PRB},  
therefore we plot the scaled charge current $J_L/\Gamma$ in Fig. \ref{fig:fig1} (b).  
As can be seen, the step structure only appears in the nonadiabatic transport regime when $\Gamma\ll\omega_b$. 
For $\Gamma=0.1$ eV, level broadening smears out the fine step structure.
When the molecule-lead coupling is increased up to $\Gamma=1$ eV, the metal leads form Ohmic contacts (energy independent
transmission probability) to the molecule. 
As a result, the current exhibits an Ohmic-like behavior, that is, an almost linearly 
increasing trend as a function of voltage.  Altogether, Fig. \ref{fig:fig1} clearly demonstrates 
that the GIOM-PoTER is able to qualitatively capture all the essential features of 
current-voltage characteristics as revealed, for instance, 
by the HEOM method 
\cite{Schinabeck.16.PRB,Schinabeck.18.PRB,Schinabeck2019}.
We note that the HEOM calculations only include a primary vibrational mode, without a secondary bath, therefore 
we do not attempt to perform a direct comparison.


\subsubsection{Cavity-induced suppression of charge current}

As the GIOM builds upon the quantum optical input-output method \cite{Gardiner.85.PRA}, 
it naturally permits to include an additional optical environment that consists modes of an electromagnetic field,
thereby uncover the effect of nano-cavities on charge transfer \cite{Schafer.19.PNAS}. 
A cavity-enhanced charge transfer process has been revealed both experimentally and theoretically 
in mesoscopic molecular-based systems  \cite{Orgiu.15.NM,Hagenmuller.17.PRL,Hagenmuller.18.PRB}.  
Here, we complement previous studies by considering a cavity-coupled single-site 
MJ in the strong light-matter coupling regime. 
Instead of an enhancement, we observe a cavity-induced suppression of the charge current.

The single-site junction is placed within a plasmonic nanocavity supporting a 
single dispersionless electromagnetic mode \cite{Chikkaraddy.16.N,Liu.17.PRL}. 
Consequently, the molecular part $H_M$ in Eq. (\ref{eq:totalH}) is extended to comprise
the cavity photon and a light-matter interaction term,
\begin{equation}\label{eq:cavity}
H_{M+C}=H_M+
\omega_ca^{\dagger}a+\Omega_R(a^{\dagger}+a)d_{1}^{\dagger}d_1.
\end{equation}
Here, a cavity mode with frequency $\omega_c$ is described by the annihilation operator $a$.
The cavity photon directly couples to an electronic excited-state with the vacuum Rabi frequency 
$\Omega_R$ measuring the light-matter coupling strength. We adopt this light-matter interaction form by taking into account the facts that (i) no cavity effect presents when there is no electron on the single-site MJ and (ii) charge conservation should be preserved. 

To mimic experimental conditions \cite{Orgiu.15.NM}, we restrict the problem to cavity-vacuum effects 
and assume that the cavity is prepared in the vacuum state. 
Loss processes from the cavity are included by adding a far-field  
photon bath to the environmental Hamiltonian,
\bea
H_{E+C}= H_E+ \sum_l\omega_lf_l^{\dagger}f_l.
\eea
%
The associated interaction in Eq. (\ref{eq:totalH}) is therefore generalized to
\bea
H_{I+C}=H_I+ i\sum_l\eta_l(a-a^{\dagger})(f_l^{\dagger}+f_l).
\eea
%
The far-field photon modes are described by frequencies $\omega_l$ and annihilation operators $f_l$. 
This coupling is characterized by the spectral density for the far-field photon bath,
 $F(\omega)=\pi\sum_l\eta_l^2\delta(\omega-\omega_l)$. 
The form of light-matter interaction indicates that one can
 diagonalize the molecular part $H_{M+C}$ by
applying the unitary transformation 
$G\equiv(\mathcal{D}_b\mathcal{D}_a)^{d_1^{\dagger}d_1}$ with $\mathcal{D}_a=\exp[\Omega_R(a^{\dagger}-a)/\omega_c]$ 
and  $\mathcal{D}_b$ as defined in  Eq. (\ref{eq:dispb}).
By doing so, the GIOM incorporates nonperturbative (strong-coupling) effects of both
the electron-vibration coupling and light-matter interaction. 
The previously-introduced renormalized on-site electronic energy is now modified 
into $\tilde{\omega}_e=\omega_e-\lambda^2\omega_b-\Omega_R^2/\omega_c$, 
and the vibrational-polariton operator becomes $\tilde{d}_1= \mathcal{D}_b^{\dagger}\mathcal{D}_a^{\dagger}d_{1}$.

By extending the GIOM, as described in Sec. \ref{sec:1},  to include the optical field,
we write down the Heisenberg EOM for the
far-field photon modes, $\dot{f}_l=-i\omega_lf_l+\eta_la$ within the rotation wave approximation. 
An input field $a_{in}(t)\equiv\frac{1}{\sqrt{2\pi}} \sum_l\eta_le^{-i\omega_l(t-t_0)}f_l(t_0)$ 
for the far-field photon bath can then be defined through the relation
\begin{equation}
\label{eq:finals_2}
\sum_l\eta_lf_l(t)~=~\sqrt{2\pi}a_{in}(t)+\kappa a(t),
\end{equation}
where $\kappa\equiv F(\omega_c)$ sets the bare cavity decay rate. 
For plasmonic cavities,  
we take $\kappa=0.05$ eV \cite{Del.18.PRL}. From the initial vacuum state we get the nonvanishing correlation 
of the input field $\langle a_{in}(t)a_{in}^{\dagger}(t')\rangle=
\int d\omega\frac{F(\omega)}{2\pi^2}e^{-i\omega(t-t')}$. 
Taking into account the contribution from the cavity photon, Eq. (\ref{eq:eom_o}) is modified into
\begin{equation}\label{eq:eom_o_a}
\dot{\mathcal{O}}~=~ i[\tilde{H}_M,\mathcal{O}]_{-}-i\sum_v\mathbb{L}_{\pm}^v-i\sum_n\mathbb{X}_n+\mathbb{C}_a,
\end{equation}
where 
\bea
\mathbb{C}_a=\left(\kappa a^{\dagger}+\sqrt{2\pi}a_{in}^{\dagger}\right)[\mathcal{O},a]_{-}-[\mathcal{O},a^{\dagger}]_{-}\left(\kappa a+\sqrt{2\pi}a_{in}\right). 
\nonumber\\
\eea
It is interesting to point out that the contribution from the cavity and the associated photonic environment 
is always incorporated in an additive manner when considering dynamics at the level of operators, in direct contrast to the emergent environmental nonadditivity in density matrix formalism \cite{Maguire.19.PRL}. 

Recall that the charge current of a single-site MJ (\ref{eq:jl1})
is written in terms of the first order transfer rate $\chi_1^v$, as defined in equation (\ref{eq:chinv}).
Generalizing this rate to include the effect of the cavity photon,
we get 
\begin{equation}\label{eq:mchi}
\chi_1^v~=~2\Gamma_v\int\,\frac{d\epsilon}{2\pi}n_F^v(\epsilon)\int_{0}^{\infty}\,d\tau e^{i(\epsilon-\tilde{\omega}_e)\tau}e^{-\Gamma\tau}B(\tau)A(\tau),
\end{equation}
where the cavity photon correlation function $A(t-t')=\langle \mathcal{D}_a^{\dagger}(t)\mathcal{D}_a(t')\rangle$ takes the form (see appendix \ref{a:2} for details),
\begin{equation}
A(\tau)~=~\exp\left[-\frac{\Omega_R^2}{\omega_c^2}\int d\omega\frac{F(\omega)}{\pi}\frac{1-e^{-i\omega\tau}}{\kappa^2+(\omega_c-\omega)^2}\right].
\end{equation}
The exponential form of $A(\tau)$ clearly indicates that the charge current is largely suppressed 
in the strong coupling regime of light-matter interaction.

\begin{figure}[tbh!]
\centering
 \includegraphics[width=1\columnwidth]{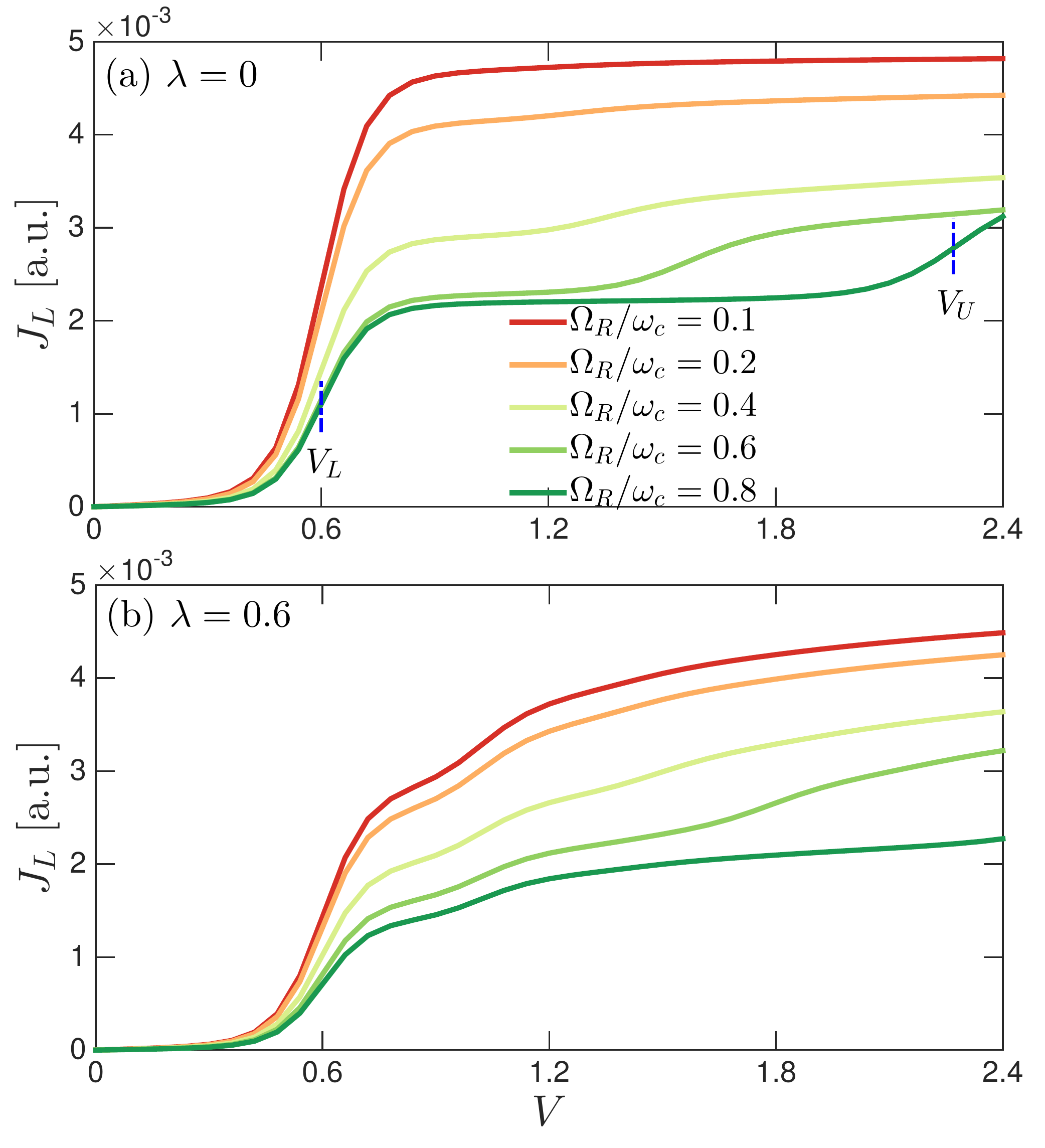} 
\caption{Current-voltage characteristics in cavity-coupled single-site 
MJ, varying the light-matter coupling strengths, 
$\Omega_R/\omega_c$ for (a) $\lambda=0$, and (b) $\lambda=0.6$. 
In (a), $V_{L,U}$ mark two estimated voltage thresholds at which resonant tunneling through the 
polariton states for $\Omega_R/\omega_c=0.8$ begins. 
Other parameters are $\mu_L=-\mu_R=V/2$, $\tilde{\omega}_e=0.3$ eV, $\Gamma=0.01$ eV, $\omega_b=0.2$ eV, $\nu=0.02$ eV, $\kappa=0.05$ eV and $T=300$ K.}
\label{fig:fig2}
\end{figure}

To visualize such a cavity-induced suppression of charge current in single-site junctions, 
we present a representative set of numerical results based on Eqs. (\ref{eq:jl1}) and (\ref{eq:mchi}) 
in Fig. \ref{fig:fig2}. In calculations, we take $\omega_c=\omega_e$, i.e. the cavity mode is made resonant
with the bare electronic level, and
we use a fixed renormalized on-site energy $\tilde{\omega}_e>0$; 
the bare value $\omega_{e,c}$ can be inferred from the relation 
$\omega_{e,c}=(\tilde{\omega}_e+\lambda^2\omega_b)/(1-\alpha^2)$ with the parameter 
$\alpha\equiv\Omega_R/\omega_c$ satisfying $0\le \alpha<1$ (noting that $\alpha$ 
can be larger than 1, which simply implies 
that the renormalized electronic level sits below the Fermi level, namely, $\tilde{\omega}_e<0$). 
A super-Ohmic spectrum $F(\omega)=\kappa\frac{\omega^3}{\omega_{c,cut}^3}e^{-(\omega-\omega_{c,cut})/\omega_{c,cut}}$ with a cut-off frequency $\omega_{c,cut}=1$ eV is adopted for the far-field photon bath \cite{Del.18.PRL}.

%

In Fig. \ref{fig:fig2} (a)  we turn off electron-vibration coupling and study the effect of the cavity mode on charge transport. 
Subsequently, in panel (b) we study the vibrationally-coupled polariton case.
In both scenarios, we observe a cavity-induced suppression of charge current upon 
increasing the light-matter interaction. 

More intriguingly, we notice that a well-resolved step structure emerges
when the system enters into the strong coupling regime of light-matter interaction (see panel (a)).
To pinpoint the origin of these steps, we note that strong light-matter interaction can give rise to 
hybrid states, the so-called polariton states \cite{Sanvitto.16.NM,Ebbesen.16.ACR}. 
In analogy with the case where a cavity mode couples to an excitonic transition in 
molecules \cite{Galego.15.PRX,Herrera.16.PRL,Del.18.PRL}, we expect that two hybrid states that correspond to 
the upper (U) and lower (L) polariton will form. 
However, in contrast to the picture of Refs. \cite{Galego.15.PRX,Herrera.16.PRL,Del.18.PRL}, 
here the cavity photon couples to the electronic excited state in the molecule,
rather than to a transition, implying that the cavity photon actually shifts the 
manifold of excited state.  Hence, 
 the energies of the two polaritons (in the absence of heat baths and vibrations) should be 
$\omega_L=\tilde{\omega}_e$ and $\omega_U=\tilde{\omega}_e+\omega_c$ where $\tilde{\omega}_e$ 
incorporates the vacuum Rabi frequency in the polariton frame.  

Since we apply the bias voltage symmetrically, 
we find the voltage thresholds $V_L=2 \tilde{\omega}_e=0.6$ and $V_U= 2(\tilde{\omega}_e+\omega_c) \approx2.27$ for 
$\Omega_R/\omega_c=0.8$. 
These values are marked by vertical blue dashed-dotted lines in Fig. \ref{fig:fig2} (a). 
As can be seen, these two voltage values well describe the thresholds for the onset of resonant tunneling with
these two polariton states---if one further takes into account level broadening induced by the couplings of the 
polaritons to  the different environments.
Altogether, our results demonstrate that current-voltage characteristics in a cavity-coupled junction 
provide a valuable tool to probe the formation of polariton states in nano-cavities.

We further allow nonzero electron-vibration coupling in Fig. \ref{fig:fig2} (b). 
In this case, the charge current suppression occurs due to the combined effect 
of the primary vibrational modes and the cavity photon, 
as can be inferred from the modified transfer rate in Eq. (\ref{eq:mchi}). 
The interplay between electron-vibration and light-matter interactions generates polariton states as a 
threefold mixture of photonic, electronic and phononic degrees of freedom 
\cite{Cwik.14.EPL,Herrera.17.PRL,Herrera.18.ACSP,Zeb.18.ACSP,Del.18.PRL}, 
leading to the breakdown of the previous simple picture of two branches of polaritons. 
Hence, we observe more sophisticated step structures in Fig. \ref{fig:fig2} (b) as compared with 
Fig. \ref{fig:fig2} (a). 
Nevertheless, our results suggest that a cavity-coupled junction offers a powerful platform for studying 
nonequilibrium polaritonics. 

The suppression effect of the current due to the formation of polaritons
as observed here is not at odds with previous findings in extended systems 
\cite{Orgiu.15.NM,Hagenmuller.17.PRL,Hagenmuller.18.PRB}, where current enhancement at strong light-matter coupling was observed.
Rather, this is a natural consequence of size reduction of the conductor.
In fact,  consider a single cavity mode that couples to a tight-binding system of $N>1$ sites with
the interaction term
$\frac{\Omega_R}{\sqrt{N}}(a^{\dagger}+a)\sum_{n=1}^Nd_{n}^{\dagger}d_n$.
By performing the unitary displacement transformation, as we did before, it
can be shown that the 
cavity induces an effective attractive electron-electron correlation,
 $-(\Omega_R^2/\omega_c)(\sum_nd_{n}^{\dagger}d_n)^2$ (different from a renormalization
$-(\Omega_R^2/\omega_c)\sum_nd_{n}^{\dagger}d_n$ when $N>1$).
We argue that this cavity-induced long-range interaction underlies the {\it cavity-enhanced} charge transfer process
observed in experiments of extended systems \cite{Orgiu.15.NM}.
This phenomenon will be further investigated in future works.

\subsection{Two-site junction}\label{ssec:2}

Continuing our investigation with building blocks of extended systems, 
we focus here on the two-site MJ.
Remarkably, our formalism allows us to derive (we believe for the first time) 
a closed-form expression for the vibrationally-coupled
charge current in the system [cf. Eq.
(\ref{eq:jl2}) below], which depends on the two types of rates, Eqs. (\ref{eq:chinv}) and (\ref{eq:eta_rate}).
The two-site MJ is described by
\begin{equation}
\tilde{H}_e~=~\sum_n\tilde{\omega}_{e,n}d_n^{\dagger}d_n+g(\tilde{d}_1^{\dagger}\tilde{d}_2+\tilde{d}_2^{\dagger}\tilde{d}_1).
\end{equation}
For the sake of simplicity, we proceed with the analytic treatment assuming a symmetric junction,
$\Gamma_L=\Gamma_R=\Gamma/2$. The corresponding drift matrix $\boldsymbol{M}$ takes a 
$2\times2$ form
\begin{equation}
\boldsymbol{M}~=~\left(
\begin{array}{cc}
\Gamma/2+i\tilde{\omega}_{e,1} & ig\mathcal{D}_{1,b}\mathcal{D}_{2,b}^{\dagger}\\
ig\mathcal{D}_{2,b}\mathcal{D}_{1,b}^{\dagger} & \Gamma/2+i\tilde{\omega}_{e,2}
\end{array}
\right).
\label{eq:M2}
\end{equation}
Diagonalizing this matrix yields two eigenvalues
\begin{equation}
\Lambda_{\pm}~=~\frac{\Gamma}{2}+\frac{i}{2}\left[\tilde{\omega}_{e,1}+\tilde{\omega}_{e,2}\pm\sqrt{(\tilde{\omega}_{e,1}-\tilde{\omega}_{e,2})^2+4g^2}\right],
\end{equation}
which are the resonance states of the electronic system,
together with the following explicit forms for the matrices 
$\boldsymbol{R}$ and $\boldsymbol{R}^{-1}$ (see Appendix \ref{a:3} for more details),
\begin{eqnarray}\label{eq:tt}
&&\boldsymbol{R}~=~ \left(
\begin{array}{cc}
\frac{\Lambda_{+}-\Gamma/2-i\tilde{\omega}_{e,2}}{i\sqrt{Z}} & \frac{g}{\sqrt{Z}}\mathcal{D}_{1,b}\mathcal{D}_{2,b}^{\dagger}\\
-\frac{g}{\sqrt{Z}}\mathcal{D}_{2,b}\mathcal{D}_{1,b}^{\dagger} & -\frac{\Lambda_{-}-\Gamma/2-i\tilde{\omega}_{e,1}}{i\sqrt{Z}}
\end{array}
\right),\nonumber\\
&&\boldsymbol{R}^{-1}~=~ \left(
\begin{array}{cc}
1 & \frac{\Lambda_{-}-\Gamma/2-i\tilde{\omega}_{e,2}}{ig}\mathcal{D}_{1,b}\mathcal{D}_{2,b}^{\dagger}\\
\frac{\Lambda_{+}-\Gamma/2-i\tilde{\omega}_{e,1}}{ig}\mathcal{D}_{2,b}\mathcal{D}_{1,b}^{\dagger} & 1
\end{array}
\right).
\nonumber\\
\end{eqnarray}
Here $Z=(\tilde{\omega}_{e,1}-\tilde{\omega}_{e,2})^2+4g^2$. 
Inserting these eigenvalues and matrix elements 
into the general charge current expression, Eq. (\ref{eq:jL}), 
we obtain a closed-form formula for the charge current,  describing the VCET in a two-site MJ,
\begin{equation}
\label{eq:jl2}
J_L~=~\frac{2g^2}{Z}\mathrm{Re}[\chi_+^L+\chi_-^L-\chi_+^R-\chi_-^R+\eta_{+-}^R-\eta_{+-}^L].
\end{equation}
Explicitly,
\begin{widetext}
\bea\label{eq:jl2_d}
J_L &=&
\frac{2g^2\Gamma}{Z} 
\int \frac{d\epsilon}{2\pi}
\left[n_F^L(\epsilon)-n_F^L(\epsilon)\right]
\sum_{n=\pm}\mathrm{Re}\int_{0}^{\infty}\,d\tau e^{i\epsilon\tau}e^{-\Lambda_{n}\tau}B_n(\tau) 
\nonumber\\
&-&
\frac{2g^2\Gamma^2}{Z}
\mathrm{Re}\int\,\frac{d\epsilon}{2\pi}\left[ n_F^L(\epsilon)-n_F^R(\epsilon)\right]\Big(\int_{0}^{\infty}d\tau e^{i\epsilon\tau}e^{-\Lambda_{+}\tau}B_+(\tau)\Big)^{\ast}\Big(\int_{0}^{\infty}d\tau e^{i\epsilon\tau}e^{-\Lambda_{-}\tau}B_-(\tau)\Big).
\eea
\end{widetext}
This analytic result is one of the central results of our work.
We note that the specific matrix forms for $\boldsymbol{R}$ and $\boldsymbol{R}^{-1}$ 
may vary due to certain equalities for eigenvalues (see Eq. (\ref{eq:equality})). 
However, we confirmed  that the above charge current form is independent of such a (normalization) freedom. 
Appendix \ref{a:3} describes the generalization to the $\Gamma_L\neq \Gamma_R$ case.
%

\begin{figure}[tbh!]
\centering
\includegraphics[width=1\columnwidth] {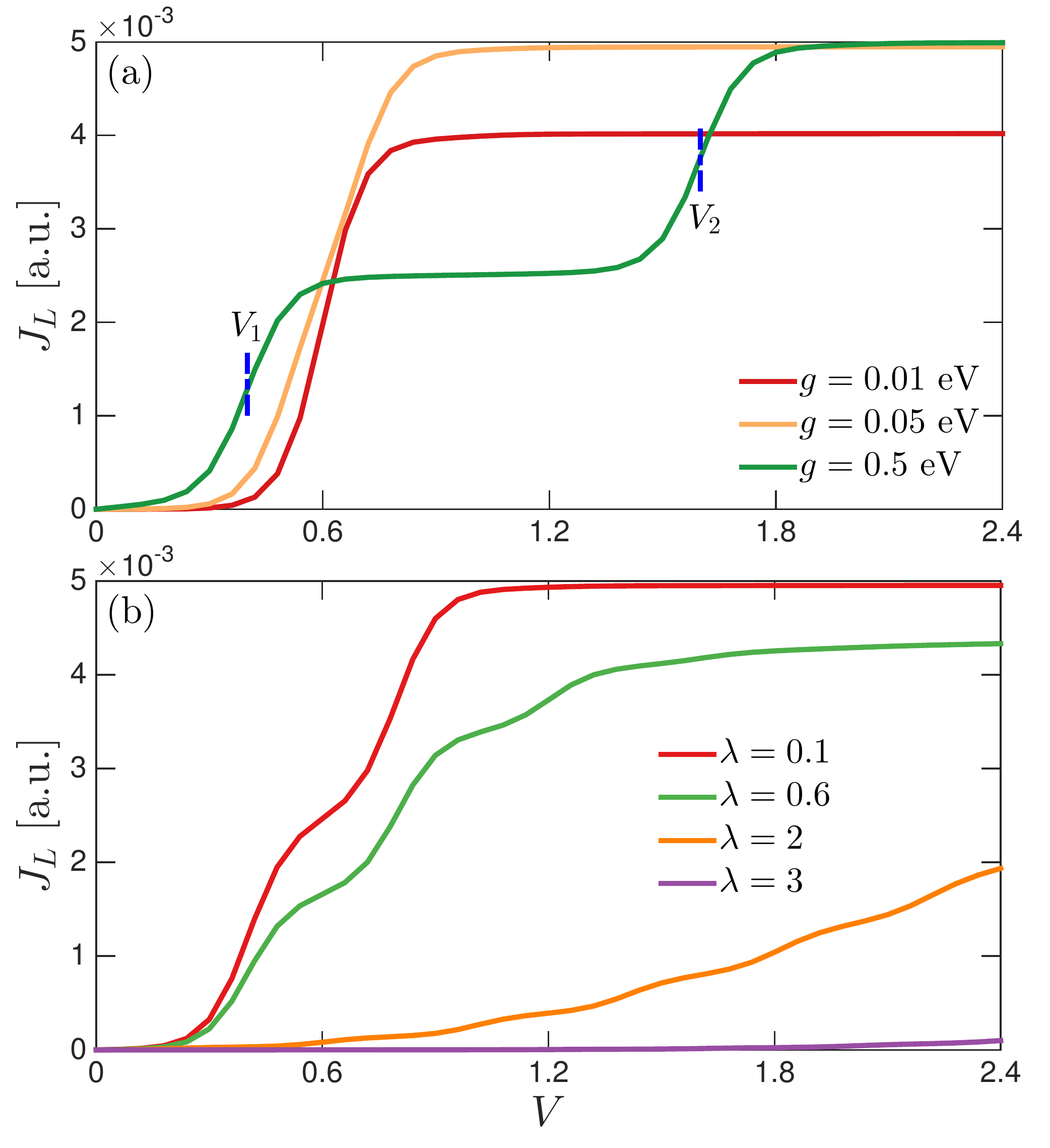} 
\caption{Current-voltage characteristics in two-site junctions for
(a) various hopping amplitude $g$ with a fixed electron-vibration coupling, $\lambda=0.1$ and
(b) different electron-vibration coupling strengths $\lambda$ at a fixed $g=0.1$ eV. 
In panel (a), $V_{1,2}$ mark the values of the onset voltages for resonant tunneling through the 
two electronic states for the case $g=0.5$ eV. 
Model parameters are $\mu_L=-\mu_R=V/2$, 
$\Gamma=0.01$ eV, $\tilde{\omega}_{e}=0.3$ eV, $\omega_{b,1}=\omega_{b,2}=0.2$ eV, $\nu_1=\nu_2=0.005$ eV and $T=300$ K.}
\label{fig:fig3}
\end{figure}

In the coherent limit of $\lambda_{1,2}=0$,  we get 
\bea
&&\mathrm{Re}[\chi_+^v+\chi_-^v-\eta_{+-}^v]=
\nonumber\\
&&\int\,\frac{d\epsilon}{2\pi}n_F^v(\epsilon)\frac{\Gamma^2 Z/2}{|(\epsilon-\omega_{e,1}+i\Gamma/2)(\epsilon-\omega_{e,2}+i\Gamma/2)-g^2|^2}.
\nonumber\\
\eea 
Inserting it into Eq. (\ref{eq:jl2}) yields  exactly the Landauer-B\"uttiker expression 
for noninteracting double quantum dot system \cite{Hartle.13.PRB,Simine.15.JPCC,Bijay.18.PRB},
\begin{equation}
J_{L}^{LB}~=~\int\,\frac{d\epsilon}{2\pi}\frac{\Gamma^2g^2[n_F^L(\epsilon)-n_F^R(\epsilon)]}{|(\epsilon-\omega_{e,1}+i\Gamma/2)(\epsilon-\omega_{e,2}+i\Gamma/2)-g^2|^2}.
\end{equation}
In the high bias limit in which $\mathrm{Re}[\chi_+^L+\chi_-^L-\eta_{+-}^L]=\Gamma(2-\Gamma^2/(\Gamma^2+Z))$ and $\mathrm{Re}[\chi_+^R+\chi_-^R-\eta_{+-}^R]=0$, we get (keeping the lowest order of $\Gamma$)
\begin{equation}\label{eq:jqme1}
J_L^{QME,1}~=~\frac{4\Gamma g^2}{\Gamma^2+(\omega_{e,1}-\omega_{e,2})^2+4g^2}, 
\end{equation}
which is the quantum master equation result obtained by 
\cite{Gurvitz.96.PRB,Gurvitz.98.PRB}. 
Notably, if we neglect the contribution from the second order (two-resonance) term determined by $\eta_{+-}^L$, 
we achieve instead another quantum master equation result \cite{Schaller.09.PRB},
\begin{equation}
J_L^{QME,2}~=~\frac{2\Gamma g^2}{(\omega_{e,1}-\omega_{e,2})^2+4g^2}.
\end{equation}
However, for degenerate on-site energies $\omega_{e,1}=\omega_{e,2}$, 
this latter charge current expression becomes independent on the hopping rate $g$, 
implying nonzero current in the limit of $g\to0$. 
This unphysical artifact highlights the importance of keeping second order terms, even in the weak coupling regime.

\begin{figure}[htb]
\centering
\includegraphics[width=1\columnwidth] {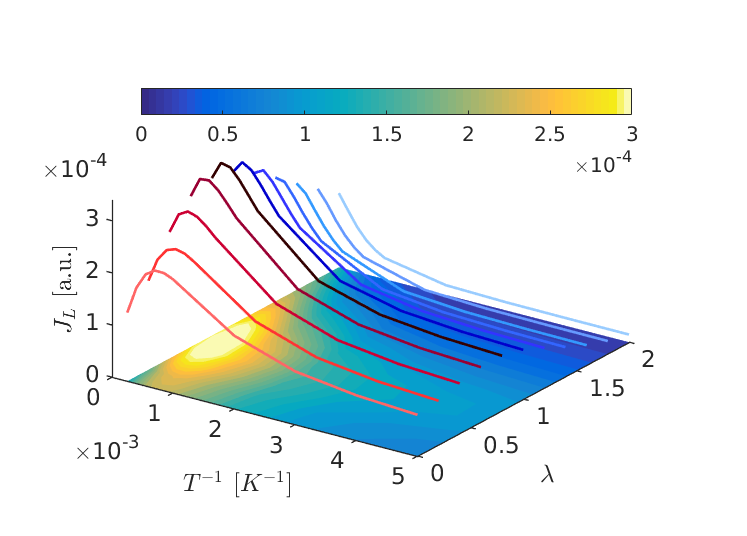} 
\caption{Contour map of charge current for a two-site MJ as a function of electron-vibration couplings 
$\lambda$ and inverse temperature $T^{-1}$.
Representative curves with fixed $\lambda$ are depicted.
Model parameters are $\mu_L=-\mu_R=V/2$, $g=0.05$ eV, $V=0.01$, 
$\Gamma=0.2$ eV, $\tilde{\omega}_{e}=0.2$ eV, $\omega_{b,1}=\omega_{b,2}=0.2$ eV, and $\nu_1=\nu_2=0.005$ eV.}
\label{fig:fig4}
\end{figure}

We now explore different limits of the current, Eq. (\ref{eq:jl2_d}), based on numerical simulations.
Representative numerical results are summarized in  Figs. \ref{fig:fig3} and \ref{fig:fig4}. 
Without loss of generality, 
we study a dimer made of identical building blocks with identical electron-vibration coupling, 
$\lambda_1=\lambda_2=\lambda$, and equal (dressed) site energies
 $\tilde{\omega}_{e,1}=\tilde{\omega}_{e,2}=\tilde{\omega}_e$. 
A super-ohmic spectrum $I_n(\omega)=\nu_n\frac{\omega^3}{\omega_{b,cut}^3}e^{-(\omega-\omega_{b,cut})/\omega_{b,cut}}$ 
with a cut-off frequency $\omega_{b,cut}=0.5$ eV is assumed for the secondary vibrational baths. 
The asymmetric two-site case 
can realize the Aviram-Ratner donor-acceptor charge rectifier \cite{Aviram.74.CPL} and its recent extensions, e.g.
\cite{Zantdiode,Kilgourdiode}. This setup, a  primary application of molecular transport junctions, is not explored in this work.

\subsubsection{Current-voltage characteristics}

In Fig. \ref{fig:fig3} (a) and (b), we display
the current-voltage characteristics for  
various hopping amplitude $g$ and electron-vibration coupling $\lambda$, respectively. 
We first focus on the effect of the hopping rate $g$ in Fig. \ref{fig:fig3} (a). 
Obviously, the charge current should vanish in the limit of $g\to0$, and
increasing $g$ from zero should favor charge transport. Indeed, we observe an enhancement of the current when 
$g$ is raised from $0.01$ eV to $0.05$ eV.
However, for larger values of $g$, different trends are observed.
Particularly, at high voltage
a two-step structure forms before reaching saturation, and the current is no longer monotonously growing in $g$.

Since the electron-vibration interaction and the molecule-lead couplings are relatively weak in this example, 
we analyze the observed behavior from the perspective of the closed electronic system. 
In this case, the eigenfrequencies for the two eigenstates are $\omega_e\pm g$. 
When $g$ is small (relative to $\omega_e-\epsilon_F$), 
these two states are almost degenerate (given level broadening),
and the onset of resonant tunneling occurs just around $V=0.6$ (noting $\omega_e=0.3$ eV) as confirmed by the result 
for $g=0.01$ eV in Fig. \ref{fig:fig3} (a). 
However, once $g$ is large enough such that the two eigenstates are well separated, we observe  the
 two-step structure in current-voltage curve
with the corresponding voltage thresholds, $V_{1,2}=2|\omega_e\pm g|$;
the absolute value is a direct consequence of the symmetric voltage drop applied. 
For instance, $\omega_e$ and $-\omega_e$ have the same voltage thresholds for resonant transport.


In Fig. \ref{fig:fig3} (b), we vary the electron-vibration coupling up  
to a moderate value, while fixing $g$.
Similarly to the single-site MJ described in Sec. \ref{ssec:1},
we observe steps in the current, corresponding to Franck-Condon transitions. 
Meanwhile, the charge current is suppressed for large $\lambda$ due to the FCB.  

\begin{figure*}[tbh!]
  \centering
  \includegraphics[width=2\columnwidth]{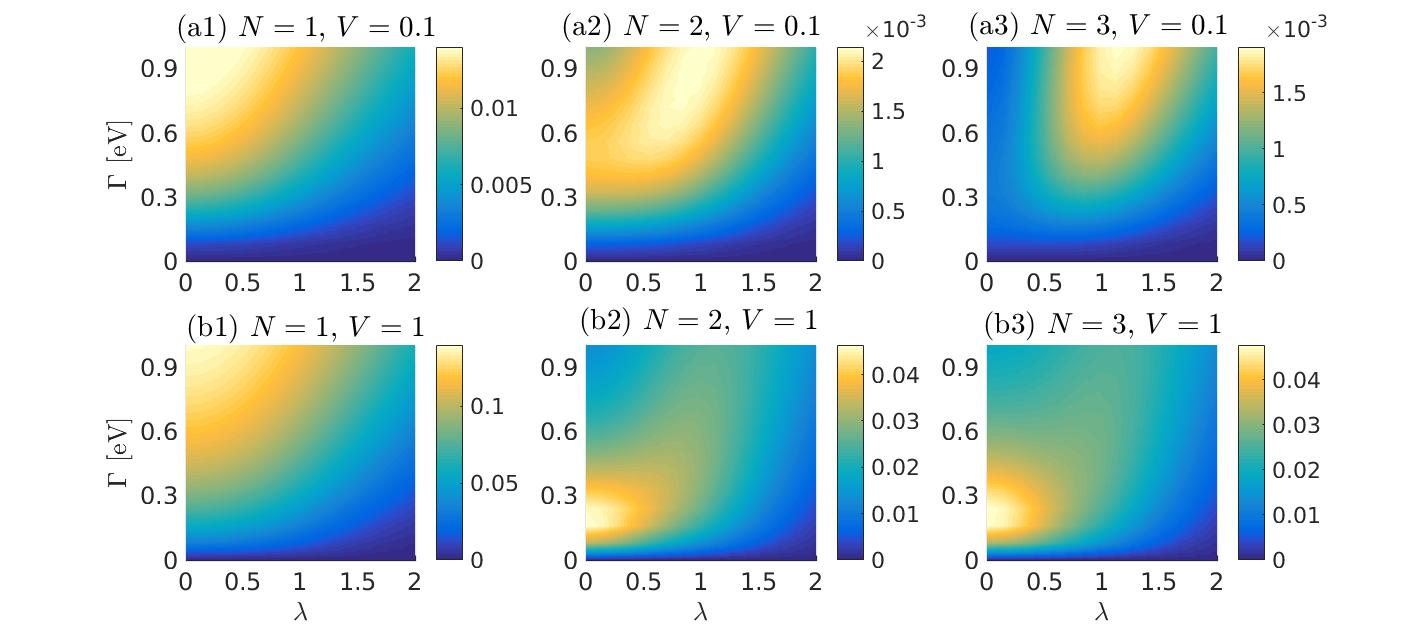} 
\caption{Contour map of charge current as a function of electron-vibration coupling $\lambda$ and
metal-molecule hybridization $\Gamma$
for (a1)-(b1) $N=1$-site junction at low and high voltage, and similarly for
(a2)-(b2) $N=2$-site junction 
and (a3)-(b3) $N=3$-site junction. 
Model parameters are $\mu_L=-\mu_R=V/2$,
$\tilde{\omega}_e$=0.3 eV, $\omega_b$=0.2 eV, $\nu$=0.005 eV, $T$=300 K,  and $g=0.1$ eV for
the multi-site cases.
}
\label{fig:fig5}
\end{figure*}

\subsubsection{Thermal effects}

Thermal effects in molecular transport junction may realize different transport mechanisms
\cite{Segal.00.JPCB,Selzer.04.JACS,Selzer.04.NN,Selzer.05.NL,Weiss.05.JACS,Selzer.06.ARPC,Poot.06.NL,Goldsmith.08.JPCA,Hines.10.JACS,Sedghi.11.NN,Luo.11.CM,Lee.12.ACSN,Kilgour.15.JCP,Garrigues.16.DT,Garrigues.16.SR,Kim.17.JCP}.
Arrhenius-type behavior of the charge current with the ambient temperature has been often considered as a guidepost
for the sequential hopping mechanism. 
Nevertheless, in a metal-molecule-metal geometry
charge current may strongly depend on temperature even in the quantum coherent limit given the broadening of the electron occupation
functions (Fermi distribution) at nonzero temperature \cite{Kilgour.15.JCP}. Thermal electrons 
contribute to the current a resonant-ballistic component,
which is often difficult to separate from the contribution of incoherent scattering effects.

Inspecting the transfer rates defined in Eqs. (\ref{eq:chinv}) and (\ref{eq:eta_rate}), 
we note that temperature enters the charge current through (i) the broadening of the 
Fermi distribution of electrons at the leads 
and (ii) the impact on vibrations, encoded in the vibrational correlation functions. 
To gauge thermal effects, in Fig. \ref{fig:fig4} we study the behavior of the charge current as a function of
temperature and electron-vibration coupling energy.
To exclude the effect of energy renormalization, induced by nonzero electron-vibration couplings, 
we fix $\tilde{\omega}_e$ and consider the off-resonant transport regime setting $\tilde{\omega}_e\gg V$. 

Based on the different curves in Fig. \ref{fig:fig4}, we identify the following trends:
(i) At low temperature, the coherent, deep tunneling (superexchange) mechanism dominates the charge transfer process, 
showing up as a temperature-independent charge current. 
(ii) As the temperature increases, the Fermi distributions broaden. 
When the exponential tail of the Fermi distribution of the left lead (source electrode) sufficiently overlaps with the molecular electronic states, 
the MJ enters into a thermally activated coherent resonant transport (ballistic) regime and the 
charge current increases with temperature showing an Arrhenius-like law. 
(iii) If one further increases the temperature, 
the exponential tail of the Fermi function at the {\it right} lead (drain electrode) starts to overlap with the molecular electronic levels,
the number of unoccupied electronic states that are accessible in the right lead (drain) decreases, and the charge current is
suppressed \cite{Poot.06.NL}. 
(iv) The charge current is a non-monotonic function of $\lambda$ 
over the whole range of temperatures, indicating  on a phonon-assisted transport behavior in the off-resonant 
tunneling regime. This should be contrasted with 
the monotonic tendency in the resonant tunneling regime revealed in Fig. \ref{fig:fig3} (b), in which the current was systematically suppressed with 
an increasing $\lambda$.

Altogether, these results  expose that the coupling of electrons to local vibrations may lead to either an
the enhancement of the charge current or to its suppression relative to the $\lambda=0$, rigid-molecule case. 
We elaborate on this effect in the discussion of Fig. \ref{fig:fig5}.

\subsubsection{Vibrational-enhancement and suppression of charge current}

The impact of vibrational effects on charge current is expected to display distinct signatures in the resonant and off-resonant transport regimes.
In the off-resonant case, molecular electronic levels are well-separated from the Fermi energy of the leads.
Beginning from such a scenario with $\tilde \omega_e-\epsilon_F >T$, the resonant regime can be approached
by raising the bias voltage, increasing the metal-molecule hybridization energy and increasing temperature.

In  Fig. \ref{fig:fig5}, we study the behavior of the charge current as a function  of electron-vibration coupling $\lambda$ and the metal-molecule hybridization $\Gamma$ at low (top panels) and high (bottom panels) voltage.
Furthermore, we depict results for single-site, two-site, and three-site MJs 
(the latter is simulated by combining a numerical diagonalization of matrix $\boldsymbol{M}_0$ and Eq. (\ref{eq:js})).
%
These results highlight a critical difference between single-site and multi-site junctions.
In the former case, a phonon-assisted transport behavior is clearly absent across all parameters, at low and high voltage,
 as can be seen from Fig. \ref{fig:fig5} (a1), (b1). 
In contrast, the two-site and three-site models display nonlinear behaviors with a crossover behavior of current as a function $\Gamma$ and $
\lambda$.

We first focus on single-site MJs. In this case, the charge current builds only upon first order transfer rates, Eq. (\ref{eq:rate_site1}).
The observed monotonic behavior as a function of both parameters, $\Gamma$ and $\lambda$, 
can be fully attributed to the properties of the first order transfer rate:
For a fixed $\Gamma$, the rate is systematically suppressed by increasing $\lambda$ due to the exponential form of the 
vibrational correlation function as given in Eq. (\ref{eq:corr_bb}). 
Fixing $\lambda$ and increasing $\Gamma$ enhances the transfer rate. 
Particularly, large $\Gamma$ hinders the suppression effect of electron-vibration coupling, as the exponential function in the
integrand $e^{i(\epsilon-\tilde{\omega}_e)\tau}e^{-\Gamma\tau}$ resembles the delta function $\delta(\tau)$ in the large $\Gamma$ limit. 
Hence, charge transport is maximal for large $\Gamma$ and vanishing $\lambda$. 
We also show that the bias voltage only change the magnitude of the charge current, by comparing Fig. \ref{fig:fig5} (a1) with (b1).

In two-site junctions, charge transport is dictated by the interplay between the first order transfer rates and the second order counterparts, 
defined in Eq. (\ref{eq:eta_rate}). 
First order rate have the $\Gamma$ as a prefactor, along with a decay function, which depends on $\lambda$.
In contrast, the prefactor in second order rates is quadratic in  $\Gamma$, 
and the rate has a more pronounced decay with $\lambda$.
Combined with the fact that
second order rates negatively contribute to the current 
[see the prefactor of $n_F^L(\epsilon)-n_F^R(\epsilon)$ in Eq. (\ref{eq:jl2_d})],
the current may go through an extremum point with respect to $\lambda$ and $\Gamma$, as opposed to the single-site case.
This observation, rooted in the analytic structure of the current, is one of the main results of this work.


Comparing Figs. \ref{fig:fig5} (a2) with (b2), we find that such a crossover persists in both voltage regimes.
However,  in the resonant tunneling regime (at high voltage)
the optimal transport region appears at weak $\Gamma$ and vanishing $\lambda$.
At low voltage, Fig. \ref{fig:fig5} (a2), the charge current is high at large $\Gamma$ as the lower electronic eigenstate can become a resonant channel for electrons. 
At high voltage,  Fig. \ref{fig:fig5} (b2), both states are already in resonance with the left lead, even for small $\Gamma$.
An interesting outcome of Eq. (\ref{eq:jqme1}), which is valid at high voltage and vanishing electron-vibration coupling,
 is that the optimal charge transport occurs at $\Gamma_{op}=2g=0.2$ eV. This value  agrees very well with the results shown in 
Fig. \ref{fig:fig5} (b2). 

Three-site MJs involve more electronic resonances, as well as additional combinations for second order transfer rates,
but, the qualitative picture resembles the  two-site MJs, as can be seen from by comparing Fig. \ref{fig:fig5} (a2) and (a3) to 
Fig. \ref{fig:fig5} (b2) and (b3). This observation further confirms that the competition between the two types of rates 
underlies the nonlinear vibrational effects. 
Therefore, we conclude that the observed enhancement-suppression crossover of charge current as a function of $\lambda$ in two-site MJs 
(and possibly in more extended systems based on the resemblance between results for $N=2$ and $N=3$) is rooted 
in the competition between the two kinds of transfer rates.
 More intriguingly, our results indicate that charge transport in extended systems can be optimized by coordinating the interaction of 
electrons with different degrees of freedom, vibrational and electronic.

\begin{figure*}[tbh!]
\centering
\includegraphics[width=2\columnwidth] {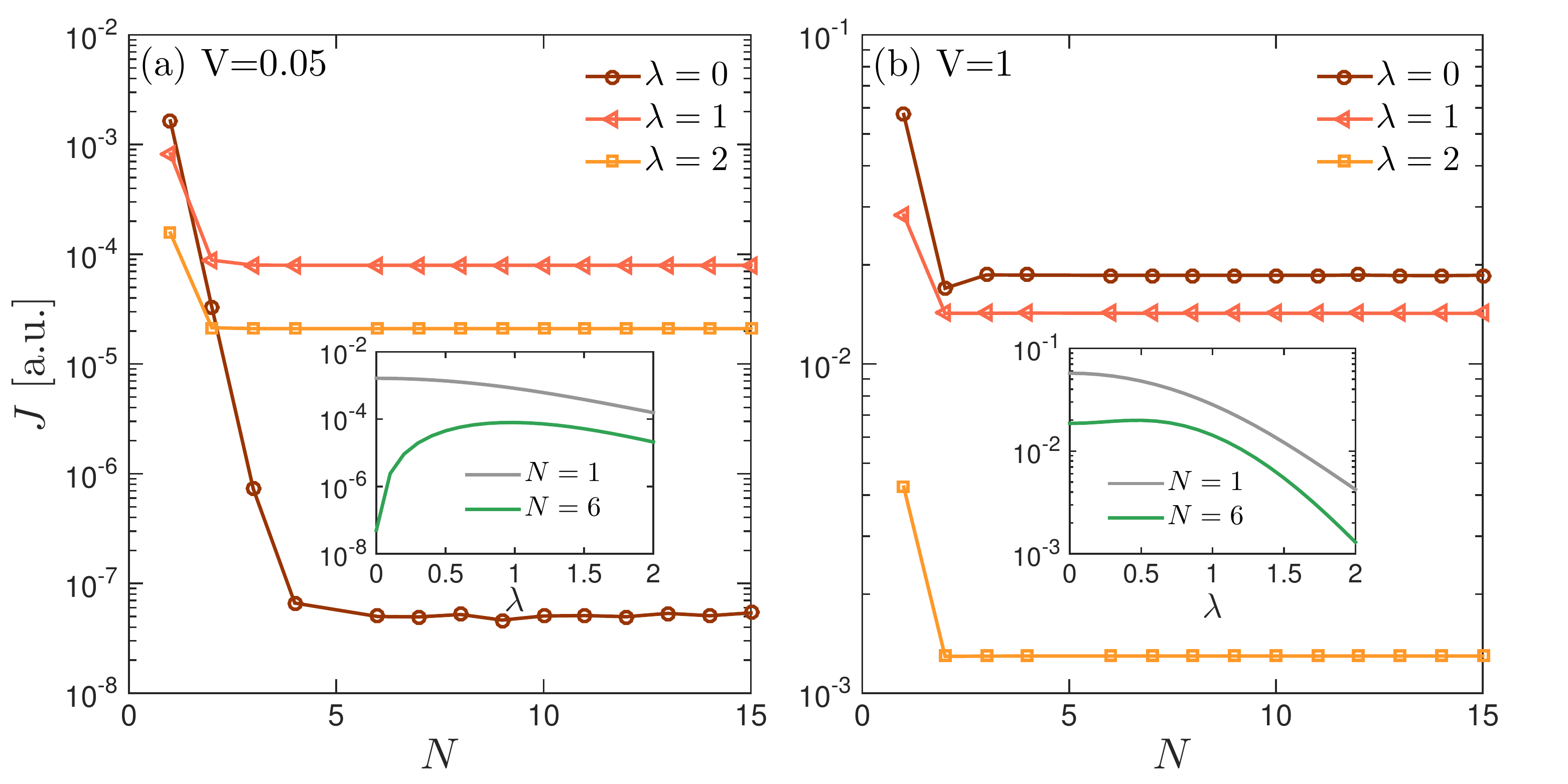} 
\caption{Charge current in an $N$-site junction as a function of size at
(a) low and (b) high bias voltage,  illustrating the crossover of the transport mechanism from off-resonant 
tunneling to resonant conduction.
Model parameters are $\mu_L=-\mu_R=V/2$, $g=0.05$ eV, 
$\Gamma=0.2$ eV, $\tilde{\omega}_{e}=0.4$ eV, $\omega_{b}=0.2$ eV, and $\nu=0.01$ eV, 
$T=300$ K.
We further present in the small panels the current with respect to the electron-vibration coupling, displaying a turnover behavior for $N>1$. 
}
\label{fig:fig6}
\end{figure*}

\subsection{$N$-site junction}\label{ssec:3}

What are the signatures of electron-vibration coupling on charge transfer mechanism at the nanoscale?
In the off-resonant case, when the voltage is small and molecular electronic levels are placed outside the bias-voltage window, transport is dominated by quantum tunneling.
In extended but ordered systems, transport is ballistic in the quantum resonant limit.
Inelastic scatterings of electrons are responsible for the development of 
an ohmic-like electrical conductance \cite{Kilgour.15.JCP}. 
These transport mechanisms, and the crossover behavior between them, have been experimentally observed 
in, e.g., conjugated molecular wires \cite{Selzer.04.NN,Selzer.04.JACS,Weiss.05.JACS,Selzer.06.ARPC,Choi.S.08,Goldsmith.08.JPCA,Lu.09.ACSN,Luo.10.JACS,Choi.10.JACS,Hines.10.JACS,Luo.11.CM,Sedghi.11.NN,Li.12.NL,Taherinia.16.ACSN}. 

We now ask the following basic question: What is the expected length dependence of charge current in 
 an extended tight-binding one-dimensional system with local electron-vibration coupling?
While many descriptions of ohmic conduction
are based on the phenomenological introduction of dissipative dynamics for electrons, here we ask this question 
with regards to a first-principle modelling of electron-vibration coupling.
In fact,  methods that have been used so far to address this question involve major simplifications,
such as QME \cite{Segal.00.JPCB} or the phenomenological B\"uttiker's probe method \cite{Kilgour.15.JCP}.
Aside from the exact solution, what is the prediction of the GIOM-PoTER for the current in extended systems?

We recall that the GIOM-PoTER scheme involves several critical assumptions on the electron-vibration coupling: 
the dynamics of electrons and vibrations (within the polaron picture) are decoupled,
vibrational corrections to electronic resonances are disregarded, and nonlocal vibrational correlations are ignored. 
The resulting picture is of polarons moving coherently across the junction. 
To what extent then can the GIOM-PoTER scheme capture different mechanisms (deep tunneling, ballistic transport, ohmic behavior), and  the crossover between them?

To this end, we adopt Eq. (\ref{eq:jL}) derived for tight-binding models,
and investigate the dependence of the charge current on the number of electronic sites $N$. 
In what follows, we consider molecular wires made of identical units,
$\{g_n\}=g$, $\{\omega_{e,n}\}=\omega_e$, $\{\omega_{b,n}\}=\omega_b$ and $\{\nu_n\}=\nu$. 
We also choose symmetric hybridization $\Gamma_L=\Gamma_R=\Gamma/2$ in simulations. 
The complex {\it c}-numbers $s_{nm}$ and $\tilde{s}_{nm}$ are obtained by diagonalizing the matrix $\boldsymbol{M}_0$, 
as illustrated in Sec. \ref{subsec:21}.

The length dependence of the current in the off-resonant (low voltage) and resonant (high voltage) tunneling regimes
are depicted in Fig. \ref{fig:fig6} (a) and (b), respectively. 
The results with different electron-vibration coupling strengths clearly illustrate the crossover of the 
transport mechanism from deep tunneling for short junctions to ballistic (distance independent) transport for long chains.
This latter mechanism corresponds to the contribution of resonant electrons, arriving at
the tail of the Fermi functions, in resonance with molecular electronic levels.
Nevertheless, given the PoTER approximations, 
the method does not capture the development of ohmic (hopping) conduction for long chains.

We further display in Fig. \ref{fig:fig6} (insets) the behavior of the current as a function of 
electron-vibration coupling strength $\lambda$ for $N=1, 6$,  complementing 
results in Fig. \ref{fig:fig5}. 
As we discussed above, for $N=1$ the current is monotonic---always suppressed---with increasing coupling $\lambda$, regardless of the voltage value. 
In contrast, in long chains we resolve a crossover behavior with $\lambda$, which confirms our expectation that this crossover behavior
is a  generic feature of extended systems. 

Altogether, while the current for e.g. $N=6$ sites clearly displays the involvement of vibrations in 
the charge transport process, an ohmic (dissipative) behavior is not developed under the PoTER scheme.
Junction with 1-20 sites are extremely important for understanding electronic conducting in nanoscale systems, yet
there is a lack of methods 
that could tackle this problem from first principles.
It is our hope that results presented in this section
will inspire and trigger developments of accurate and feasible methods for treating VCET in extended systems.

\section{Summary}
\label{sec:4}

\subsection{Discussion}
We presented an original framework for studying electronic current in vibrationally-coupled 
molecular junctions,  an alternative approach to standard (often costly) 
nonequilibrium Green's function and perturbative quantum master equation methods.
There are four main contributions to this paper:

(i) The first achievement lies in the generalization of the quantum optical input-output method to treat
molecular electronic problems culminating with the formally-exact equation of motion for electronic operators [cf. Eq. (\ref{eq:m1})].
A central advantage of our generalized input-output method (GIOM) lies in its transparency:
The formal steady state solution (at the level of operators), Eq. (\ref{eq:exact}),
clearly displays the effect of electron-vibration couplings on electron transport:
This interaction leads to the formation of polarons that
are transported through vibrationally-modified electronic resonances.

(ii) We devise an approximation scheme to the GIOM, 
that is, the ``polaron transport in electronic resonance" (PoTER) solution,
 Eq. (\ref{eq:ddd}), which describes polaron transport through purely-electronic resonances. 
Using this approach, we avoid perturbation expansions of parameters, achieve computationally
manageable closed-form expressions for the charge current, and explore challenging parameter regimes. 
Notably, in the GIOM scheme,
the charge current expression for a generic tight-binding model involves only two types of transfer rates, which are
easily computed under the PoTER approximation.

(iii) The GIOM framework exposes that the single-site vibrationally-coupled electron transfer model, though involved and rich,
is in fact deceptively simple  
 relative to the two-site model (and naturally beyond). This is because the single-site
case does not necessitate the main part of the PoTER approximation, 
which concerns correcting electronic resonances by local and nonlocal vibrational effects. 
To the most part, numerical methods developed to study VCET effects in MJs
have been benchmarked only on the single-site model. 
We argue that approximate methods that correctly reproduce vibrational effects for
the single site junction may be still defective, missing fundamental VCET effects. 
Future development efforts should therefore focus on both models.

(iv) Using the GIOM-PoTER framework we were able to obtain improved results or new predictions
regarding VCET across MJs. 
Signatures of such effects were distinctively observed in the current-voltage characteristics as steps in the current,
the Franck-Condon current blockade at low voltage with strong electron-vibration couplings, 
and current suppression or enhancement as a function of the electron-vibration couplings. 
Specifically, we showed that in multi-site ($N>1$) chains 
there is a crossover behavior of the current with electron-vibration coupling and metal-molecule hybridization
as opposed to the single-site case. 
Furthermore, cooperative metal-electron-vibration effects allow to optimize electronic performance.
The temperature and length
dependence of the charge current indicate on the role of thermal effects in transport. Nevertheless,  
discerning the specific
roles of vibrational modes (as opposed to electronic thermal effects) require further advancements.

%




\subsection{Outlook}

Realizing the tunneling-to-hopping crossover in the electrical conductance
of long molecular wires 
\cite{Engelkes.04.JACS,Choi.S.08,Lu.09.ACSN,Tuccitto.09.NM,Lafferentz.09.S,Luo.10.JACS,Choi.10.JACS,Luo.11.CM,Li.12.NL,Taherinia.16.ACSN} based on a explicit microscopic, first-principle treatment of VCET remains a fundamental, difficult challenge.
In fact, ohmic-like conduction has not been conclusively realized in single-molecule 
junctions (see e.g. \cite{Hines.10.JACS,Tao16});
typical conductance measurements convincingly displaying this transition have been performed 
on a self-assembled monolayer (SAM)
including over $100$ parallel aligned conjugated molecular wires 
\cite{Engelkes.04.JACS,Choi.S.08,Choi.10.JACS}, with the resulting normalized current reflecting 
an averaged behavior of the monolayer.
It remains an open question as to whether such an ensemble functions as a collection of 
independent parallel wires, or whether cooperative effects between chains determine charge transport through a SAM.
Indeed, experiments \cite{Salomon.03.AM,Selzer.05.NL} indicate that there may be 
a significant difference in the conduction of isolated individual molecules and the behavior of the SAM.
Furthermore, the role of electronic disorder on the development of hopping conduction in MJs is still unclear.
%
In order to address these questions, the GIOM-PoTER scheme needs to be further advanced in two ways:
(i) The Kernel (\ref{eq:K}) should accommodate local dissipation effects
by including the contribution of vibrations to the electronic resonances.
As a basic improvement, this could be done at a mean-field level.
More sophisticated approaches would rely on self-consistently
solving for the electronic and nuclear degrees of freedom to improve the PoTER Kernel.
(ii) The method should be generalized to treat 
more complex models, specifically to go beyond the present one-dimensional tight-binding model. 


Beyond the calculation of charge current,
shot noise, the second moment of the
electron transport probability distribution in steady state, provides 
additional, fundamental information on the transport behavior. 
Signatures of vibrational effects show up in careful measurements, see for example \cite{Tal08, Kumar12}, 
but there is still discrepancy between experiments and theoretical predictions 
\cite{Galperin.06.PRB,Yeyati09, Haupt09, Haupt10,Haupt11}. 
The GIOM-PoTER method, which directly constructs electronic operators,
could be advanced to provide the frequency-dependent noise signal in MJs, 
and bring concrete predictions on signatures of VCET in the current noise.


Finally, the GIOM framework is naturally fitting to study electron transport in the context of
molecular polaritonics--molecules placed in optical cavities leading to hybridized light-matter states.
The cavity can cultivate effective attractive interactions between electrons, thus realizing new transport behavior
in MJs. Work in this direction is in progress.

\begin{acknowledgments}
The authors acknowledge support from the Natural Sciences and Engineering Research Council (NSERC) of Canada Discovery Grant
and the Canada Research Chairs Program.
\end{acknowledgments}

\appendix
\renewcommand{\theequation}{A\arabic{equation}}
\setcounter{equation}{0}  
\section{Statistics of input fields and input-output relations}
\label{a:1}

From the definitions of the input fields we easily identify the following commutation/anticommutation relations
\begin{subequations}
\begin{align}
[b_{in}^n(t),b_{in}^{m,\dagger}(t')] &= \delta_{nm}\int d\omega\frac{I_n(\omega)}{2\pi^2}e^{-i\omega(t-t')},\label{eq:a1}\\
\{d_{in}^v(t),d_{in}^{v',\dagger}(t')\} &= \delta_{vv'}\Gamma_v\int \frac{d\epsilon}{2\pi^2}e^{-i\epsilon(t-t')},
\end{align}
\end{subequations}
from which we obtain the following correlation functions for input fields 
\begin{eqnarray}
\langle b_{in}^{n,\dagger}(t')b_{in}^m(t)\rangle &=& \delta_{nm}\int d\omega\frac{I_n(\omega)}{2\pi^2}e^{-i\omega(t-t')}n_B(\omega),\nonumber\\
\langle b_{in}^n(t)b_{in}^{m,\dagger}(t')\rangle &=& \delta_{nm}\int d\omega\frac{I_n(\omega)}{2\pi^2}e^{-i\omega(t-t')}\left[1+n_B(\omega)\right],\nonumber\\
\langle d_{in}^{v,\dagger}(t')d_{in}^{v'}(t)\rangle &=& \delta_{vv'}\Gamma_v\int \frac{d\epsilon}{2\pi^2}e^{-i\epsilon(t-t')}n_F^v(\epsilon),\nonumber\\
\langle d_{in}^v(t)d_{in}^{v',\dagger}(t')\rangle &=& \delta_{vv'}\Gamma_v\int \frac{d\epsilon}{2\pi^2}e^{-i\epsilon(t-t')}\left[1-n_F^v(\epsilon)\right].
\end{eqnarray}
Here $n_B(\omega)=\{\exp[\omega/T]-1\}^{-1}$ and $n_F^v(\epsilon)=\{\exp[(\epsilon-\mu_v)/T]+1\}^{-1}$ are the Bose-Einstein distribution and the Dirac-Fermi distribution, respectively.

To define output fields, we express the formal solution of Eqs. (\ref{eq:eom1}) and (\ref{eq:eom3}) in terms of the final conditions $r_j(t_1)$ and $c_{kv}(t_1)$ at a later time $t_1>t$,
\begin{subequations}
\begin{align}
r_{n,j}(t) &= e^{-i\omega_{n,j}(t-t_1)}r_{n,j}(t_1)+i\gamma_{n,j}\int_{t}^{t_1}\,e^{-i\omega_{n,j}(t-\tau)}b_n(\tau)d\tau,\\
c_{kv}(t) &= e^{-i\epsilon_{kv}(t-t_1)}c_{kv}(t_1)+it_{kv}\int^{t_1}_t\,e^{-i\epsilon_{kv}(t-\tau)}\tilde{d}_{\sigma}(\tau)d\tau.
\end{align}
\end{subequations}
Introducing output fields
\begin{subequations}
\begin{align}
b_{out}^n(t) &\equiv \frac{1}{\sqrt{2\pi}} \sum_j\gamma_{n,j}e^{-i\omega_{n,j}(t-t_1)}r_{n,j}(t_1),\\
d_{out}^v(t) &\equiv \frac{1}{\sqrt{2\pi}} \sum_kt_{kv}e^{-i\epsilon_{kv}(t-t_1)}c_{kv}(t_1),
\end{align}
\end{subequations}
we  obtain the input-output relations,
\begin{subequations}
\begin{align}
& b_{out}^n(t)-b_{in}^n(t)~=~-i\sqrt{\frac{2}{\pi}}\nu_nb_n(t),\\
& d_{out}^v(t)-d_{in}^v(t)~=~-i\sqrt{\frac{2}{\pi}}\Gamma_v\tilde{d}_{\sigma}(t).
\end{align}
\end{subequations}

\renewcommand{\theequation}{B\arabic{equation}}
\setcounter{equation}{0}  
\section{Charge conservation}
\label{a:4}

The total charge should be conserved in the junction, that is in the molecule + metals.
Given the approximate solution, Eq. (\ref{eq:ddd}), its validity in this respect requires a careful examination. 
To this end, we focus on the time derivative of the total number operator in the system,
$\sum_nd_n^{\dagger}d_n$, 
which is just $\boldsymbol{\dot{d}}^{\dagger}\boldsymbol{d}+\boldsymbol{d}^{\dagger}\boldsymbol{\dot{d}}$ in a matrix form. 
From Eq. (\ref{eq:ddd}), we find that 
\bea
\boldsymbol{\dot{d}}(t)&=&-i\sqrt{2\pi}\boldsymbol{R}^{-1}(t)
[-\boldsymbol{\Lambda}_M\int_{-\infty}^td\tau e^{-\boldsymbol{\Lambda}_M(t-\tau)}\boldsymbol{R}(\tau)\boldsymbol{\tilde d}_{in}(\tau)
\nonumber\\
&+&\boldsymbol{R}(t)\boldsymbol{\tilde d}_{in}(t)]=-\boldsymbol{M}(t)\cdot\boldsymbol{d}(t)-i\sqrt{2\pi}\boldsymbol{\tilde d}_{in}(t),
\eea
by noting that $\boldsymbol{M}(t)=\boldsymbol{R}^{-1}(t)\cdot\boldsymbol{\Lambda}_M\cdot\boldsymbol{R}(t)$. Hence, we have
\begin{eqnarray}
\label{eq:sum_dd}
\frac{d}{dt}\sum_nd_n^{\dagger}d_n 
&=& 
-\boldsymbol{d}^{\dagger}\cdot(\boldsymbol{M}+\boldsymbol{M}^{\dagger})\cdot\boldsymbol{d}
-i\sqrt{2\pi}\boldsymbol{d}^{\dagger}\cdot\boldsymbol{\tilde d}_{in}
\nonumber\\
&&+i\sqrt{2\pi}\boldsymbol{\tilde d}_{in}^{\dagger}\cdot\boldsymbol{d}.
\end{eqnarray}
The right-hand-side of the above equation is compensated 
by $d(\sum_{kv}c_{kv}^{\dagger}c_{kv})/dt=\sum_v(i\sqrt{2\pi}\tilde{d}_{\sigma}^{\dagger}d_{in}^v-i\sqrt{2\pi}d_{in}^{v,\dagger}\tilde{d}_{\sigma}+2\Gamma_vd_{\sigma}^{\dagger}d_{\sigma})$.
Charge conservation therefore implies that
\bea
 -\boldsymbol{d}^{\dagger}\cdot(\boldsymbol{M}+\boldsymbol{M}^{\dagger})\cdot\boldsymbol{d}
+2\Gamma_vd_{\sigma}^{\dagger}d_{\sigma} =0,
\eea
which is true regardless of the PoTER approximation.
We conclude that the approximate solution Eq. (\ref{eq:ddd}) conserves the total charge.

\renewcommand{\theequation}{C\arabic{equation}}
\setcounter{equation}{0}  
\section{Charge current expression in the GIOM}
\label{a:5}

In this Appendix we derive the charge current expression under the GIOM, and show that it is given by two types of rates.
Performing the PoTER approximation, we reach simple integral forms for these rates that allow facile calculations.
In fact, if we only make the second PoTER approximation, that is we propagate the vibrations without backaction of electrons, we
reach a closed form expression for the current, which allows extensions beyond results presented in this work.

We begin with Equation (\ref{eq:js}), an exact form for the charge current,
and focus on
\begin{widetext}
\bea
C_1(t) &\equiv& -\sqrt{2\pi} \mathrm{Im} \left\langle d_{in}^{L,\dagger}(t) \tilde{d}_{1}(t)\right\rangle
\nonumber\\
&=&
 2\pi \mathrm{Re}  
   \sum_{nm}\int_{-\infty}^td\tau s_{1n}\tilde{s}_{m1}  \left\langle d_{in}^{L,\dagger}(t) D_1^{\dagger}(t) K_{nm}(t,\tau)d_{in}^L(\tau) \right\rangle
\nonumber\\
&+&
 2\pi \mathrm{Re}  
   \sum_{nm}\int_{-\infty}^t d\tau
s_{1n}\tilde{s}_{mN} \left\langle   d_{in}^{L,\dagger}(t) D_1^{\dagger}(t)K_{nm}(t,\tau)d_{in}^R(\tau)  \right\rangle,
\nonumber\\
\eea
with the kernel  defined in Eq. (\ref{eq:K}) and the operator of the system given by Eq. (\ref{eq:exactd1}).
To proceed, we make the second PoTER approximation, which allows us to separate the nuclear and electronic correlation functions.  Mixed correlation functions of the left and right lead electrons are zero, and we get
\bea
C_1
=
2\pi \mathrm{Re}  
   \sum_{nm}\int_{-\infty}^td\tau s_{1n}\tilde{s}_{m1}
\left\langle d_{in}^{L,\dagger}(t) d_{in}^L(\tau) \rangle  \langle D_1^{\dagger}(t) K_{nm}(t,\tau) \right\rangle.
\eea
We reorganize the result as 
\bea
C_1= \mathrm{Re}   \sum_{nm}  s_{1n}\tilde{s}_{m1} \chi_{nm}^L,
\eea
where
\bea
\chi_{nm}^v=
2\Gamma_v\int\,\frac{d\epsilon}{2\pi}n_F^v(\epsilon)\int_{0}^{\infty}\,d\tau e^{i\epsilon\tau}e^{-\Lambda_{n}\tau}B_{nm}(\tau),
\eea
with
\bea
B_{nm}(t-\tau)=\Big\langle \mathcal{D}_{n,b}^{\dagger}(t)\Big[\exp\Big(\int_{\tau}^t(\boldsymbol{\dot{R}}\boldsymbol{R}^{-1})_{\tau'}d\tau'\Big)\Big]_{nm}\mathcal{D}_{m,b}(\tau)\Big\rangle.
\eea
We identify $\chi_{nm}^v$ as the first-order rate; it depends on a single electronic resonance.
The correlation function $B_{nm}$ only involves vibrational degrees of freedom, and one should be able to simulate it
under certain assumptions.
Here, we perform the PoTER and replace
$\Big[\exp\Big(\int_{\tau}^t(\boldsymbol{\dot{R}}\boldsymbol{R}^{-1})_{\tau'}d\tau'\Big)\Big]_{nm}$ with $\delta_{nm}$.
This allows us to reach
\bea
C_1= \mathrm{Re} \sum_n s_{1n}\tilde s_{n1} \chi_n^L,
\eea
with
\bea
\chi_{n}^v=
2\Gamma_v\int\,\frac{d\epsilon}{2\pi}n_F^v(\epsilon)\int_{0}^{\infty}\,d\tau e^{i\epsilon\tau}e^{-\Lambda_{n}\tau}
\langle \mathcal{D}_{n,b}^{\dagger}(\tau)\mathcal{D}_{n,b}(0)\rangle.
\eea
Next, we analyze the second contribution in the charge current expression,
\bea
C_2\equiv  \langle d_1^{\dagger} d_1 \rangle.
\eea
We decouple the electronic and vibrational degrees of freedom and get
\bea
C_2&=&2 \pi  \sum_{nm}\sum_{n'm'}s_{1n}^*\tilde s_{m1}^* s_{1n'}\tilde s_{m'1}
\int_{-\infty}^t d\tau \int_{-\infty}^t d\tau'  
\left\langle  d_{in}^{L,\dagger}(\tau) d_{in}^L(\tau')\rangle\langle  K_{nm}^{\dagger}(t,\tau) K_{n'm'}(t,\tau')\right\rangle
\nonumber\\
&+&
  2 \pi  \sum_{nm}\sum_{n'm'}s_{1n}^*\tilde s_{mN}^* s_{1n'}\tilde s_{m'N} 
\int_{-\infty}^t d\tau \int_{-\infty}^t d\tau'  
\left\langle  d_{in}^{R,\dagger}(\tau) d_{in}^R(\tau')\rangle\langle  K_{nm}^{\dagger}(t,\tau) K_{n'm'}(t,\tau')\right\rangle.
\eea
Using the PoTER kernel this simplifies to 
 \bea
C_2&=&2 \pi  \sum_{n}\sum_{n'}s_{1n}^*\tilde s_{n1}^* s_{1n'}\tilde s_{n'1}
\int_{-\infty}^t d\tau \int_{-\infty}^t d\tau'
\left\langle  d_{in}^{L,\dagger}(\tau) d_{in}^L(\tau')\right\rangle
e^{-\Lambda_n^*(t-\tau)}  e^{-\Lambda_{n'}(t-\tau')} 
\left\langle   \mathcal{D}_{n,b}^{\dagger}(\tau) \mathcal{D}_{n,b}(t)    \mathcal{D}_{n',b}^{\dagger}(t) \mathcal{D}_{n',b}(\tau')
\right\rangle
\nonumber\\
&+&
 2 \pi  \sum_{n}\sum_{n'}s_{1n}^*\tilde s_{nN}^* s_{1n'}\tilde s_{n'N}        
\int_{-\infty}^t d\tau \int_{-\infty}^t d\tau'
\left\langle  d_{in}^{R,\dagger}(\tau) d_{in}^R(\tau')\right\rangle
e^{-\Lambda_n^*(t-\tau)}  e^{-\Lambda_{n'}(t-\tau')} 
\left\langle     \mathcal{D}_{n,b}^{\dagger}(\tau) \mathcal{D}_{n,b}(t)    \mathcal{D}_{n',b}^{\dagger}(t) \mathcal{D}_{n',b}(\tau')
\right\rangle
\nonumber\\
\eea
This expression comprises the so-called second-order rate, since two resonances act together to build the transfer rate.
Noting that the correlation functions on different sites are uncorrelated, we recover Eqs. (\ref{eq:n11})-(\ref{eq:eta_rate}) 
in the main text. 

\end{widetext}

\renewcommand{\theequation}{D\arabic{equation}}
\setcounter{equation}{0}  
\section{Evaluation of bath correlation functions}
\label{a:2}

In this Appendix we calculate the following two correlation functions: The polaron correlation function,
\bea
B_n(t,t')=\langle \mathcal{D}_{n,b}^{\dagger}(t)\mathcal{D}_{n,b}(t')\rangle,
\eea
and the photon correlation function,
\bea
A(t,t')=\langle \mathcal{D}_{a}^{\dagger}(t)\mathcal{D}_{a}(t')\rangle.
\eea
Recall that the primary vibrations are coupled to a secondary phonon bath, leading to level broadening $\nu_n$.
Similarly, the cavity mode couples to far-field modes, resulting in the decay rate $\kappa$. 

We begin with the correlation function of displacement operators.
For the sake of simplicity, we omit the subscript $n$. Neglecting the back-action of electrons, the GIOM EOM
for the vibrations satisfy,
\begin{equation}
\dot{b}~=~-(\nu+i\omega_b)b-i\sqrt{2\pi}b_{in}.
\end{equation}
The formal solution of $b$, together with Eq. (\ref{eq:a1}) lead to the commutation relation
\begin{equation}
[b(t),b^{\dagger}(t')]~=~\int\,d\omega\frac{I(\omega)}{\pi}\frac{e^{-i\omega(t-t')}}{\nu^2+(\omega-\omega_b)^2}.
\end{equation}
Considering a two-time correlation function $B(t,t')=\langle \mathcal{D}_b^{\dagger}(t)\mathcal{D}_b(t')\rangle$, we have
\begin{eqnarray}
B(t,t') &=& \exp\Big[-\lambda^2\int\,d\omega\frac{I(\omega)}{\pi}\frac{1-e^{-i\omega(t-t')}}{\nu^2+(\omega-\omega_b)^2}\Big]\nonumber\\
&&\times\underbrace{\langle e^{-\lambda b^{\dagger}(t)}e^{\lambda b^{\dagger}(t')}e^{\lambda b(t)}e^{-\lambda b(t')}\rangle}_{C(t,t')}.
\end{eqnarray}
The ensemble average of $C(t,t')$ is performed with respect to the thermal equilibrium state of thermal bath. 

Adopting the technique of Feynman disentangling of operators \cite{Mahan.00.NULL} and the formal definition of the input field $b_{in}(t)$, we find
\begin{equation}
C(t,t')~=~\exp\Big[-\sum_j\frac{|\mu_j(t,t')|^2}{e^{\beta\omega_j}-1}\Big]
\end{equation}
with $\mu_j(t,t')=-i\lambda\gamma_j\frac{e^{i\omega_j(t-t_0)}-e^{i\omega_j(t'-t_0)}}{\nu-i\omega_b+i\omega_j}$. Inserting $C(t,t')$ into $B(t,t')$, we recover Eq. (\ref{eq:corr_bb}) in the main text.

As for the cavity photon correlation function, we  note that the ensemble average is performed with respect to an initial vacuum state for the cavity mode, as well as for all far-field photon modes encompassed in the input noise for the cavity. The GIOM EOM of the cavity mode is (after neglecting back-action of electrons) 
$\dot{a} = -(\kappa+i\omega_c)a-\sqrt{2\pi}a_{in}$,
and its solution leads to ($t_0\to-\infty$ is taken)
\begin{equation}\label{eq:commutator_at}
[a(t),a^{\dagger}(t')]~=~\int d\omega\frac{F(\omega)}{\pi}\frac{e^{-i\omega(t-t')}}{\kappa^2+(\omega_c-\omega)^2}.
\end{equation}
The cavity correlation function $A(t,t')$ can be proceeded as follows:
\begin{eqnarray}
A(t,t')
&=& \exp\Big[-\frac{\Omega_R^2}{\omega_c^2}\int d\omega\frac{F(\omega)}{\pi[\kappa^2+(\omega_c-\omega)^2]}\Big]\nonumber\\
&&\times\langle 0|e^{\Omega_Ra(t)}e^{\Omega_Ra^{\dagger}(t')}|0\rangle\nonumber\\
&=& \exp\left[-\frac{\Omega_R^2}{\omega_c^2}\int d\omega\frac{F(\omega)}{\pi}\frac{1-e^{-i\omega(t-t')}}{\kappa^2+(\omega_c-\omega)^2}\right].
\end{eqnarray}

\renewcommand{\theequation}{E\arabic{equation}}
\setcounter{equation}{0}  
\section{Determining matrix forms for molecular dimer model}
\label{a:3} 

We discuss here the diagonalization of $\boldsymbol{M}$ for the two-site model, Eq.
(\ref{eq:M2}). 
The left eigenvectors $\boldsymbol{B}_{\pm}$ of the $\boldsymbol{M}$ matrix are determined by the following equations
\begin{equation}\label{eq:c1}
\boldsymbol{B}_{\pm}\left(
\begin{array}{cc}
\Gamma/2+i\tilde{\omega}_{e,1} & ig\mathcal{D}_{1,b}\mathcal{D}_{2,b}^{\dagger}\\
ig\mathcal{D}_{2,b}\mathcal{D}_{1,b}^{\dagger} & \Gamma/2+i\tilde{\omega}_{e,2}
\end{array}
\right)~=~\Lambda_{\pm}\boldsymbol{B}_{\pm}.
\end{equation}
We introduce the following ansatz for the eigenvectors 
\begin{eqnarray}
\boldsymbol{B}_{+}~=~\left(
\begin{array}{cc}
\tilde{s}_{11} & \tilde{s}_{12}\mathcal{D}_{1,b}\mathcal{D}_{2,b}^{\dagger}
\end{array}
\right),\boldsymbol{B}_{-}=\left(
\begin{array}{cc}
\tilde{s}_{21}\mathcal{D}_{2,b}\mathcal{D}_{1,b}^{\dagger} & 
\tilde{s}_{22}
\end{array}
\right). \nonumber\\
\end{eqnarray}
Here, $\tilde{s}_{11,12,21,22}$ are complex {\it c}-numbers. Inserting the above ansatz into Eq. (\ref{eq:c1}), we identify
\begin{eqnarray}
&&\boldsymbol{B}_{+}~=~\left(
\begin{array}{cc}
\frac{\Lambda_+-\Gamma/2-i\tilde{\omega}_{e,2}}{i\sqrt{Z}} & \frac{g}{\sqrt{Z}}\mathcal{D}_{1,b}\mathcal{D}_{2,b}^{\dagger}
\end{array}
\right),\nonumber\\
&&\boldsymbol{B}_{-}~=~\left(
\begin{array}{cc}
-\frac{g}{\sqrt{Z}}\mathcal{D}_{2,b}\mathcal{D}_{1,b}^{\dagger} & 
-\frac{\Lambda_{-}-\Gamma/2-i\tilde{\omega}_{e,1}}{i\sqrt{Z}}
\end{array}
\right),
\end{eqnarray}
where $Z=(\tilde{\omega}_{e,1}-\tilde{\omega}_{e,2})^2+4g^2$.
The matrix $\boldsymbol{R}$ can be expressed as 
\begin{equation}
\boldsymbol{R}~=~\left(
\begin{array}{c}
\boldsymbol{B}_{+} \\
\boldsymbol{B}_{-}
\end{array}
\right).
\end{equation}
Its inverse matrix $\boldsymbol{R}^{-1}$ takes the form
\begin{equation}
\boldsymbol{R}^{-1}~=~\left(
\begin{array}{cc}
\boldsymbol{A}_{+} & \boldsymbol{A}_{-}
\end{array}
\right)
\end{equation}
with
\begin{eqnarray}
&&\boldsymbol{A}_{+}~=~\left(
\begin{array}{c}
1\\
\frac{\Lambda_+-\Gamma/2-i\tilde{\omega}_{e,1}}{ig}\mathcal{D}_{2,b}\mathcal{D}_{1,b}^{\dagger}
\end{array}
\right),\nonumber\\
&&\boldsymbol{A}_{-}=\left(
\begin{array}{c}
\frac{\Lambda_{-}-\Gamma/2-i\tilde{\omega}_{e,2}}{ig}\mathcal{D}_{1,b}\mathcal{D}_{2,b}^{\dagger}\\
1
\end{array}
\right).
\end{eqnarray}
One can easily check that $\boldsymbol{R}^{-1}\boldsymbol{R}=\boldsymbol{I}$ and $\boldsymbol{R}\boldsymbol{R}^{-1}=\boldsymbol{I}$ by noting the following identities for eigenvalues
\begin{eqnarray}\label{eq:equality}
(\Lambda_+-\Gamma/2-i\tilde{\omega}_{e,2})(\Lambda_{-}-\Gamma/2-i\tilde{\omega}_{e,2}) &=& g^2,\nonumber\\
(\Lambda_{+}-\Gamma/2-i\tilde{\omega}_{e,1})(\Lambda_{-}-\Gamma/2-i\tilde{\omega}_{e,1}) &=& g^2,\nonumber\\
(\Lambda_+-\Gamma/2-i\tilde{\omega}_{e,1})(\Lambda_{+}-\Gamma/2-i\tilde{\omega}_{e,2}) &=& -g^2,\nonumber\\
(\Lambda_{-}-\Gamma/2-i\tilde{\omega}_{e,1})(\Lambda_{-}-\Gamma/2-i\tilde{\omega}_{e,2}) &=& -g^2.
\end{eqnarray} 

If $\Gamma_L\neq\Gamma_R$, the eigenvalues of $\boldsymbol{M}$ become
\begin{eqnarray}
\Lambda_{\pm} &=& \frac{1}{2}\Big[\Gamma_L+\Gamma_R+i(\tilde{\omega}_1+\tilde{\omega}_2)\nonumber\\
&&\pm\sqrt{[\Gamma_L-\Gamma_R+i(\tilde{\omega}_1-\tilde{\omega}_2)]^2-4g^2}\Big].
\end{eqnarray}
The corresponding matrices $\boldsymbol{R}$ and $\boldsymbol{R}^{-1}$ can be identified as
\begin{eqnarray}
\boldsymbol{R}^{-1} &=& \left(
\begin{array}{cc}
1 & \frac{\Lambda_{-}-\Gamma_R-i\tilde{\omega}_2}{ig}\mathcal{D}_{1,b}\mathcal{D}_{2,b}^{\dagger}\\
\frac{\Lambda_{+}-\Gamma_L-i\tilde{\omega}_1}{ig}\mathcal{D}_{2,b}\mathcal{D}_{1,b}^{\dagger} & 1
\end{array}
\right),\nonumber\\
\boldsymbol{R} &=& \left(
\begin{array}{cc}
\frac{\Lambda_{+}-\Gamma_R-i\tilde{\omega}_2}{\sqrt{Z}} & \frac{ig}{\sqrt{Z}}\mathcal{D}_{1,b}\mathcal{D}_{2,b}^{\dagger}\\
-\frac{ig}{\sqrt{Z}}\mathcal{D}_{2,b}\mathcal{D}_{1,b}^{\dagger} & -\frac{\Lambda_{-}-\Gamma_L-i\tilde{\omega}_1}{\sqrt{Z}}
\end{array}
\right),
\end{eqnarray}
 where $Z=[\Gamma_L-\Gamma_R+i(\tilde{\omega}_1-\tilde{\omega}_2)]^2-4g^2$. Equalities in Eq. (\ref{eq:equality}) are replaced by more general ones
 \begin{eqnarray}
(\Lambda_+-\Gamma_R-i\tilde{\omega}_2)(\Lambda_{-}-\Gamma_R-i\tilde{\omega}_2) &=& g^2,\nonumber\\
(\Lambda_{+}-\Gamma_L-i\tilde{\omega}_1)(\Lambda_{-}-\Gamma_L-i\tilde{\omega}_1) &=& g^2,\nonumber\\
(\Lambda_+-\Gamma_L-i\tilde{\omega}_1)(\Lambda_{+}-\Gamma_R-i\tilde{\omega}_2) &=& -g^2,\nonumber\\
(\Lambda_{-}-\Gamma_L-i\tilde{\omega}_1)(\Lambda_{-}-\Gamma_R-i\tilde{\omega}_2) &=& -g^2.
\end{eqnarray}


\begin{thebibliography}{171}%
\makeatletter
\providecommand \@ifxundefined [1]{%
 \@ifx{#1\undefined}
}%
\providecommand \@ifnum [1]{%
 \ifnum #1\expandafter \@firstoftwo
 \else \expandafter \@secondoftwo
 \fi
}%
\providecommand \@ifx [1]{%
 \ifx #1\expandafter \@firstoftwo
 \else \expandafter \@secondoftwo
 \fi
}%
\providecommand \natexlab [1]{#1}%
\providecommand \enquote  [1]{``#1''}%
\providecommand \bibnamefont  [1]{#1}%
\providecommand \bibfnamefont [1]{#1}%
\providecommand \citenamefont [1]{#1}%
\providecommand \href@noop [0]{\@secondoftwo}%
\providecommand \href [0]{\begingroup \@sanitize@url \@href}%
\providecommand \@href[1]{\@@startlink{#1}\@@href}%
\providecommand \@@href[1]{\endgroup#1\@@endlink}%
\providecommand \@sanitize@url [0]{\catcode `\\12\catcode `\$12\catcode
  `\&12\catcode `\#12\catcode `\^12\catcode `\_12\catcode `\%12\relax}%
\providecommand \@@startlink[1]{}%
\providecommand \@@endlink[0]{}%
\providecommand \url  [0]{\begingroup\@sanitize@url \@url }%
\providecommand \@url [1]{\endgroup\@href {#1}{\urlprefix }}%
\providecommand \urlprefix  [0]{URL }%
\providecommand \Eprint [0]{\href }%
\providecommand \doibase [0]{http://dx.doi.org/}%
\providecommand \selectlanguage [0]{\@gobble}%
\providecommand \bibinfo  [0]{\@secondoftwo}%
\providecommand \bibfield  [0]{\@secondoftwo}%
\providecommand \translation [1]{[#1]}%
\providecommand \BibitemOpen [0]{}%
\providecommand \bibitemStop [0]{}%
\providecommand \bibitemNoStop [0]{.\EOS\space}%
\providecommand \EOS [0]{\spacefactor3000\relax}%
\providecommand \BibitemShut  [1]{\csname bibitem#1\endcsname}%
\let\auto@bib@innerbib\@empty
\bibitem [{\citenamefont {Cuevas}\ and\ \citenamefont
  {Scheer}(2010)}]{Cuevas.10.NULL}%
  \BibitemOpen
  \bibfield  {author} {\bibinfo {author} {\bibfnamefont {J.~C.}\ \bibnamefont
  {Cuevas}}\ and\ \bibinfo {author} {\bibfnamefont {E.}~\bibnamefont
  {Scheer}},\ }\href@noop {} {\emph {\bibinfo {title} {Molecular Electronics:
  An Introduction To Theory And Experiment}}}\ (\bibinfo  {publisher} {World
  Scientific, Singapore},\ \bibinfo {year} {2010})\BibitemShut {NoStop}%
\bibitem [{\citenamefont {Aviram}\ and\ \citenamefont
  {Ratner}(1974)}]{Aviram.74.CPL}%
  \BibitemOpen
  \bibfield  {author} {\bibinfo {author} {\bibfnamefont {A.}~\bibnamefont
  {Aviram}}\ and\ \bibinfo {author} {\bibfnamefont {M.~A.}\ \bibnamefont
  {Ratner}},\ }\href {\doibase https://doi.org/10.1016/0009-2614(74)85031-1}
  {\bibfield  {journal} {\bibinfo  {journal} {Chem. Phys. Lett.}\ }\textbf
  {\bibinfo {volume} {29}},\ \bibinfo {pages} {277} (\bibinfo {year}
  {1974})}\BibitemShut {NoStop}%
\bibitem [{\citenamefont {Wold}\ and\ \citenamefont
  {Frisbie}(2001)}]{Wold.01.JACS}%
  \BibitemOpen
  \bibfield  {author} {\bibinfo {author} {\bibfnamefont {D.~J.}\ \bibnamefont
  {Wold}}\ and\ \bibinfo {author} {\bibfnamefont {C.~D.}\ \bibnamefont
  {Frisbie}},\ }\href {\doibase 10.1021/ja0101532} {\bibfield  {journal}
  {\bibinfo  {journal} {J. Am. Chem. Soc.}\ }\textbf {\bibinfo {volume}
  {123}},\ \bibinfo {pages} {5549} (\bibinfo {year} {2001})}\BibitemShut
  {NoStop}%
\bibitem [{\citenamefont {Holmlin}\ \emph {et~al.}(2001)\citenamefont
  {Holmlin}, \citenamefont {Haag}, \citenamefont {Chabinyc}, \citenamefont
  {Ismagilov}, \citenamefont {Cohen}, \citenamefont {Terfort}, \citenamefont
  {Rampi},\ and\ \citenamefont {Whitesides}}]{Holmlin.01.JACS}%
  \BibitemOpen
  \bibfield  {author} {\bibinfo {author} {\bibfnamefont {R.~E.}\ \bibnamefont
  {Holmlin}}, \bibinfo {author} {\bibfnamefont {R.}~\bibnamefont {Haag}},
  \bibinfo {author} {\bibfnamefont {M.~L.}\ \bibnamefont {Chabinyc}}, \bibinfo
  {author} {\bibfnamefont {R.~F.}\ \bibnamefont {Ismagilov}}, \bibinfo {author}
  {\bibfnamefont {A.~E.}\ \bibnamefont {Cohen}}, \bibinfo {author}
  {\bibfnamefont {A.}~\bibnamefont {Terfort}}, \bibinfo {author} {\bibfnamefont
  {M.}~\bibnamefont {Rampi}}, \ and\ \bibinfo {author} {\bibfnamefont {G.~M.}\
  \bibnamefont {Whitesides}},\ }\href {\doibase 10.1021/ja004055c} {\bibfield
  {journal} {\bibinfo  {journal} {J. Am. Chem. Soc.}\ }\textbf {\bibinfo
  {volume} {123}},\ \bibinfo {pages} {5075} (\bibinfo {year}
  {2001})}\BibitemShut {NoStop}%
\bibitem [{\citenamefont {Zhitenev}\ \emph {et~al.}(2002)\citenamefont
  {Zhitenev}, \citenamefont {Meng},\ and\ \citenamefont
  {Bao}}]{Zhitenev.02.PRL}%
  \BibitemOpen
  \bibfield  {author} {\bibinfo {author} {\bibfnamefont {N.~B.}\ \bibnamefont
  {Zhitenev}}, \bibinfo {author} {\bibfnamefont {H.}~\bibnamefont {Meng}}, \
  and\ \bibinfo {author} {\bibfnamefont {Z.}~\bibnamefont {Bao}},\ }\href
  {\doibase 10.1103/PhysRevLett.88.226801} {\bibfield  {journal} {\bibinfo
  {journal} {Phys. Rev. Lett.}\ }\textbf {\bibinfo {volume} {88}},\ \bibinfo
  {pages} {226801} (\bibinfo {year} {2002})}\BibitemShut {NoStop}%
\bibitem [{\citenamefont {Agrait}\ \emph {et~al.}(2003)\citenamefont {Agrait},
  \citenamefont {Yeyati},\ and\ \citenamefont {van Ruitenbeek}}]{Agrait.03.PR}%
  \BibitemOpen
  \bibfield  {author} {\bibinfo {author} {\bibfnamefont {N.}~\bibnamefont
  {Agrait}}, \bibinfo {author} {\bibfnamefont {A.}~\bibnamefont {Yeyati}}, \
  and\ \bibinfo {author} {\bibfnamefont {J.~M.}\ \bibnamefont {van
  Ruitenbeek}},\ }\href
  {http://www.sciencedirect.com/science/article/pii/S0370157302006336}
  {\bibfield  {journal} {\bibinfo  {journal} {Phys. Rep.}\ }\textbf {\bibinfo
  {volume} {377}},\ \bibinfo {pages} {81} (\bibinfo {year} {2003})}\BibitemShut
  {NoStop}%
\bibitem [{\citenamefont {Tao}(2006)}]{Tao.06.NN}%
  \BibitemOpen
  \bibfield  {author} {\bibinfo {author} {\bibfnamefont {N.}~\bibnamefont
  {Tao}},\ }\href {https://doi.org/10.1038/nnano.2006.130} {\bibfield
  {journal} {\bibinfo  {journal} {Nat. Nanotech.}\ }\textbf {\bibinfo {volume}
  {1}},\ \bibinfo {pages} {173} (\bibinfo {year} {2006})}\BibitemShut {NoStop}%
\bibitem [{\citenamefont {Venkataraman}\ \emph {et~al.}(2006)\citenamefont
  {Venkataraman}, \citenamefont {Klare}, \citenamefont {Nuckolls},
  \citenamefont {Hybertsen},\ and\ \citenamefont
  {Steigerwald}}]{Venkataraman.06.N}%
  \BibitemOpen
  \bibfield  {author} {\bibinfo {author} {\bibfnamefont {L.}~\bibnamefont
  {Venkataraman}}, \bibinfo {author} {\bibfnamefont {J.~E.}\ \bibnamefont
  {Klare}}, \bibinfo {author} {\bibfnamefont {C.}~\bibnamefont {Nuckolls}},
  \bibinfo {author} {\bibfnamefont {M.~S.}\ \bibnamefont {Hybertsen}}, \ and\
  \bibinfo {author} {\bibfnamefont {M.~L.}\ \bibnamefont {Steigerwald}},\
  }\href {https://doi.org/10.1038/nature05037} {\bibfield  {journal} {\bibinfo
  {journal} {Nature}\ }\textbf {\bibinfo {volume} {442}},\ \bibinfo {pages}
  {904} (\bibinfo {year} {2006})}\BibitemShut {NoStop}%
\bibitem [{\citenamefont {Ward}\ \emph {et~al.}(2008)\citenamefont {Ward},
  \citenamefont {Scott}, \citenamefont {Keane}, \citenamefont {Halas},\ and\
  \citenamefont {Natelson}}]{Ward.08.JP}%
  \BibitemOpen
  \bibfield  {author} {\bibinfo {author} {\bibfnamefont {D.}~\bibnamefont
  {Ward}}, \bibinfo {author} {\bibfnamefont {G.}~\bibnamefont {Scott}},
  \bibinfo {author} {\bibfnamefont {Z.}~\bibnamefont {Keane}}, \bibinfo
  {author} {\bibfnamefont {N.}~\bibnamefont {Halas}}, \ and\ \bibinfo {author}
  {\bibfnamefont {D.}~\bibnamefont {Natelson}},\ }\href {\doibase
  10.1088/0953-8984/20/37/374118} {\bibfield  {journal} {\bibinfo  {journal}
  {J. Phys.: Condens. Matter}\ }\textbf {\bibinfo {volume} {20}},\ \bibinfo
  {pages} {374118} (\bibinfo {year} {2008})}\BibitemShut {NoStop}%
\bibitem [{\citenamefont {McCreery}\ and\ \citenamefont
  {Bergren}(2009)}]{McCreery.09.AM}%
  \BibitemOpen
  \bibfield  {author} {\bibinfo {author} {\bibfnamefont {R.~L.}\ \bibnamefont
  {McCreery}}\ and\ \bibinfo {author} {\bibfnamefont {A.~J.}\ \bibnamefont
  {Bergren}},\ }\href {\doibase 10.1002/adma.200802850} {\bibfield  {journal}
  {\bibinfo  {journal} {Adv. Mater.}\ }\textbf {\bibinfo {volume} {21}},\
  \bibinfo {pages} {4303} (\bibinfo {year} {2009})}\BibitemShut {NoStop}%
\bibitem [{\citenamefont {Moth-Poulsen}\ and\ \citenamefont
  {Bjornholm}(2009)}]{Poulsen.09.NN}%
  \BibitemOpen
  \bibfield  {author} {\bibinfo {author} {\bibfnamefont {K.}~\bibnamefont
  {Moth-Poulsen}}\ and\ \bibinfo {author} {\bibfnamefont {T.}~\bibnamefont
  {Bjornholm}},\ }\href {https://doi.org/10.1038/nnano.2009.176} {\bibfield
  {journal} {\bibinfo  {journal} {Nat. Nanotech.}\ }\textbf {\bibinfo {volume}
  {4}},\ \bibinfo {pages} {551} (\bibinfo {year} {2009})}\BibitemShut {NoStop}%
\bibitem [{\citenamefont {Tuccitto}\ \emph {et~al.}(2008)\citenamefont
  {Tuccitto}, \citenamefont {Ferri}, \citenamefont {Cavazzini}, \citenamefont
  {Quici}, \citenamefont {Zhavnerko}, \citenamefont {Licciardello},\ and\
  \citenamefont {Rampi}}]{Tuccitto.09.NM}%
  \BibitemOpen
  \bibfield  {author} {\bibinfo {author} {\bibfnamefont {N.}~\bibnamefont
  {Tuccitto}}, \bibinfo {author} {\bibfnamefont {V.}~\bibnamefont {Ferri}},
  \bibinfo {author} {\bibfnamefont {M.}~\bibnamefont {Cavazzini}}, \bibinfo
  {author} {\bibfnamefont {S.}~\bibnamefont {Quici}}, \bibinfo {author}
  {\bibfnamefont {G.}~\bibnamefont {Zhavnerko}}, \bibinfo {author}
  {\bibfnamefont {A.}~\bibnamefont {Licciardello}}, \ and\ \bibinfo {author}
  {\bibfnamefont {M.~A.}\ \bibnamefont {Rampi}},\ }\href
  {https://doi.org/10.1038/nmat2332} {\bibfield  {journal} {\bibinfo  {journal}
  {Nat. Mater.}\ }\textbf {\bibinfo {volume} {8}},\ \bibinfo {pages} {41}
  (\bibinfo {year} {2008})}\BibitemShut {NoStop}%
\bibitem [{\citenamefont {Lafferentz}\ \emph {et~al.}(2009)\citenamefont
  {Lafferentz}, \citenamefont {Ample}, \citenamefont {Yu}, \citenamefont
  {Hecht}, \citenamefont {Joachim},\ and\ \citenamefont
  {Grill}}]{Lafferentz.09.S}%
  \BibitemOpen
  \bibfield  {author} {\bibinfo {author} {\bibfnamefont {L.}~\bibnamefont
  {Lafferentz}}, \bibinfo {author} {\bibfnamefont {F.}~\bibnamefont {Ample}},
  \bibinfo {author} {\bibfnamefont {H.}~\bibnamefont {Yu}}, \bibinfo {author}
  {\bibfnamefont {S.}~\bibnamefont {Hecht}}, \bibinfo {author} {\bibfnamefont
  {C.}~\bibnamefont {Joachim}}, \ and\ \bibinfo {author} {\bibfnamefont
  {L.}~\bibnamefont {Grill}},\ }\href {\doibase 10.1126/science.1168255}
  {\bibfield  {journal} {\bibinfo  {journal} {Science}\ }\textbf {\bibinfo
  {volume} {323}},\ \bibinfo {pages} {1193} (\bibinfo {year}
  {2009})}\BibitemShut {NoStop}%
\bibitem [{\citenamefont {Aradhya}\ and\ \citenamefont
  {Venkataraman}(2013)}]{Aradhya.13.NN}%
  \BibitemOpen
  \bibfield  {author} {\bibinfo {author} {\bibfnamefont {S.~V.}\ \bibnamefont
  {Aradhya}}\ and\ \bibinfo {author} {\bibfnamefont {L.}~\bibnamefont
  {Venkataraman}},\ }\href {http://dx.doi.org/10.1038/nnano.2013.91} {\bibfield
   {journal} {\bibinfo  {journal} {Nat. Nanotech.}\ }\textbf {\bibinfo {volume}
  {8}},\ \bibinfo {pages} {399} (\bibinfo {year} {2013})}\BibitemShut {NoStop}%
\bibitem [{\citenamefont {Xiang}\ \emph {et~al.}(2016)\citenamefont {Xiang},
  \citenamefont {Wang}, \citenamefont {Jia}, \citenamefont {Lee},\ and\
  \citenamefont {Guo}}]{Xiang.16.CR}%
  \BibitemOpen
  \bibfield  {author} {\bibinfo {author} {\bibfnamefont {D.}~\bibnamefont
  {Xiang}}, \bibinfo {author} {\bibfnamefont {X.}~\bibnamefont {Wang}},
  \bibinfo {author} {\bibfnamefont {C.}~\bibnamefont {Jia}}, \bibinfo {author}
  {\bibfnamefont {T.}~\bibnamefont {Lee}}, \ and\ \bibinfo {author}
  {\bibfnamefont {X.}~\bibnamefont {Guo}},\ }\href {\doibase
  10.1021/acs.chemrev.5b00680} {\bibfield  {journal} {\bibinfo  {journal}
  {Chem. Rev.}\ }\textbf {\bibinfo {volume} {116}},\ \bibinfo {pages} {4318}
  (\bibinfo {year} {2016})}\BibitemShut {NoStop}%
\bibitem [{\citenamefont {Akkerman}\ and\ \citenamefont
  {de~Boer}(2007)}]{Akkerman.08.JP}%
  \BibitemOpen
  \bibfield  {author} {\bibinfo {author} {\bibfnamefont {H.~B.}\ \bibnamefont
  {Akkerman}}\ and\ \bibinfo {author} {\bibfnamefont {B.}~\bibnamefont
  {de~Boer}},\ }\href {https://doi.org/10.1088%2F0953-8984%2F20%2F01%2F013001}
  {\bibfield  {journal} {\bibinfo  {journal} {J. Phys.: Condens. Matter}\
  }\textbf {\bibinfo {volume} {20}},\ \bibinfo {pages} {013001} (\bibinfo
  {year} {2007})}\BibitemShut {NoStop}%
\bibitem [{\citenamefont {Su}\ \emph {et~al.}(2016)\citenamefont {Su},
  \citenamefont {Neupane}, \citenamefont {Steigerwald}, \citenamefont
  {Venkataraman},\ and\ \citenamefont {Nuckolls}}]{Su.16.NRM}%
  \BibitemOpen
  \bibfield  {author} {\bibinfo {author} {\bibfnamefont {T.~A.}\ \bibnamefont
  {Su}}, \bibinfo {author} {\bibfnamefont {M.}~\bibnamefont {Neupane}},
  \bibinfo {author} {\bibfnamefont {M.~L.}\ \bibnamefont {Steigerwald}},
  \bibinfo {author} {\bibfnamefont {L.}~\bibnamefont {Venkataraman}}, \ and\
  \bibinfo {author} {\bibfnamefont {C.}~\bibnamefont {Nuckolls}},\ }\href
  {https://doi.org/10.1038/natrevmats.2016.2} {\bibfield  {journal} {\bibinfo
  {journal} {Nat. Rev. Mater.}\ }\textbf {\bibinfo {volume} {1}},\ \bibinfo
  {pages} {16002} (\bibinfo {year} {2016})}\BibitemShut {NoStop}%
\bibitem [{\citenamefont {Cui}\ \emph {et~al.}(2017)\citenamefont {Cui},
  \citenamefont {Miao}, \citenamefont {Jiang}, \citenamefont {Meyhofer},\ and\
  \citenamefont {Reddy}}]{Cui.17.JCP}%
  \BibitemOpen
  \bibfield  {author} {\bibinfo {author} {\bibfnamefont {L.}~\bibnamefont
  {Cui}}, \bibinfo {author} {\bibfnamefont {R.}~\bibnamefont {Miao}}, \bibinfo
  {author} {\bibfnamefont {C.}~\bibnamefont {Jiang}}, \bibinfo {author}
  {\bibfnamefont {E.}~\bibnamefont {Meyhofer}}, \ and\ \bibinfo {author}
  {\bibfnamefont {P.}~\bibnamefont {Reddy}},\ }\href {\doibase
  10.1063/1.4976982} {\bibfield  {journal} {\bibinfo  {journal} {J. Chem.
  Phys.}\ }\textbf {\bibinfo {volume} {146}},\ \bibinfo {pages} {092201}
  (\bibinfo {year} {2017})}\BibitemShut {NoStop}%
\bibitem [{\citenamefont {Thoss}\ and\ \citenamefont
  {Evers}(2018)}]{Thoss.18.JCP}%
  \BibitemOpen
  \bibfield  {author} {\bibinfo {author} {\bibfnamefont {M.}~\bibnamefont
  {Thoss}}\ and\ \bibinfo {author} {\bibfnamefont {F.}~\bibnamefont {Evers}},\
  }\href {\doibase 10.1063/1.5003306} {\bibfield  {journal} {\bibinfo
  {journal} {J. Chem. Phys.}\ }\textbf {\bibinfo {volume} {148}},\ \bibinfo
  {pages} {030901} (\bibinfo {year} {2018})}\BibitemShut {NoStop}%
\bibitem [{\citenamefont {Selzer}\ \emph
  {et~al.}(2004{\natexlab{a}})\citenamefont {Selzer}, \citenamefont {Cabassi},
  \citenamefont {Mayer},\ and\ \citenamefont {Allara}}]{Selzer.04.JACS}%
  \BibitemOpen
  \bibfield  {author} {\bibinfo {author} {\bibfnamefont {Y.}~\bibnamefont
  {Selzer}}, \bibinfo {author} {\bibfnamefont {M.~A.}\ \bibnamefont {Cabassi}},
  \bibinfo {author} {\bibfnamefont {T.~S.}\ \bibnamefont {Mayer}}, \ and\
  \bibinfo {author} {\bibfnamefont {D.~L.}\ \bibnamefont {Allara}},\ }\href
  {\doibase 10.1021/ja039015y} {\bibfield  {journal} {\bibinfo  {journal} {J.
  Am. Chem. Soc.}\ }\textbf {\bibinfo {volume} {126}},\ \bibinfo {pages} {4052}
  (\bibinfo {year} {2004}{\natexlab{a}})}\BibitemShut {NoStop}%
\bibitem [{\citenamefont {Selzer}\ \emph
  {et~al.}(2004{\natexlab{b}})\citenamefont {Selzer}, \citenamefont {Cabassi},
  \citenamefont {Mayer},\ and\ \citenamefont {Allara}}]{Selzer.04.NN}%
  \BibitemOpen
  \bibfield  {author} {\bibinfo {author} {\bibfnamefont {Y.}~\bibnamefont
  {Selzer}}, \bibinfo {author} {\bibfnamefont {M.~A.}\ \bibnamefont {Cabassi}},
  \bibinfo {author} {\bibfnamefont {T.~S.}\ \bibnamefont {Mayer}}, \ and\
  \bibinfo {author} {\bibfnamefont {D.~L.}\ \bibnamefont {Allara}},\
  }\href@noop {} {\bibfield  {journal} {\bibinfo  {journal} {Nanotechnology}\
  }\textbf {\bibinfo {volume} {15}},\ \bibinfo {pages} {S483} (\bibinfo {year}
  {2004}{\natexlab{b}})}\BibitemShut {NoStop}%
\bibitem [{\citenamefont {Weiss}\ \emph {et~al.}(2005)\citenamefont {Weiss},
  \citenamefont {Tauber}, \citenamefont {Kelley}, \citenamefont {Ahrens},
  \citenamefont {Ratner},\ and\ \citenamefont {Wasielewski}}]{Weiss.05.JACS}%
  \BibitemOpen
  \bibfield  {author} {\bibinfo {author} {\bibfnamefont {E.~A.}\ \bibnamefont
  {Weiss}}, \bibinfo {author} {\bibfnamefont {M.~J.}\ \bibnamefont {Tauber}},
  \bibinfo {author} {\bibfnamefont {R.~F.}\ \bibnamefont {Kelley}}, \bibinfo
  {author} {\bibfnamefont {M.~J.}\ \bibnamefont {Ahrens}}, \bibinfo {author}
  {\bibfnamefont {M.~A.}\ \bibnamefont {Ratner}}, \ and\ \bibinfo {author}
  {\bibfnamefont {M.~R.}\ \bibnamefont {Wasielewski}},\ }\href {\doibase
  10.1021/ja052901j} {\bibfield  {journal} {\bibinfo  {journal} {J. Am. Chem.
  Soc.}\ }\textbf {\bibinfo {volume} {127}},\ \bibinfo {pages} {11842}
  (\bibinfo {year} {2005})}\BibitemShut {NoStop}%
\bibitem [{\citenamefont {Selzer}\ and\ \citenamefont
  {Allara}(2006)}]{Selzer.06.ARPC}%
  \BibitemOpen
  \bibfield  {author} {\bibinfo {author} {\bibfnamefont {Y.}~\bibnamefont
  {Selzer}}\ and\ \bibinfo {author} {\bibfnamefont {D.~L.}\ \bibnamefont
  {Allara}},\ }\href {\doibase 10.1146/annurev.physchem.57.032905.104709}
  {\bibfield  {journal} {\bibinfo  {journal} {Annu. Rev. Phys. Chem.}\ }\textbf
  {\bibinfo {volume} {57}},\ \bibinfo {pages} {593} (\bibinfo {year}
  {2006})}\BibitemShut {NoStop}%
\bibitem [{\citenamefont {Poot}\ \emph {et~al.}(2006)\citenamefont {Poot},
  \citenamefont {Osorio}, \citenamefont {O'Neill}, \citenamefont {Thijssen},
  \citenamefont {Vanmaekelbergh}, \citenamefont {van Walree}, \citenamefont
  {Jenneskens},\ and\ \citenamefont {van~der Zant}}]{Poot.06.NL}%
  \BibitemOpen
  \bibfield  {author} {\bibinfo {author} {\bibfnamefont {M.}~\bibnamefont
  {Poot}}, \bibinfo {author} {\bibfnamefont {E.}~\bibnamefont {Osorio}},
  \bibinfo {author} {\bibfnamefont {K.}~\bibnamefont {O'Neill}}, \bibinfo
  {author} {\bibfnamefont {J.~M.}\ \bibnamefont {Thijssen}}, \bibinfo {author}
  {\bibfnamefont {D.}~\bibnamefont {Vanmaekelbergh}}, \bibinfo {author}
  {\bibfnamefont {C.~A.}\ \bibnamefont {van Walree}}, \bibinfo {author}
  {\bibfnamefont {L.~W.}\ \bibnamefont {Jenneskens}}, \ and\ \bibinfo {author}
  {\bibfnamefont {H.~S.~J.}\ \bibnamefont {van~der Zant}},\ }\href {\doibase
  10.1021/nl0604513} {\bibfield  {journal} {\bibinfo  {journal} {Nano Lett.}\
  }\textbf {\bibinfo {volume} {6}},\ \bibinfo {pages} {1031} (\bibinfo {year}
  {2006})}\BibitemShut {NoStop}%
\bibitem [{\citenamefont {Goldsmith}\ \emph {et~al.}(2008)\citenamefont
  {Goldsmith}, \citenamefont {DeLeon}, \citenamefont {Wilson}, \citenamefont
  {Finkelstein-Shapiro}, \citenamefont {Ratner},\ and\ \citenamefont
  {Wasielewski}}]{Goldsmith.08.JPCA}%
  \BibitemOpen
  \bibfield  {author} {\bibinfo {author} {\bibfnamefont {R.~H.}\ \bibnamefont
  {Goldsmith}}, \bibinfo {author} {\bibfnamefont {O.}~\bibnamefont {DeLeon}},
  \bibinfo {author} {\bibfnamefont {T.~M.}\ \bibnamefont {Wilson}}, \bibinfo
  {author} {\bibfnamefont {D.}~\bibnamefont {Finkelstein-Shapiro}}, \bibinfo
  {author} {\bibfnamefont {M.~A.}\ \bibnamefont {Ratner}}, \ and\ \bibinfo
  {author} {\bibfnamefont {M.~R.}\ \bibnamefont {Wasielewski}},\ }\href
  {\doibase 10.1021/jp801084v} {\bibfield  {journal} {\bibinfo  {journal} {J.
  Phys. Chem. A}\ }\textbf {\bibinfo {volume} {112}},\ \bibinfo {pages} {4410}
  (\bibinfo {year} {2008})}\BibitemShut {NoStop}%
\bibitem [{\citenamefont {Ho~Choi}\ \emph {et~al.}(2008)\citenamefont
  {Ho~Choi}, \citenamefont {Kim},\ and\ \citenamefont {Frisbie}}]{Choi.S.08}%
  \BibitemOpen
  \bibfield  {author} {\bibinfo {author} {\bibfnamefont {S.}~\bibnamefont
  {Ho~Choi}}, \bibinfo {author} {\bibfnamefont {B.}~\bibnamefont {Kim}}, \ and\
  \bibinfo {author} {\bibfnamefont {C.~D.}\ \bibnamefont {Frisbie}},\ }\href
  {\doibase 10.1126/science.1156538} {\bibfield  {journal} {\bibinfo  {journal}
  {Science}\ }\textbf {\bibinfo {volume} {320}},\ \bibinfo {pages} {1482}
  (\bibinfo {year} {2008})}\BibitemShut {NoStop}%
\bibitem [{\citenamefont {Lu}\ \emph {et~al.}(2009)\citenamefont {Lu},
  \citenamefont {Liu}, \citenamefont {Zhang}, \citenamefont {Du}, \citenamefont
  {Wang},\ and\ \citenamefont {Wang}}]{Lu.09.ACSN}%
  \BibitemOpen
  \bibfield  {author} {\bibinfo {author} {\bibfnamefont {Q.}~\bibnamefont
  {Lu}}, \bibinfo {author} {\bibfnamefont {K.}~\bibnamefont {Liu}}, \bibinfo
  {author} {\bibfnamefont {H.}~\bibnamefont {Zhang}}, \bibinfo {author}
  {\bibfnamefont {Z.}~\bibnamefont {Du}}, \bibinfo {author} {\bibfnamefont
  {X.}~\bibnamefont {Wang}}, \ and\ \bibinfo {author} {\bibfnamefont
  {F.}~\bibnamefont {Wang}},\ }\href {\doibase 10.1021/nn9012687} {\bibfield
  {journal} {\bibinfo  {journal} {ACS Nano}\ }\textbf {\bibinfo {volume} {3}},\
  \bibinfo {pages} {3861} (\bibinfo {year} {2009})}\BibitemShut {NoStop}%
\bibitem [{\citenamefont {Luo}\ and\ \citenamefont
  {Frisbie}(2010)}]{Luo.10.JACS}%
  \BibitemOpen
  \bibfield  {author} {\bibinfo {author} {\bibfnamefont {L.}~\bibnamefont
  {Luo}}\ and\ \bibinfo {author} {\bibfnamefont {C.~D.}\ \bibnamefont
  {Frisbie}},\ }\bibfield  {booktitle} {\emph {\bibinfo {booktitle} {J. Am.
  Chem. Soc.}},\ }\href {\doibase 10.1021/ja103239b} {\bibfield  {journal}
  {\bibinfo  {journal} {J. Am. Chem. Soc.}\ }\textbf {\bibinfo {volume}
  {132}},\ \bibinfo {pages} {8854} (\bibinfo {year} {2010})}\BibitemShut
  {NoStop}%
\bibitem [{\citenamefont {Choi}\ \emph {et~al.}(2010)\citenamefont {Choi},
  \citenamefont {Risko}, \citenamefont {Delgado}, \citenamefont {Kim},
  \citenamefont {Brédas},\ and\ \citenamefont {Frisbie}}]{Choi.10.JACS}%
  \BibitemOpen
  \bibfield  {author} {\bibinfo {author} {\bibfnamefont {S.}~\bibnamefont
  {Choi}}, \bibinfo {author} {\bibfnamefont {C.}~\bibnamefont {Risko}},
  \bibinfo {author} {\bibfnamefont {M.}~\bibnamefont {Delgado}}, \bibinfo
  {author} {\bibfnamefont {B.}~\bibnamefont {Kim}}, \bibinfo {author}
  {\bibfnamefont {J.}~\bibnamefont {Brédas}}, \ and\ \bibinfo {author}
  {\bibfnamefont {C.~D.}\ \bibnamefont {Frisbie}},\ }\href {\doibase
  10.1021/ja910547c} {\bibfield  {journal} {\bibinfo  {journal} {J. Am. Chem.
  Soc.}\ }\textbf {\bibinfo {volume} {132}},\ \bibinfo {pages} {4358} (\bibinfo
  {year} {2010})}\BibitemShut {NoStop}%
\bibitem [{\citenamefont {Hines}\ \emph {et~al.}(2010)\citenamefont {Hines},
  \citenamefont {Diez-Perez}, \citenamefont {Hihath}, \citenamefont {Liu},
  \citenamefont {Wang}, \citenamefont {Zhao}, \citenamefont {Zhou},
  \citenamefont {M\"ullen},\ and\ \citenamefont {Tao}}]{Hines.10.JACS}%
  \BibitemOpen
  \bibfield  {author} {\bibinfo {author} {\bibfnamefont {T.}~\bibnamefont
  {Hines}}, \bibinfo {author} {\bibfnamefont {I.}~\bibnamefont {Diez-Perez}},
  \bibinfo {author} {\bibfnamefont {J.}~\bibnamefont {Hihath}}, \bibinfo
  {author} {\bibfnamefont {H.}~\bibnamefont {Liu}}, \bibinfo {author}
  {\bibfnamefont {Z.}~\bibnamefont {Wang}}, \bibinfo {author} {\bibfnamefont
  {J.}~\bibnamefont {Zhao}}, \bibinfo {author} {\bibfnamefont {G.}~\bibnamefont
  {Zhou}}, \bibinfo {author} {\bibfnamefont {K.}~\bibnamefont {M\"ullen}}, \
  and\ \bibinfo {author} {\bibfnamefont {N.}~\bibnamefont {Tao}},\ }\href
  {\doibase 10.1021/ja1040946} {\bibfield  {journal} {\bibinfo  {journal} {J.
  Am. Chem. Soc.}\ }\textbf {\bibinfo {volume} {132}},\ \bibinfo {pages}
  {11658} (\bibinfo {year} {2010})}\BibitemShut {NoStop}%
\bibitem [{\citenamefont {Luo}\ \emph {et~al.}(2011)\citenamefont {Luo},
  \citenamefont {Choi},\ and\ \citenamefont {Frisbie}}]{Luo.11.CM}%
  \BibitemOpen
  \bibfield  {author} {\bibinfo {author} {\bibfnamefont {L.}~\bibnamefont
  {Luo}}, \bibinfo {author} {\bibfnamefont {S.}~\bibnamefont {Choi}}, \ and\
  \bibinfo {author} {\bibfnamefont {C.~D.}\ \bibnamefont {Frisbie}},\ }\href
  {\doibase 10.1021/cm102402t} {\bibfield  {journal} {\bibinfo  {journal}
  {Chem. Mater.}\ }\textbf {\bibinfo {volume} {23}},\ \bibinfo {pages} {631}
  (\bibinfo {year} {2011})}\BibitemShut {NoStop}%
\bibitem [{\citenamefont {Sedghi}\ \emph {et~al.}(2011)\citenamefont {Sedghi},
  \citenamefont {Garc\'ia-Su\'arez}, \citenamefont {Esdaile}, \citenamefont
  {Anderson}, \citenamefont {Lambert}, \citenamefont {Mart\'in}, \citenamefont
  {Bethell}, \citenamefont {Higgins}, \citenamefont {Elliott}, \citenamefont
  {Bennett}, \citenamefont {Macdonald},\ and\ \citenamefont
  {Nichols}}]{Sedghi.11.NN}%
  \BibitemOpen
  \bibfield  {author} {\bibinfo {author} {\bibfnamefont {G.}~\bibnamefont
  {Sedghi}}, \bibinfo {author} {\bibfnamefont {V.}~\bibnamefont
  {Garc\'ia-Su\'arez}}, \bibinfo {author} {\bibfnamefont {L.~J.}\ \bibnamefont
  {Esdaile}}, \bibinfo {author} {\bibfnamefont {H.~L.}\ \bibnamefont
  {Anderson}}, \bibinfo {author} {\bibfnamefont {C.~J.}\ \bibnamefont
  {Lambert}}, \bibinfo {author} {\bibfnamefont {S.}~\bibnamefont {Mart\'in}},
  \bibinfo {author} {\bibfnamefont {D.}~\bibnamefont {Bethell}}, \bibinfo
  {author} {\bibfnamefont {S.~J.}\ \bibnamefont {Higgins}}, \bibinfo {author}
  {\bibfnamefont {M.}~\bibnamefont {Elliott}}, \bibinfo {author} {\bibfnamefont
  {N.}~\bibnamefont {Bennett}}, \bibinfo {author} {\bibfnamefont {J.~E.}\
  \bibnamefont {Macdonald}}, \ and\ \bibinfo {author} {\bibfnamefont {R.~J.}\
  \bibnamefont {Nichols}},\ }\href {https://doi.org/10.1038/nnano.2011.111}
  {\bibfield  {journal} {\bibinfo  {journal} {Nat. Nanotech.}\ }\textbf
  {\bibinfo {volume} {6}},\ \bibinfo {pages} {517} (\bibinfo {year}
  {2011})}\BibitemShut {NoStop}%
\bibitem [{\citenamefont {Li}\ \emph {et~al.}(2012{\natexlab{a}})\citenamefont
  {Li}, \citenamefont {Park}, \citenamefont {Rawson}, \citenamefont {Therien},\
  and\ \citenamefont {Borguet}}]{Li.12.NL}%
  \BibitemOpen
  \bibfield  {author} {\bibinfo {author} {\bibfnamefont {Z.}~\bibnamefont
  {Li}}, \bibinfo {author} {\bibfnamefont {T.}~\bibnamefont {Park}}, \bibinfo
  {author} {\bibfnamefont {J.}~\bibnamefont {Rawson}}, \bibinfo {author}
  {\bibfnamefont {M.~J.}\ \bibnamefont {Therien}}, \ and\ \bibinfo {author}
  {\bibfnamefont {E.}~\bibnamefont {Borguet}},\ }\href {\doibase
  10.1021/nl2043216} {\bibfield  {journal} {\bibinfo  {journal} {Nano Lett.}\
  }\textbf {\bibinfo {volume} {12}},\ \bibinfo {pages} {2722} (\bibinfo {year}
  {2012}{\natexlab{a}})}\BibitemShut {NoStop}%
\bibitem [{\citenamefont {Taherinia}\ \emph {et~al.}(2016)\citenamefont
  {Taherinia}, \citenamefont {Smith}, \citenamefont {Ghosh}, \citenamefont
  {Odoh}, \citenamefont {Balhorn}, \citenamefont {Gagliardi}, \citenamefont
  {Cramer},\ and\ \citenamefont {Frisbie}}]{Taherinia.16.ACSN}%
  \BibitemOpen
  \bibfield  {author} {\bibinfo {author} {\bibfnamefont {D.}~\bibnamefont
  {Taherinia}}, \bibinfo {author} {\bibfnamefont {C.~E.}\ \bibnamefont
  {Smith}}, \bibinfo {author} {\bibfnamefont {S.}~\bibnamefont {Ghosh}},
  \bibinfo {author} {\bibfnamefont {S.~O.}\ \bibnamefont {Odoh}}, \bibinfo
  {author} {\bibfnamefont {L.}~\bibnamefont {Balhorn}}, \bibinfo {author}
  {\bibfnamefont {L.}~\bibnamefont {Gagliardi}}, \bibinfo {author}
  {\bibfnamefont {C.~J.}\ \bibnamefont {Cramer}}, \ and\ \bibinfo {author}
  {\bibfnamefont {C.~D.}\ \bibnamefont {Frisbie}},\ }\href {\doibase
  10.1021/acsnano.5b08126} {\bibfield  {journal} {\bibinfo  {journal} {ACS
  Nano}\ }\textbf {\bibinfo {volume} {10}},\ \bibinfo {pages} {4372} (\bibinfo
  {year} {2016})}\BibitemShut {NoStop}%
\bibitem [{\citenamefont {Thomas}\ \emph {et~al.}(2019)\citenamefont {Thomas},
  \citenamefont {Limburg}, \citenamefont {Sowa}, \citenamefont {Willick},
  \citenamefont {Baugh}, \citenamefont {Briggs}, \citenamefont {Gauger},
  \citenamefont {Anderson},\ and\ \citenamefont {Mol}}]{Thomas.19.NC}%
  \BibitemOpen
  \bibfield  {author} {\bibinfo {author} {\bibfnamefont {J.~O.}\ \bibnamefont
  {Thomas}}, \bibinfo {author} {\bibfnamefont {B.}~\bibnamefont {Limburg}},
  \bibinfo {author} {\bibfnamefont {J.~K.}\ \bibnamefont {Sowa}}, \bibinfo
  {author} {\bibfnamefont {K.}~\bibnamefont {Willick}}, \bibinfo {author}
  {\bibfnamefont {J.}~\bibnamefont {Baugh}}, \bibinfo {author} {\bibfnamefont
  {G.~A.~D.}\ \bibnamefont {Briggs}}, \bibinfo {author} {\bibfnamefont {E.~M.}\
  \bibnamefont {Gauger}}, \bibinfo {author} {\bibfnamefont {H.~L.}\
  \bibnamefont {Anderson}}, \ and\ \bibinfo {author} {\bibfnamefont {J.~A.}\
  \bibnamefont {Mol}},\ }\href {https://doi.org/10.1038/s41467-019-12625-4}
  {\bibfield  {journal} {\bibinfo  {journal} {Nat. Commun.}\ }\textbf {\bibinfo
  {volume} {10}},\ \bibinfo {pages} {4628} (\bibinfo {year}
  {2019})}\BibitemShut {NoStop}%
\bibitem [{\citenamefont {Solomon}\ \emph {et~al.}(2010)\citenamefont
  {Solomon}, \citenamefont {Herrmann}, \citenamefont {Hansen}, \citenamefont
  {Mujica},\ and\ \citenamefont {Ratner}}]{Solomon.10.NC}%
  \BibitemOpen
  \bibfield  {author} {\bibinfo {author} {\bibfnamefont {G.~C.}\ \bibnamefont
  {Solomon}}, \bibinfo {author} {\bibfnamefont {C.}~\bibnamefont {Herrmann}},
  \bibinfo {author} {\bibfnamefont {T.}~\bibnamefont {Hansen}}, \bibinfo
  {author} {\bibfnamefont {V.}~\bibnamefont {Mujica}}, \ and\ \bibinfo {author}
  {\bibfnamefont {M.~A.}\ \bibnamefont {Ratner}},\ }\href
  {https://doi.org/10.1038/nchem.546} {\bibfield  {journal} {\bibinfo
  {journal} {Nat. Chem.}\ }\textbf {\bibinfo {volume} {2}},\ \bibinfo {pages}
  {223} (\bibinfo {year} {2010})}\BibitemShut {NoStop}%
\bibitem [{\citenamefont {Gu\'edon}\ \emph {et~al.}(2012)\citenamefont
  {Gu\'edon}, \citenamefont {Valkenier}, \citenamefont {Markussen},
  \citenamefont {Thygesen}, \citenamefont {Hummelen},\ and\ \citenamefont
  {van~der Molen}}]{Guedon.12.NN}%
  \BibitemOpen
  \bibfield  {author} {\bibinfo {author} {\bibfnamefont {C.~M.}\ \bibnamefont
  {Gu\'edon}}, \bibinfo {author} {\bibfnamefont {H.}~\bibnamefont {Valkenier}},
  \bibinfo {author} {\bibfnamefont {T.}~\bibnamefont {Markussen}}, \bibinfo
  {author} {\bibfnamefont {K.~S.}\ \bibnamefont {Thygesen}}, \bibinfo {author}
  {\bibfnamefont {J.~C.}\ \bibnamefont {Hummelen}}, \ and\ \bibinfo {author}
  {\bibfnamefont {S.~J.}\ \bibnamefont {van~der Molen}},\ }\href
  {http://dx.doi.org/10.1038/nnano.2012.37} {\bibfield  {journal} {\bibinfo
  {journal} {Nat. Nanotech.}\ }\textbf {\bibinfo {volume} {7}},\ \bibinfo
  {pages} {305} (\bibinfo {year} {2012})}\BibitemShut {NoStop}%
\bibitem [{\citenamefont {Vazquez}\ \emph {et~al.}(2012)\citenamefont
  {Vazquez}, \citenamefont {Skouta}, \citenamefont {Schneebeli}, \citenamefont
  {Kamenetska}, \citenamefont {Breslow}, \citenamefont {Venkataraman},\ and\
  \citenamefont {Hybertsen}}]{Vazquez.12.NN}%
  \BibitemOpen
  \bibfield  {author} {\bibinfo {author} {\bibfnamefont {H.}~\bibnamefont
  {Vazquez}}, \bibinfo {author} {\bibfnamefont {R.}~\bibnamefont {Skouta}},
  \bibinfo {author} {\bibfnamefont {S.}~\bibnamefont {Schneebeli}}, \bibinfo
  {author} {\bibfnamefont {M.}~\bibnamefont {Kamenetska}}, \bibinfo {author}
  {\bibfnamefont {R.}~\bibnamefont {Breslow}}, \bibinfo {author} {\bibfnamefont
  {L.}~\bibnamefont {Venkataraman}}, \ and\ \bibinfo {author} {\bibfnamefont
  {M.~S.}\ \bibnamefont {Hybertsen}},\ }\href
  {http://dx.doi.org/10.1038/nnano.2012.147} {\bibfield  {journal} {\bibinfo
  {journal} {Nat Nanotech.}\ }\textbf {\bibinfo {volume} {7}},\ \bibinfo
  {pages} {663} (\bibinfo {year} {2012})}\BibitemShut {NoStop}%
\bibitem [{\citenamefont {Garner}\ \emph {et~al.}(2018)\citenamefont {Garner},
  \citenamefont {Li}, \citenamefont {Chen}, \citenamefont {Su}, \citenamefont
  {Shangguan}, \citenamefont {Paley}, \citenamefont {Liu}, \citenamefont {Ng},
  \citenamefont {Li}, \citenamefont {Xiao}, \citenamefont {Nuckolls},
  \citenamefont {Venkataraman},\ and\ \citenamefont {Solomon}}]{Garner.18.N}%
  \BibitemOpen
  \bibfield  {author} {\bibinfo {author} {\bibfnamefont {M.~H.}\ \bibnamefont
  {Garner}}, \bibinfo {author} {\bibfnamefont {H.}~\bibnamefont {Li}}, \bibinfo
  {author} {\bibfnamefont {Y.}~\bibnamefont {Chen}}, \bibinfo {author}
  {\bibfnamefont {T.~A.}\ \bibnamefont {Su}}, \bibinfo {author} {\bibfnamefont
  {Z.}~\bibnamefont {Shangguan}}, \bibinfo {author} {\bibfnamefont {D.~W.}\
  \bibnamefont {Paley}}, \bibinfo {author} {\bibfnamefont {T.}~\bibnamefont
  {Liu}}, \bibinfo {author} {\bibfnamefont {F.}~\bibnamefont {Ng}}, \bibinfo
  {author} {\bibfnamefont {H.}~\bibnamefont {Li}}, \bibinfo {author}
  {\bibfnamefont {S.}~\bibnamefont {Xiao}}, \bibinfo {author} {\bibfnamefont
  {C.}~\bibnamefont {Nuckolls}}, \bibinfo {author} {\bibfnamefont
  {L.}~\bibnamefont {Venkataraman}}, \ and\ \bibinfo {author} {\bibfnamefont
  {G.~C.}\ \bibnamefont {Solomon}},\ }\href
  {https://doi.org/10.1038/s41586-018-0197-9} {\bibfield  {journal} {\bibinfo
  {journal} {Nature}\ }\textbf {\bibinfo {volume} {558}},\ \bibinfo {pages}
  {415} (\bibinfo {year} {2018})}\BibitemShut {NoStop}%
\bibitem [{\citenamefont {Evers}\ \emph {et~al.}(2019)\citenamefont {Evers},
  \citenamefont {Koryt\'ar}, \citenamefont {Tewari},\ and\ \citenamefont {van
  Ruitenbeek}}]{Evers.19.NULL}%
  \BibitemOpen
  \bibfield  {author} {\bibinfo {author} {\bibfnamefont {F.}~\bibnamefont
  {Evers}}, \bibinfo {author} {\bibfnamefont {R.}~\bibnamefont {Koryt\'ar}},
  \bibinfo {author} {\bibfnamefont {S.}~\bibnamefont {Tewari}}, \ and\ \bibinfo
  {author} {\bibfnamefont {J.~M.}\ \bibnamefont {van Ruitenbeek}},\ }\href
  {https://arxiv.org/abs/1906.10449} {\  (\bibinfo {year} {2019})},\ \bibinfo
  {note} {arXiv:1906.10449}\BibitemShut {NoStop}%
\bibitem [{\citenamefont {Reddy}\ \emph {et~al.}(2007)\citenamefont {Reddy},
  \citenamefont {Jang}, \citenamefont {Segalman},\ and\ \citenamefont
  {Majumdar}}]{Reddy.07.S}%
  \BibitemOpen
  \bibfield  {author} {\bibinfo {author} {\bibfnamefont {P.}~\bibnamefont
  {Reddy}}, \bibinfo {author} {\bibfnamefont {S.}~\bibnamefont {Jang}},
  \bibinfo {author} {\bibfnamefont {R.~A.}\ \bibnamefont {Segalman}}, \ and\
  \bibinfo {author} {\bibfnamefont {A.}~\bibnamefont {Majumdar}},\ }\href
  {\doibase 10.1126/science.1137149} {\bibfield  {journal} {\bibinfo  {journal}
  {Science}\ }\textbf {\bibinfo {volume} {315}},\ \bibinfo {pages} {1568}
  (\bibinfo {year} {2007})}\BibitemShut {NoStop}%
\bibitem [{\citenamefont {Malen}\ \emph {et~al.}(2009)\citenamefont {Malen},
  \citenamefont {Doak}, \citenamefont {Baheti}, \citenamefont {Tilley},
  \citenamefont {Segalman},\ and\ \citenamefont {Majumdar}}]{Malen.09.NL}%
  \BibitemOpen
  \bibfield  {author} {\bibinfo {author} {\bibfnamefont {J.~A.}\ \bibnamefont
  {Malen}}, \bibinfo {author} {\bibfnamefont {P.}~\bibnamefont {Doak}},
  \bibinfo {author} {\bibfnamefont {K.}~\bibnamefont {Baheti}}, \bibinfo
  {author} {\bibfnamefont {T.~D.}\ \bibnamefont {Tilley}}, \bibinfo {author}
  {\bibfnamefont {R.~A.}\ \bibnamefont {Segalman}}, \ and\ \bibinfo {author}
  {\bibfnamefont {A.}~\bibnamefont {Majumdar}},\ }\href {\doibase
  10.1021/nl803814f} {\bibfield  {journal} {\bibinfo  {journal} {Nano Lett.}\
  }\textbf {\bibinfo {volume} {9}},\ \bibinfo {pages} {1164} (\bibinfo {year}
  {2009})}\BibitemShut {NoStop}%
\bibitem [{\citenamefont {Zimbovskaya}(2016)}]{Zimbovskaya.16.JCP}%
  \BibitemOpen
  \bibfield  {author} {\bibinfo {author} {\bibfnamefont {N.~A.}\ \bibnamefont
  {Zimbovskaya}},\ }\href {\doibase 10.1063/1.4972002} {\bibfield  {journal}
  {\bibinfo  {journal} {J. Chem. Phys.}\ }\textbf {\bibinfo {volume} {145}},\
  \bibinfo {pages} {221101} (\bibinfo {year} {2016})}\BibitemShut {NoStop}%
\bibitem [{\citenamefont {Rinc\'on-Garc\'ia}\ \emph {et~al.}(2016)\citenamefont
  {Rinc\'on-Garc\'ia}, \citenamefont {Evangeli}, \citenamefont
  {Rubio-Bollinger},\ and\ \citenamefont {Agra\"it}}]{Garcia.16.CSR}%
  \BibitemOpen
  \bibfield  {author} {\bibinfo {author} {\bibfnamefont {L.}~\bibnamefont
  {Rinc\'on-Garc\'ia}}, \bibinfo {author} {\bibfnamefont {C.}~\bibnamefont
  {Evangeli}}, \bibinfo {author} {\bibfnamefont {G.}~\bibnamefont
  {Rubio-Bollinger}}, \ and\ \bibinfo {author} {\bibfnamefont {N.}~\bibnamefont
  {Agra\"it}},\ }\href {\doibase 10.1039/C6CS00141F} {\bibfield  {journal}
  {\bibinfo  {journal} {Chem. Soc. Rev.}\ }\textbf {\bibinfo {volume} {45}},\
  \bibinfo {pages} {4285} (\bibinfo {year} {2016})}\BibitemShut {NoStop}%
\bibitem [{\citenamefont {Schmaus}\ \emph {et~al.}(2011)\citenamefont
  {Schmaus}, \citenamefont {Bagrets}, \citenamefont {Nahas}, \citenamefont
  {Yamada}, \citenamefont {Bork}, \citenamefont {Bowen}, \citenamefont
  {Beaurepaire}, \citenamefont {Evers},\ and\ \citenamefont
  {Wulfhekel}}]{Schmaus.11.NN}%
  \BibitemOpen
  \bibfield  {author} {\bibinfo {author} {\bibfnamefont {S.}~\bibnamefont
  {Schmaus}}, \bibinfo {author} {\bibfnamefont {A.}~\bibnamefont {Bagrets}},
  \bibinfo {author} {\bibfnamefont {Y.}~\bibnamefont {Nahas}}, \bibinfo
  {author} {\bibfnamefont {T.~K.}\ \bibnamefont {Yamada}}, \bibinfo {author}
  {\bibfnamefont {A.}~\bibnamefont {Bork}}, \bibinfo {author} {\bibfnamefont
  {M.}~\bibnamefont {Bowen}}, \bibinfo {author} {\bibfnamefont
  {E.}~\bibnamefont {Beaurepaire}}, \bibinfo {author} {\bibfnamefont
  {F.}~\bibnamefont {Evers}}, \ and\ \bibinfo {author} {\bibfnamefont
  {W.}~\bibnamefont {Wulfhekel}},\ }\href
  {https://doi.org/10.1038/nnano.2011.11} {\bibfield  {journal} {\bibinfo
  {journal} {Nat. Nanotech.}\ }\textbf {\bibinfo {volume} {6}},\ \bibinfo
  {pages} {185} (\bibinfo {year} {2011})}\BibitemShut {NoStop}%
\bibitem [{\citenamefont {Liang}\ \emph {et~al.}(2002)\citenamefont {Liang},
  \citenamefont {Shores}, \citenamefont {Bockrath}, \citenamefont {Long},\ and\
  \citenamefont {Park}}]{Liang.02.N}%
  \BibitemOpen
  \bibfield  {author} {\bibinfo {author} {\bibfnamefont {W.}~\bibnamefont
  {Liang}}, \bibinfo {author} {\bibfnamefont {M.~P.}\ \bibnamefont {Shores}},
  \bibinfo {author} {\bibfnamefont {M.}~\bibnamefont {Bockrath}}, \bibinfo
  {author} {\bibfnamefont {J.~R.}\ \bibnamefont {Long}}, \ and\ \bibinfo
  {author} {\bibfnamefont {H.}~\bibnamefont {Park}},\ }\href
  {https://doi.org/10.1038/nature00790} {\bibfield  {journal} {\bibinfo
  {journal} {Nature}\ }\textbf {\bibinfo {volume} {417}},\ \bibinfo {pages}
  {725} (\bibinfo {year} {2002})}\BibitemShut {NoStop}%
\bibitem [{\citenamefont {Scott}\ and\ \citenamefont
  {Natelson}(2010)}]{Scott.10.ACSN}%
  \BibitemOpen
  \bibfield  {author} {\bibinfo {author} {\bibfnamefont {G.}~\bibnamefont
  {Scott}}\ and\ \bibinfo {author} {\bibfnamefont {D.}~\bibnamefont
  {Natelson}},\ }\href {\doibase 10.1021/nn100793s} {\bibfield  {journal}
  {\bibinfo  {journal} {ACS Nano}\ }\textbf {\bibinfo {volume} {4}},\ \bibinfo
  {pages} {3560} (\bibinfo {year} {2010})}\BibitemShut {NoStop}%
\bibitem [{\citenamefont {G{\"o}hler}\ \emph {et~al.}(2011)\citenamefont
  {G{\"o}hler}, \citenamefont {Hamelbeck}, \citenamefont {Markus},
  \citenamefont {Kettner}, \citenamefont {Hanne}, \citenamefont {Vager},
  \citenamefont {Naaman},\ and\ \citenamefont {Zacharias}}]{Gohler.S.11}%
  \BibitemOpen
  \bibfield  {author} {\bibinfo {author} {\bibfnamefont {B.}~\bibnamefont
  {G{\"o}hler}}, \bibinfo {author} {\bibfnamefont {V.}~\bibnamefont
  {Hamelbeck}}, \bibinfo {author} {\bibfnamefont {T.~Z.}\ \bibnamefont
  {Markus}}, \bibinfo {author} {\bibfnamefont {M.}~\bibnamefont {Kettner}},
  \bibinfo {author} {\bibfnamefont {G.~F.}\ \bibnamefont {Hanne}}, \bibinfo
  {author} {\bibfnamefont {Z.}~\bibnamefont {Vager}}, \bibinfo {author}
  {\bibfnamefont {R.}~\bibnamefont {Naaman}}, \ and\ \bibinfo {author}
  {\bibfnamefont {H.}~\bibnamefont {Zacharias}},\ }\href {\doibase
  10.1126/science.1199339} {\bibfield  {journal} {\bibinfo  {journal}
  {Science}\ }\textbf {\bibinfo {volume} {331}},\ \bibinfo {pages} {894}
  (\bibinfo {year} {2011})}\BibitemShut {NoStop}%
\bibitem [{\citenamefont {Naaman}\ and\ \citenamefont
  {Waldeck}(2015)}]{Naaman.15.ARPC}%
  \BibitemOpen
  \bibfield  {author} {\bibinfo {author} {\bibfnamefont {R.}~\bibnamefont
  {Naaman}}\ and\ \bibinfo {author} {\bibfnamefont {D.~H.}\ \bibnamefont
  {Waldeck}},\ }\href {\doibase 10.1146/annurev-physchem-040214-121554}
  {\bibfield  {journal} {\bibinfo  {journal} {Ann. Rev. Phys. Chem.}\ }\textbf
  {\bibinfo {volume} {66}},\ \bibinfo {pages} {263} (\bibinfo {year}
  {2015})}\BibitemShut {NoStop}%
\bibitem [{\citenamefont {Naaman}\ \emph {et~al.}(2019)\citenamefont {Naaman},
  \citenamefont {Paltiel},\ and\ \citenamefont {Waldeck}}]{Naaman.19.NCR}%
  \BibitemOpen
  \bibfield  {author} {\bibinfo {author} {\bibfnamefont {R.}~\bibnamefont
  {Naaman}}, \bibinfo {author} {\bibfnamefont {Y.}~\bibnamefont {Paltiel}}, \
  and\ \bibinfo {author} {\bibfnamefont {D.~H.}\ \bibnamefont {Waldeck}},\
  }\href {https://doi.org/10.1038/s41570-019-0087-1} {\bibfield  {journal}
  {\bibinfo  {journal} {Nat. Rev. Chem.}\ }\textbf {\bibinfo {volume} {3}},\
  \bibinfo {pages} {250} (\bibinfo {year} {2019})}\BibitemShut {NoStop}%
\bibitem [{\citenamefont {Koch}\ and\ \citenamefont {von
  Oppen}(2005)}]{Koch.05.PRL}%
  \BibitemOpen
  \bibfield  {author} {\bibinfo {author} {\bibfnamefont {J.}~\bibnamefont
  {Koch}}\ and\ \bibinfo {author} {\bibfnamefont {F.}~\bibnamefont {von
  Oppen}},\ }\href {\doibase 10.1103/PhysRevLett.94.206804} {\bibfield
  {journal} {\bibinfo  {journal} {Phys. Rev. Lett.}\ }\textbf {\bibinfo
  {volume} {94}},\ \bibinfo {pages} {206804} (\bibinfo {year}
  {2005})}\BibitemShut {NoStop}%
\bibitem [{\citenamefont {Koch}\ \emph
  {et~al.}(2006{\natexlab{a}})\citenamefont {Koch}, \citenamefont {von Oppen},\
  and\ \citenamefont {Andreev}}]{Koch.06.PRB}%
  \BibitemOpen
  \bibfield  {author} {\bibinfo {author} {\bibfnamefont {J.}~\bibnamefont
  {Koch}}, \bibinfo {author} {\bibfnamefont {F.}~\bibnamefont {von Oppen}}, \
  and\ \bibinfo {author} {\bibfnamefont {A.~V.}\ \bibnamefont {Andreev}},\
  }\href {\doibase 10.1103/PhysRevB.74.205438} {\bibfield  {journal} {\bibinfo
  {journal} {Phys. Rev. B}\ }\textbf {\bibinfo {volume} {74}},\ \bibinfo
  {pages} {205438} (\bibinfo {year} {2006}{\natexlab{a}})}\BibitemShut
  {NoStop}%
\bibitem [{\citenamefont {Leturcq}\ \emph {et~al.}(2009)\citenamefont
  {Leturcq}, \citenamefont {Stampfer}, \citenamefont {Inderbitzin},
  \citenamefont {Durrer}, \citenamefont {Hierold}, \citenamefont {Mariani},
  \citenamefont {Schultz}, \citenamefont {von Oppen},\ and\ \citenamefont
  {Ensslin}}]{Leturcq.09.NP}%
  \BibitemOpen
  \bibfield  {author} {\bibinfo {author} {\bibfnamefont {R.}~\bibnamefont
  {Leturcq}}, \bibinfo {author} {\bibfnamefont {C.}~\bibnamefont {Stampfer}},
  \bibinfo {author} {\bibfnamefont {K.}~\bibnamefont {Inderbitzin}}, \bibinfo
  {author} {\bibfnamefont {L.}~\bibnamefont {Durrer}}, \bibinfo {author}
  {\bibfnamefont {C.}~\bibnamefont {Hierold}}, \bibinfo {author} {\bibfnamefont
  {E.}~\bibnamefont {Mariani}}, \bibinfo {author} {\bibfnamefont {M.~G.}\
  \bibnamefont {Schultz}}, \bibinfo {author} {\bibfnamefont {F.}~\bibnamefont
  {von Oppen}}, \ and\ \bibinfo {author} {\bibfnamefont {K.}~\bibnamefont
  {Ensslin}},\ }\href {https://doi.org/10.1038/nphys1234} {\bibfield  {journal}
  {\bibinfo  {journal} {Nat. Phys.}\ }\textbf {\bibinfo {volume} {5}},\
  \bibinfo {pages} {327} (\bibinfo {year} {2009})}\BibitemShut {NoStop}%
\bibitem [{\citenamefont {Burzur\'i}\ \emph {et~al.}(2014)\citenamefont
  {Burzur\'i}, \citenamefont {Yamamoto}, \citenamefont {Warnock}, \citenamefont
  {Zhong}, \citenamefont {Park}, \citenamefont {Cornia},\ and\ \citenamefont
  {van~der Zant}}]{Burzuri.14.NL}%
  \BibitemOpen
  \bibfield  {author} {\bibinfo {author} {\bibfnamefont {E.}~\bibnamefont
  {Burzur\'i}}, \bibinfo {author} {\bibfnamefont {Y.}~\bibnamefont {Yamamoto}},
  \bibinfo {author} {\bibfnamefont {M.}~\bibnamefont {Warnock}}, \bibinfo
  {author} {\bibfnamefont {X.}~\bibnamefont {Zhong}}, \bibinfo {author}
  {\bibfnamefont {K.}~\bibnamefont {Park}}, \bibinfo {author} {\bibfnamefont
  {A.}~\bibnamefont {Cornia}}, \ and\ \bibinfo {author} {\bibfnamefont
  {H.~S.~J.}\ \bibnamefont {van~der Zant}},\ }\href {\doibase
  10.1021/nl500524w} {\bibfield  {journal} {\bibinfo  {journal} {Nano Lett.}\
  }\textbf {\bibinfo {volume} {14}},\ \bibinfo {pages} {3191} (\bibinfo {year}
  {2014})}\BibitemShut {NoStop}%
\bibitem [{\citenamefont {Lau}\ \emph {et~al.}(2016)\citenamefont {Lau},
  \citenamefont {Sadeghi}, \citenamefont {Rogers}, \citenamefont {Sangtarash},
  \citenamefont {Dallas}, \citenamefont {Porfyrakis}, \citenamefont {Warner},
  \citenamefont {Lambert}, \citenamefont {Briggs},\ and\ \citenamefont
  {Mol}}]{Lau.16.NL}%
  \BibitemOpen
  \bibfield  {author} {\bibinfo {author} {\bibfnamefont {C.}~\bibnamefont
  {Lau}}, \bibinfo {author} {\bibfnamefont {H.}~\bibnamefont {Sadeghi}},
  \bibinfo {author} {\bibfnamefont {G.}~\bibnamefont {Rogers}}, \bibinfo
  {author} {\bibfnamefont {S.}~\bibnamefont {Sangtarash}}, \bibinfo {author}
  {\bibfnamefont {P.}~\bibnamefont {Dallas}}, \bibinfo {author} {\bibfnamefont
  {K.}~\bibnamefont {Porfyrakis}}, \bibinfo {author} {\bibfnamefont
  {J.}~\bibnamefont {Warner}}, \bibinfo {author} {\bibfnamefont {C.~J.}\
  \bibnamefont {Lambert}}, \bibinfo {author} {\bibfnamefont {G.~A.~D.}\
  \bibnamefont {Briggs}}, \ and\ \bibinfo {author} {\bibfnamefont {J.~A.}\
  \bibnamefont {Mol}},\ }\href {\doibase 10.1021/acs.nanolett.5b03434}
  {\bibfield  {journal} {\bibinfo  {journal} {Nano Lett.}\ }\textbf {\bibinfo
  {volume} {16}},\ \bibinfo {pages} {170} (\bibinfo {year} {2016})}\BibitemShut
  {NoStop}%
\bibitem [{\citenamefont {Ness}\ \emph {et~al.}(2001)\citenamefont {Ness},
  \citenamefont {Shevlin},\ and\ \citenamefont {Fisher}}]{Ness.01.PRB}%
  \BibitemOpen
  \bibfield  {author} {\bibinfo {author} {\bibfnamefont {H.}~\bibnamefont
  {Ness}}, \bibinfo {author} {\bibfnamefont {S.~A.}\ \bibnamefont {Shevlin}}, \
  and\ \bibinfo {author} {\bibfnamefont {A.~J.}\ \bibnamefont {Fisher}},\
  }\href {\doibase 10.1103/PhysRevB.63.125422} {\bibfield  {journal} {\bibinfo
  {journal} {Phys. Rev. B}\ }\textbf {\bibinfo {volume} {63}},\ \bibinfo
  {pages} {125422} (\bibinfo {year} {2001})}\BibitemShut {NoStop}%
\bibitem [{\citenamefont {\v{C}\'{\i}\v{z}ek}\ \emph
  {et~al.}(2004)\citenamefont {\v{C}\'{\i}\v{z}ek}, \citenamefont {Thoss},\
  and\ \citenamefont {Domcke}}]{Cizek.04.PRB}%
  \BibitemOpen
  \bibfield  {author} {\bibinfo {author} {\bibfnamefont {M.}~\bibnamefont
  {\v{C}\'{\i}\v{z}ek}}, \bibinfo {author} {\bibfnamefont {M.}~\bibnamefont
  {Thoss}}, \ and\ \bibinfo {author} {\bibfnamefont {W.}~\bibnamefont
  {Domcke}},\ }\href {\doibase 10.1103/PhysRevB.70.125406} {\bibfield
  {journal} {\bibinfo  {journal} {Phys. Rev. B}\ }\textbf {\bibinfo {volume}
  {70}},\ \bibinfo {pages} {125406} (\bibinfo {year} {2004})}\BibitemShut
  {NoStop}%
\bibitem [{\citenamefont {Caspary~Toroker}\ and\ \citenamefont
  {Peskin}(2007)}]{Toroker.07.JCP}%
  \BibitemOpen
  \bibfield  {author} {\bibinfo {author} {\bibfnamefont {M.}~\bibnamefont
  {Caspary~Toroker}}\ and\ \bibinfo {author} {\bibfnamefont {U.}~\bibnamefont
  {Peskin}},\ }\href {\doibase 10.1063/1.2759916} {\bibfield  {journal}
  {\bibinfo  {journal} {J. Chem. Phys.}\ }\textbf {\bibinfo {volume} {127}},\
  \bibinfo {pages} {154706} (\bibinfo {year} {2007})}\BibitemShut {NoStop}%
\bibitem [{\citenamefont {Benesch}\ \emph {et~al.}(2008)\citenamefont
  {Benesch}, \citenamefont {Cizek}, \citenamefont {Klimes}, \citenamefont
  {Kondov}, \citenamefont {Thoss},\ and\ \citenamefont
  {Domcke}}]{Benesch.08.JPCC}%
  \BibitemOpen
  \bibfield  {author} {\bibinfo {author} {\bibfnamefont {C.}~\bibnamefont
  {Benesch}}, \bibinfo {author} {\bibfnamefont {M.}~\bibnamefont {Cizek}},
  \bibinfo {author} {\bibfnamefont {J.}~\bibnamefont {Klimes}}, \bibinfo
  {author} {\bibfnamefont {I.}~\bibnamefont {Kondov}}, \bibinfo {author}
  {\bibfnamefont {M.}~\bibnamefont {Thoss}}, \ and\ \bibinfo {author}
  {\bibfnamefont {W.}~\bibnamefont {Domcke}},\ }\href {\doibase
  10.1021/jp711940n} {\bibfield  {journal} {\bibinfo  {journal} {J. Phys. Chem.
  C}\ }\textbf {\bibinfo {volume} {112}},\ \bibinfo {pages} {9880} (\bibinfo
  {year} {2008})}\BibitemShut {NoStop}%
\bibitem [{\citenamefont {Zimbovskaya}\ and\ \citenamefont
  {Kuklja}(2009)}]{Zimbovskaya.09.JCP}%
  \BibitemOpen
  \bibfield  {author} {\bibinfo {author} {\bibfnamefont {N.~A.}\ \bibnamefont
  {Zimbovskaya}}\ and\ \bibinfo {author} {\bibfnamefont {M.~M.}\ \bibnamefont
  {Kuklja}},\ }\href {\doibase 10.1063/1.3231604} {\bibfield  {journal}
  {\bibinfo  {journal} {J. Chem. Phys.}\ }\textbf {\bibinfo {volume} {131}},\
  \bibinfo {pages} {114703} (\bibinfo {year} {2009})}\BibitemShut {NoStop}%
\bibitem [{\citenamefont {Jorn}\ and\ \citenamefont
  {Seideman}(2009)}]{Jorn.09.JCP}%
  \BibitemOpen
  \bibfield  {author} {\bibinfo {author} {\bibfnamefont {R.}~\bibnamefont
  {Jorn}}\ and\ \bibinfo {author} {\bibfnamefont {T.}~\bibnamefont
  {Seideman}},\ }\href {\doibase 10.1063/1.3276281} {\bibfield  {journal}
  {\bibinfo  {journal} {J. Chem. Phys.}\ }\textbf {\bibinfo {volume} {131}},\
  \bibinfo {pages} {244114} (\bibinfo {year} {2009})}\BibitemShut {NoStop}%
\bibitem [{\citenamefont {Koch}\ \emph
  {et~al.}(2006{\natexlab{b}})\citenamefont {Koch}, \citenamefont {Semmelhack},
  \citenamefont {von Oppen},\ and\ \citenamefont {Nitzan}}]{Koch.06.PRBa}%
  \BibitemOpen
  \bibfield  {author} {\bibinfo {author} {\bibfnamefont {J.}~\bibnamefont
  {Koch}}, \bibinfo {author} {\bibfnamefont {M.}~\bibnamefont {Semmelhack}},
  \bibinfo {author} {\bibfnamefont {F.}~\bibnamefont {von Oppen}}, \ and\
  \bibinfo {author} {\bibfnamefont {A.}~\bibnamefont {Nitzan}},\ }\href
  {\doibase 10.1103/PhysRevB.73.155306} {\bibfield  {journal} {\bibinfo
  {journal} {Phys. Rev. B}\ }\textbf {\bibinfo {volume} {73}},\ \bibinfo
  {pages} {155306} (\bibinfo {year} {2006}{\natexlab{b}})}\BibitemShut
  {NoStop}%
\bibitem [{\citenamefont {Shen}\ \emph {et~al.}(2007)\citenamefont {Shen},
  \citenamefont {Dong}, \citenamefont {Lei},\ and\ \citenamefont
  {Horing}}]{Shen.07.PRB}%
  \BibitemOpen
  \bibfield  {author} {\bibinfo {author} {\bibfnamefont {X.~Y.}\ \bibnamefont
  {Shen}}, \bibinfo {author} {\bibfnamefont {B.}~\bibnamefont {Dong}}, \bibinfo
  {author} {\bibfnamefont {X.~L.}\ \bibnamefont {Lei}}, \ and\ \bibinfo
  {author} {\bibfnamefont {N.~J.~M.}\ \bibnamefont {Horing}},\ }\href {\doibase
  10.1103/PhysRevB.76.115308} {\bibfield  {journal} {\bibinfo  {journal} {Phys.
  Rev. B}\ }\textbf {\bibinfo {volume} {76}},\ \bibinfo {pages} {115308}
  (\bibinfo {year} {2007})}\BibitemShut {NoStop}%
\bibitem [{\citenamefont {Dzhioev}\ and\ \citenamefont
  {Kosov}(2011)}]{Dzhioev.11.JCP}%
  \BibitemOpen
  \bibfield  {author} {\bibinfo {author} {\bibfnamefont {A.~A.}\ \bibnamefont
  {Dzhioev}}\ and\ \bibinfo {author} {\bibfnamefont {D.~S.}\ \bibnamefont
  {Kosov}},\ }\href {\doibase 10.1063/1.3626521} {\bibfield  {journal}
  {\bibinfo  {journal} {J. Chem. Phys.}\ }\textbf {\bibinfo {volume} {135}},\
  \bibinfo {pages} {074701} (\bibinfo {year} {2011})}\BibitemShut {NoStop}%
\bibitem [{\citenamefont {Li}\ \emph {et~al.}(2014)\citenamefont {Li},
  \citenamefont {Wilner}, \citenamefont {Thoss}, \citenamefont {Rabani},\ and\
  \citenamefont {Miller}}]{Li.14.JCP}%
  \BibitemOpen
  \bibfield  {author} {\bibinfo {author} {\bibfnamefont {B.}~\bibnamefont
  {Li}}, \bibinfo {author} {\bibfnamefont {E.~Y.}\ \bibnamefont {Wilner}},
  \bibinfo {author} {\bibfnamefont {M.}~\bibnamefont {Thoss}}, \bibinfo
  {author} {\bibfnamefont {E.}~\bibnamefont {Rabani}}, \ and\ \bibinfo {author}
  {\bibfnamefont {W.~H.}\ \bibnamefont {Miller}},\ }\href {\doibase
  10.1063/1.4867789} {\bibfield  {journal} {\bibinfo  {journal} {J. Chem.
  Phys.}\ }\textbf {\bibinfo {volume} {140}},\ \bibinfo {pages} {104110}
  (\bibinfo {year} {2014})}\BibitemShut {NoStop}%
\bibitem [{\citenamefont {Segal}\ \emph {et~al.}(2000)\citenamefont {Segal},
  \citenamefont {Nitzan}, \citenamefont {Davis}, \citenamefont {Wasielewski},\
  and\ \citenamefont {Ratner}}]{Segal.00.JPCB}%
  \BibitemOpen
  \bibfield  {author} {\bibinfo {author} {\bibfnamefont {D.}~\bibnamefont
  {Segal}}, \bibinfo {author} {\bibfnamefont {A.}~\bibnamefont {Nitzan}},
  \bibinfo {author} {\bibfnamefont {W.~B.}\ \bibnamefont {Davis}}, \bibinfo
  {author} {\bibfnamefont {M.~R.}\ \bibnamefont {Wasielewski}}, \ and\ \bibinfo
  {author} {\bibfnamefont {M.~A.}\ \bibnamefont {Ratner}},\ }\href {\doibase
  10.1021/jp993260f} {\bibfield  {journal} {\bibinfo  {journal} {J. Phys. Chem.
  B}\ }\textbf {\bibinfo {volume} {104}},\ \bibinfo {pages} {3817} (\bibinfo
  {year} {2000})}\BibitemShut {NoStop}%
\bibitem [{\citenamefont {May}(2002)}]{May.02.PRB}%
  \BibitemOpen
  \bibfield  {author} {\bibinfo {author} {\bibfnamefont {V.}~\bibnamefont
  {May}},\ }\href {\doibase 10.1103/PhysRevB.66.245411} {\bibfield  {journal}
  {\bibinfo  {journal} {Phys. Rev. B}\ }\textbf {\bibinfo {volume} {66}},\
  \bibinfo {pages} {245411} (\bibinfo {year} {2002})}\BibitemShut {NoStop}%
\bibitem [{\citenamefont {Mitra}\ \emph {et~al.}(2004)\citenamefont {Mitra},
  \citenamefont {Aleiner},\ and\ \citenamefont {Millis}}]{Mitra.04.PRB}%
  \BibitemOpen
  \bibfield  {author} {\bibinfo {author} {\bibfnamefont {A.}~\bibnamefont
  {Mitra}}, \bibinfo {author} {\bibfnamefont {I.}~\bibnamefont {Aleiner}}, \
  and\ \bibinfo {author} {\bibfnamefont {A.~J.}\ \bibnamefont {Millis}},\
  }\href {\doibase 10.1103/PhysRevB.69.245302} {\bibfield  {journal} {\bibinfo
  {journal} {Phys. Rev. B}\ }\textbf {\bibinfo {volume} {69}},\ \bibinfo
  {pages} {245302} (\bibinfo {year} {2004})}\BibitemShut {NoStop}%
\bibitem [{\citenamefont {Pedersen}\ and\ \citenamefont
  {Wacker}(2005)}]{Pedersen.05.PRB}%
  \BibitemOpen
  \bibfield  {author} {\bibinfo {author} {\bibfnamefont {J.~N.}\ \bibnamefont
  {Pedersen}}\ and\ \bibinfo {author} {\bibfnamefont {A.}~\bibnamefont
  {Wacker}},\ }\href {\doibase 10.1103/PhysRevB.72.195330} {\bibfield
  {journal} {\bibinfo  {journal} {Phys. Rev. B}\ }\textbf {\bibinfo {volume}
  {72}},\ \bibinfo {pages} {195330} (\bibinfo {year} {2005})}\BibitemShut
  {NoStop}%
\bibitem [{\citenamefont {Harbola}\ \emph {et~al.}(2006)\citenamefont
  {Harbola}, \citenamefont {Esposito},\ and\ \citenamefont
  {Mukamel}}]{Harbola.06.PRB}%
  \BibitemOpen
  \bibfield  {author} {\bibinfo {author} {\bibfnamefont {U.}~\bibnamefont
  {Harbola}}, \bibinfo {author} {\bibfnamefont {M.}~\bibnamefont {Esposito}}, \
  and\ \bibinfo {author} {\bibfnamefont {S.}~\bibnamefont {Mukamel}},\ }\href
  {\doibase 10.1103/PhysRevB.74.235309} {\bibfield  {journal} {\bibinfo
  {journal} {Phys. Rev. B}\ }\textbf {\bibinfo {volume} {74}},\ \bibinfo
  {pages} {235309} (\bibinfo {year} {2006})}\BibitemShut {NoStop}%
\bibitem [{\citenamefont {Donarini}\ \emph {et~al.}(2006)\citenamefont
  {Donarini}, \citenamefont {Grifoni},\ and\ \citenamefont
  {Richter}}]{Donarini.06.PRL}%
  \BibitemOpen
  \bibfield  {author} {\bibinfo {author} {\bibfnamefont {A.}~\bibnamefont
  {Donarini}}, \bibinfo {author} {\bibfnamefont {M.}~\bibnamefont {Grifoni}}, \
  and\ \bibinfo {author} {\bibfnamefont {K.}~\bibnamefont {Richter}},\ }\href
  {\doibase 10.1103/PhysRevLett.97.166801} {\bibfield  {journal} {\bibinfo
  {journal} {Phys. Rev. Lett.}\ }\textbf {\bibinfo {volume} {97}},\ \bibinfo
  {pages} {166801} (\bibinfo {year} {2006})}\BibitemShut {NoStop}%
\bibitem [{\citenamefont {Leijnse}\ and\ \citenamefont
  {Wegewijs}(2008)}]{Leijnse.08.PRB}%
  \BibitemOpen
  \bibfield  {author} {\bibinfo {author} {\bibfnamefont {M.}~\bibnamefont
  {Leijnse}}\ and\ \bibinfo {author} {\bibfnamefont {M.~R.}\ \bibnamefont
  {Wegewijs}},\ }\href {\doibase 10.1103/PhysRevB.78.235424} {\bibfield
  {journal} {\bibinfo  {journal} {Phys. Rev. B}\ }\textbf {\bibinfo {volume}
  {78}},\ \bibinfo {pages} {235424} (\bibinfo {year} {2008})}\BibitemShut
  {NoStop}%
\bibitem [{\citenamefont {Timm}(2008)}]{Timm.08.PRB}%
  \BibitemOpen
  \bibfield  {author} {\bibinfo {author} {\bibfnamefont {C.}~\bibnamefont
  {Timm}},\ }\href {\doibase 10.1103/PhysRevB.77.195416} {\bibfield  {journal}
  {\bibinfo  {journal} {Phys. Rev. B}\ }\textbf {\bibinfo {volume} {77}},\
  \bibinfo {pages} {195416} (\bibinfo {year} {2008})}\BibitemShut {NoStop}%
\bibitem [{\citenamefont {Esposito}\ and\ \citenamefont
  {Galperin}(2009)}]{Esposito.09.PRB}%
  \BibitemOpen
  \bibfield  {author} {\bibinfo {author} {\bibfnamefont {M.}~\bibnamefont
  {Esposito}}\ and\ \bibinfo {author} {\bibfnamefont {M.}~\bibnamefont
  {Galperin}},\ }\href {\doibase 10.1103/PhysRevB.79.205303} {\bibfield
  {journal} {\bibinfo  {journal} {Phys. Rev. B}\ }\textbf {\bibinfo {volume}
  {79}},\ \bibinfo {pages} {205303} (\bibinfo {year} {2009})}\BibitemShut
  {NoStop}%
\bibitem [{\citenamefont {Volkovich}\ \emph {et~al.}(2011)\citenamefont
  {Volkovich}, \citenamefont {H\"artle}, \citenamefont {Thoss},\ and\
  \citenamefont {Peskin}}]{Volkovich.11.PCCP}%
  \BibitemOpen
  \bibfield  {author} {\bibinfo {author} {\bibfnamefont {R.}~\bibnamefont
  {Volkovich}}, \bibinfo {author} {\bibfnamefont {R.}~\bibnamefont {H\"artle}},
  \bibinfo {author} {\bibfnamefont {M.}~\bibnamefont {Thoss}}, \ and\ \bibinfo
  {author} {\bibfnamefont {U.}~\bibnamefont {Peskin}},\ }\href {\doibase
  10.1039/C1CP21161G} {\bibfield  {journal} {\bibinfo  {journal} {Phys. Chem.
  Chem. Phys.}\ }\textbf {\bibinfo {volume} {13}},\ \bibinfo {pages} {14333}
  (\bibinfo {year} {2011})}\BibitemShut {NoStop}%
\bibitem [{\citenamefont {H\"artle}\ and\ \citenamefont
  {Thoss}(2011)}]{Hartle.11.PRB}%
  \BibitemOpen
  \bibfield  {author} {\bibinfo {author} {\bibfnamefont {R.}~\bibnamefont
  {H\"artle}}\ and\ \bibinfo {author} {\bibfnamefont {M.}~\bibnamefont
  {Thoss}},\ }\href {\doibase 10.1103/PhysRevB.83.115414} {\bibfield  {journal}
  {\bibinfo  {journal} {Phys. Rev. B}\ }\textbf {\bibinfo {volume} {83}},\
  \bibinfo {pages} {115414} (\bibinfo {year} {2011})}\BibitemShut {NoStop}%
\bibitem [{\citenamefont {Sowa}\ \emph
  {et~al.}(2017{\natexlab{a}})\citenamefont {Sowa}, \citenamefont {Mol},
  \citenamefont {Briggs},\ and\ \citenamefont {Gauger}}]{Sowa.17.PCCP}%
  \BibitemOpen
  \bibfield  {author} {\bibinfo {author} {\bibfnamefont {J.~K.}\ \bibnamefont
  {Sowa}}, \bibinfo {author} {\bibfnamefont {J.~A.}\ \bibnamefont {Mol}},
  \bibinfo {author} {\bibfnamefont {G.~A.~D.}\ \bibnamefont {Briggs}}, \ and\
  \bibinfo {author} {\bibfnamefont {E.~M.}\ \bibnamefont {Gauger}},\ }\href
  {\doibase 10.1039/C7CP06237K} {\bibfield  {journal} {\bibinfo  {journal}
  {Phys. Chem. Chem. Phys.}\ }\textbf {\bibinfo {volume} {19}},\ \bibinfo
  {pages} {29534} (\bibinfo {year} {2017}{\natexlab{a}})}\BibitemShut {NoStop}%
\bibitem [{\citenamefont {Flensberg}(2003)}]{Flensberg.03.PRB}%
  \BibitemOpen
  \bibfield  {author} {\bibinfo {author} {\bibfnamefont {K.}~\bibnamefont
  {Flensberg}},\ }\href {\doibase 10.1103/PhysRevB.68.205323} {\bibfield
  {journal} {\bibinfo  {journal} {Phys. Rev. B}\ }\textbf {\bibinfo {volume}
  {68}},\ \bibinfo {pages} {205323} (\bibinfo {year} {2003})}\BibitemShut
  {NoStop}%
\bibitem [{\citenamefont {Galperin}\ \emph {et~al.}(2004)\citenamefont
  {Galperin}, \citenamefont {Ratner},\ and\ \citenamefont
  {Nitzan}}]{Galperin.04.JCP}%
  \BibitemOpen
  \bibfield  {author} {\bibinfo {author} {\bibfnamefont {M.}~\bibnamefont
  {Galperin}}, \bibinfo {author} {\bibfnamefont {M.~A.}\ \bibnamefont
  {Ratner}}, \ and\ \bibinfo {author} {\bibfnamefont {A.}~\bibnamefont
  {Nitzan}},\ }\href {\doibase 10.1063/1.1814076} {\bibfield  {journal}
  {\bibinfo  {journal} {J. Chem. Phys.}\ }\textbf {\bibinfo {volume} {121}},\
  \bibinfo {pages} {11965} (\bibinfo {year} {2004})}\BibitemShut {NoStop}%
\bibitem [{\citenamefont {Galperin}\ \emph
  {et~al.}(2006{\natexlab{a}})\citenamefont {Galperin}, \citenamefont
  {Nitzan},\ and\ \citenamefont {Ratner}}]{Galperin.06.PRBa}%
  \BibitemOpen
  \bibfield  {author} {\bibinfo {author} {\bibfnamefont {M.}~\bibnamefont
  {Galperin}}, \bibinfo {author} {\bibfnamefont {A.}~\bibnamefont {Nitzan}}, \
  and\ \bibinfo {author} {\bibfnamefont {M.~A.}\ \bibnamefont {Ratner}},\
  }\href {\doibase 10.1103/PhysRevB.73.045314} {\bibfield  {journal} {\bibinfo
  {journal} {Phys. Rev. B}\ }\textbf {\bibinfo {volume} {73}},\ \bibinfo
  {pages} {045314} (\bibinfo {year} {2006}{\natexlab{a}})}\BibitemShut
  {NoStop}%
\bibitem [{\citenamefont {Galperin}\ \emph
  {et~al.}(2006{\natexlab{b}})\citenamefont {Galperin}, \citenamefont
  {Nitzan},\ and\ \citenamefont {Ratner}}]{Galperin.06.PRB}%
  \BibitemOpen
  \bibfield  {author} {\bibinfo {author} {\bibfnamefont {M.}~\bibnamefont
  {Galperin}}, \bibinfo {author} {\bibfnamefont {A.}~\bibnamefont {Nitzan}}, \
  and\ \bibinfo {author} {\bibfnamefont {M.~A.}\ \bibnamefont {Ratner}},\
  }\href {\doibase 10.1103/PhysRevB.74.075326} {\bibfield  {journal} {\bibinfo
  {journal} {Phys. Rev. B}\ }\textbf {\bibinfo {volume} {74}},\ \bibinfo
  {pages} {075326} (\bibinfo {year} {2006}{\natexlab{b}})}\BibitemShut
  {NoStop}%
\bibitem [{\citenamefont {Ryndyk}\ \emph {et~al.}(2006)\citenamefont {Ryndyk},
  \citenamefont {Hartung},\ and\ \citenamefont {Cuniberti}}]{Ryndyk.06.PRB}%
  \BibitemOpen
  \bibfield  {author} {\bibinfo {author} {\bibfnamefont {D.~A.}\ \bibnamefont
  {Ryndyk}}, \bibinfo {author} {\bibfnamefont {M.}~\bibnamefont {Hartung}}, \
  and\ \bibinfo {author} {\bibfnamefont {G.}~\bibnamefont {Cuniberti}},\ }\href
  {\doibase 10.1103/PhysRevB.73.045420} {\bibfield  {journal} {\bibinfo
  {journal} {Phys. Rev. B}\ }\textbf {\bibinfo {volume} {73}},\ \bibinfo
  {pages} {045420} (\bibinfo {year} {2006})}\BibitemShut {NoStop}%
\bibitem [{\citenamefont {Frederiksen}\ \emph {et~al.}(2007)\citenamefont
  {Frederiksen}, \citenamefont {Paulsson}, \citenamefont {Brandbyge},\ and\
  \citenamefont {Jauho}}]{Frederiksen.07.PRB}%
  \BibitemOpen
  \bibfield  {author} {\bibinfo {author} {\bibfnamefont {T.}~\bibnamefont
  {Frederiksen}}, \bibinfo {author} {\bibfnamefont {M.}~\bibnamefont
  {Paulsson}}, \bibinfo {author} {\bibfnamefont {M.}~\bibnamefont {Brandbyge}},
  \ and\ \bibinfo {author} {\bibfnamefont {A.-P.}\ \bibnamefont {Jauho}},\
  }\href {\doibase 10.1103/PhysRevB.75.205413} {\bibfield  {journal} {\bibinfo
  {journal} {Phys. Rev. B}\ }\textbf {\bibinfo {volume} {75}},\ \bibinfo
  {pages} {205413} (\bibinfo {year} {2007})}\BibitemShut {NoStop}%
\bibitem [{\citenamefont {Entin-Wohlman}\ \emph {et~al.}(2009)\citenamefont
  {Entin-Wohlman}, \citenamefont {Imry},\ and\ \citenamefont
  {Aharony}}]{Entin.08.PRB}%
  \BibitemOpen
  \bibfield  {author} {\bibinfo {author} {\bibfnamefont {O.}~\bibnamefont
  {Entin-Wohlman}}, \bibinfo {author} {\bibfnamefont {Y.}~\bibnamefont {Imry}},
  \ and\ \bibinfo {author} {\bibfnamefont {A.}~\bibnamefont {Aharony}},\ }\href
  {\doibase 10.1103/PhysRevB.80.035417} {\bibfield  {journal} {\bibinfo
  {journal} {Phys. Rev. B}\ }\textbf {\bibinfo {volume} {80}},\ \bibinfo
  {pages} {035417} (\bibinfo {year} {2009})}\BibitemShut {NoStop}%
\bibitem [{\citenamefont {H\"artle}\ \emph {et~al.}(2008)\citenamefont
  {H\"artle}, \citenamefont {Benesch},\ and\ \citenamefont
  {Thoss}}]{Hartle.08.PRB}%
  \BibitemOpen
  \bibfield  {author} {\bibinfo {author} {\bibfnamefont {R.}~\bibnamefont
  {H\"artle}}, \bibinfo {author} {\bibfnamefont {C.}~\bibnamefont {Benesch}}, \
  and\ \bibinfo {author} {\bibfnamefont {M.}~\bibnamefont {Thoss}},\ }\href
  {\doibase 10.1103/PhysRevB.77.205314} {\bibfield  {journal} {\bibinfo
  {journal} {Phys. Rev. B}\ }\textbf {\bibinfo {volume} {77}},\ \bibinfo
  {pages} {205314} (\bibinfo {year} {2008})}\BibitemShut {NoStop}%
\bibitem [{\citenamefont {H\"artle}\ \emph {et~al.}(2009)\citenamefont
  {H\"artle}, \citenamefont {Benesch},\ and\ \citenamefont
  {Thoss}}]{Hartle.09.PRL}%
  \BibitemOpen
  \bibfield  {author} {\bibinfo {author} {\bibfnamefont {R.}~\bibnamefont
  {H\"artle}}, \bibinfo {author} {\bibfnamefont {C.}~\bibnamefont {Benesch}}, \
  and\ \bibinfo {author} {\bibfnamefont {M.}~\bibnamefont {Thoss}},\ }\href
  {\doibase 10.1103/PhysRevLett.102.146801} {\bibfield  {journal} {\bibinfo
  {journal} {Phys. Rev. Lett.}\ }\textbf {\bibinfo {volume} {102}},\ \bibinfo
  {pages} {146801} (\bibinfo {year} {2009})}\BibitemShut {NoStop}%
\bibitem [{\citenamefont {Erpenbeck}\ \emph {et~al.}(2015)\citenamefont
  {Erpenbeck}, \citenamefont {H\"artle},\ and\ \citenamefont
  {Thoss}}]{Erpenbeck.15.PRB}%
  \BibitemOpen
  \bibfield  {author} {\bibinfo {author} {\bibfnamefont {A.}~\bibnamefont
  {Erpenbeck}}, \bibinfo {author} {\bibfnamefont {R.}~\bibnamefont {H\"artle}},
  \ and\ \bibinfo {author} {\bibfnamefont {M.}~\bibnamefont {Thoss}},\ }\href
  {\doibase 10.1103/PhysRevB.91.195418} {\bibfield  {journal} {\bibinfo
  {journal} {Phys. Rev. B}\ }\textbf {\bibinfo {volume} {91}},\ \bibinfo
  {pages} {195418} (\bibinfo {year} {2015})}\BibitemShut {NoStop}%
\bibitem [{\citenamefont {Cabra}\ \emph {et~al.}(2018)\citenamefont {Cabra},
  \citenamefont {Jensen},\ and\ \citenamefont {Galperin}}]{Cabra.18.JCP}%
  \BibitemOpen
  \bibfield  {author} {\bibinfo {author} {\bibfnamefont {G.}~\bibnamefont
  {Cabra}}, \bibinfo {author} {\bibfnamefont {A.}~\bibnamefont {Jensen}}, \
  and\ \bibinfo {author} {\bibfnamefont {M.}~\bibnamefont {Galperin}},\ }\href
  {\doibase 10.1063/1.5029252} {\bibfield  {journal} {\bibinfo  {journal} {J.
  Chem. Phys.}\ }\textbf {\bibinfo {volume} {148}},\ \bibinfo {pages} {204103}
  (\bibinfo {year} {2018})}\BibitemShut {NoStop}%
\bibitem [{\citenamefont {Galperin}\ \emph {et~al.}(2007)\citenamefont
  {Galperin}, \citenamefont {Ratner},\ and\ \citenamefont
  {Nitzan}}]{Galperin.07.JP}%
  \BibitemOpen
  \bibfield  {author} {\bibinfo {author} {\bibfnamefont {M.}~\bibnamefont
  {Galperin}}, \bibinfo {author} {\bibfnamefont {M.~A.}\ \bibnamefont
  {Ratner}}, \ and\ \bibinfo {author} {\bibfnamefont {A.}~\bibnamefont
  {Nitzan}},\ }\href {\doibase 10.1088/0953-8984/19/10/103201} {\bibfield
  {journal} {\bibinfo  {journal} {J. Phys.: Condens. Matter}\ }\textbf
  {\bibinfo {volume} {19}},\ \bibinfo {pages} {103201} (\bibinfo {year}
  {2007})}\BibitemShut {NoStop}%
\bibitem [{\citenamefont {Jin}\ \emph {et~al.}(2008)\citenamefont {Jin},
  \citenamefont {Zheng},\ and\ \citenamefont {Yan}}]{Jin.08.JCP}%
  \BibitemOpen
  \bibfield  {author} {\bibinfo {author} {\bibfnamefont {J.}~\bibnamefont
  {Jin}}, \bibinfo {author} {\bibfnamefont {X.}~\bibnamefont {Zheng}}, \ and\
  \bibinfo {author} {\bibfnamefont {Y.}~\bibnamefont {Yan}},\ }\href {\doibase
  10.1063/1.2938087} {\bibfield  {journal} {\bibinfo  {journal} {J. Chem.
  Phys.}\ }\textbf {\bibinfo {volume} {128}},\ \bibinfo {pages} {234703}
  (\bibinfo {year} {2008})}\BibitemShut {NoStop}%
\bibitem [{\citenamefont {Li}\ \emph {et~al.}(2012{\natexlab{b}})\citenamefont
  {Li}, \citenamefont {Tong}, \citenamefont {Zheng}, \citenamefont {Hou},
  \citenamefont {Wei}, \citenamefont {Hu},\ and\ \citenamefont
  {Yan}}]{Li.12.PRL}%
  \BibitemOpen
  \bibfield  {author} {\bibinfo {author} {\bibfnamefont {Z.}~\bibnamefont
  {Li}}, \bibinfo {author} {\bibfnamefont {N.}~\bibnamefont {Tong}}, \bibinfo
  {author} {\bibfnamefont {X.}~\bibnamefont {Zheng}}, \bibinfo {author}
  {\bibfnamefont {D.}~\bibnamefont {Hou}}, \bibinfo {author} {\bibfnamefont
  {J.}~\bibnamefont {Wei}}, \bibinfo {author} {\bibfnamefont {J.}~\bibnamefont
  {Hu}}, \ and\ \bibinfo {author} {\bibfnamefont {Y.}~\bibnamefont {Yan}},\
  }\href {\doibase 10.1103/PhysRevLett.109.266403} {\bibfield  {journal}
  {\bibinfo  {journal} {Phys. Rev. Lett.}\ }\textbf {\bibinfo {volume} {109}},\
  \bibinfo {pages} {266403} (\bibinfo {year} {2012}{\natexlab{b}})}\BibitemShut
  {NoStop}%
\bibitem [{\citenamefont {Jiang}\ \emph {et~al.}(2012)\citenamefont {Jiang},
  \citenamefont {Jin}, \citenamefont {Wang},\ and\ \citenamefont
  {Yan}}]{Jiang.12.PRB}%
  \BibitemOpen
  \bibfield  {author} {\bibinfo {author} {\bibfnamefont {F.}~\bibnamefont
  {Jiang}}, \bibinfo {author} {\bibfnamefont {J.}~\bibnamefont {Jin}}, \bibinfo
  {author} {\bibfnamefont {S.}~\bibnamefont {Wang}}, \ and\ \bibinfo {author}
  {\bibfnamefont {Y.}~\bibnamefont {Yan}},\ }\href {\doibase
  10.1103/PhysRevB.85.245427} {\bibfield  {journal} {\bibinfo  {journal} {Phys.
  Rev. B}\ }\textbf {\bibinfo {volume} {85}},\ \bibinfo {pages} {245427}
  (\bibinfo {year} {2012})}\BibitemShut {NoStop}%
\bibitem [{\citenamefont {H\"artle}\ \emph
  {et~al.}(2013{\natexlab{a}})\citenamefont {H\"artle}, \citenamefont {Cohen},
  \citenamefont {Reichman},\ and\ \citenamefont {Millis}}]{Hartle.13.PRBa}%
  \BibitemOpen
  \bibfield  {author} {\bibinfo {author} {\bibfnamefont {R.}~\bibnamefont
  {H\"artle}}, \bibinfo {author} {\bibfnamefont {G.}~\bibnamefont {Cohen}},
  \bibinfo {author} {\bibfnamefont {D.~R.}\ \bibnamefont {Reichman}}, \ and\
  \bibinfo {author} {\bibfnamefont {A.~J.}\ \bibnamefont {Millis}},\ }\href
  {\doibase 10.1103/PhysRevB.88.235426} {\bibfield  {journal} {\bibinfo
  {journal} {Phys. Rev. B}\ }\textbf {\bibinfo {volume} {88}},\ \bibinfo
  {pages} {235426} (\bibinfo {year} {2013}{\natexlab{a}})}\BibitemShut
  {NoStop}%
\bibitem [{\citenamefont {H\"artle}\ and\ \citenamefont
  {Millis}(2014)}]{Hartle.14.PRB}%
  \BibitemOpen
  \bibfield  {author} {\bibinfo {author} {\bibfnamefont {R.}~\bibnamefont
  {H\"artle}}\ and\ \bibinfo {author} {\bibfnamefont {A.~J.}\ \bibnamefont
  {Millis}},\ }\href {\doibase 10.1103/PhysRevB.90.245426} {\bibfield
  {journal} {\bibinfo  {journal} {Phys. Rev. B}\ }\textbf {\bibinfo {volume}
  {90}},\ \bibinfo {pages} {245426} (\bibinfo {year} {2014})}\BibitemShut
  {NoStop}%
\bibitem [{\citenamefont {H\"artle}\ \emph {et~al.}(2015)\citenamefont
  {H\"artle}, \citenamefont {Cohen}, \citenamefont {Reichman},\ and\
  \citenamefont {Millis}}]{Hartle.15.PRB}%
  \BibitemOpen
  \bibfield  {author} {\bibinfo {author} {\bibfnamefont {R.}~\bibnamefont
  {H\"artle}}, \bibinfo {author} {\bibfnamefont {G.}~\bibnamefont {Cohen}},
  \bibinfo {author} {\bibfnamefont {D.~R.}\ \bibnamefont {Reichman}}, \ and\
  \bibinfo {author} {\bibfnamefont {A.~J.}\ \bibnamefont {Millis}},\ }\href
  {\doibase 10.1103/PhysRevB.92.085430} {\bibfield  {journal} {\bibinfo
  {journal} {Phys. Rev. B}\ }\textbf {\bibinfo {volume} {92}},\ \bibinfo
  {pages} {085430} (\bibinfo {year} {2015})}\BibitemShut {NoStop}%
\bibitem [{\citenamefont {Schinabeck}\ \emph {et~al.}(2016)\citenamefont
  {Schinabeck}, \citenamefont {Erpenbeck}, \citenamefont {H\"artle},\ and\
  \citenamefont {Thoss}}]{Schinabeck.16.PRB}%
  \BibitemOpen
  \bibfield  {author} {\bibinfo {author} {\bibfnamefont {C.}~\bibnamefont
  {Schinabeck}}, \bibinfo {author} {\bibfnamefont {A.}~\bibnamefont
  {Erpenbeck}}, \bibinfo {author} {\bibfnamefont {R.}~\bibnamefont {H\"artle}},
  \ and\ \bibinfo {author} {\bibfnamefont {M.}~\bibnamefont {Thoss}},\ }\href
  {\doibase 10.1103/PhysRevB.94.201407} {\bibfield  {journal} {\bibinfo
  {journal} {Phys. Rev. B}\ }\textbf {\bibinfo {volume} {94}},\ \bibinfo
  {pages} {201407} (\bibinfo {year} {2016})}\BibitemShut {NoStop}%
\bibitem [{\citenamefont {Schinabeck}\ \emph {et~al.}(2018)\citenamefont
  {Schinabeck}, \citenamefont {H\"artle},\ and\ \citenamefont
  {Thoss}}]{Schinabeck.18.PRB}%
  \BibitemOpen
  \bibfield  {author} {\bibinfo {author} {\bibfnamefont {C.}~\bibnamefont
  {Schinabeck}}, \bibinfo {author} {\bibfnamefont {R.}~\bibnamefont
  {H\"artle}}, \ and\ \bibinfo {author} {\bibfnamefont {M.}~\bibnamefont
  {Thoss}},\ }\href {\doibase 10.1103/PhysRevB.97.235429} {\bibfield  {journal}
  {\bibinfo  {journal} {Phys. Rev. B}\ }\textbf {\bibinfo {volume} {97}},\
  \bibinfo {pages} {235429} (\bibinfo {year} {2018})}\BibitemShut {NoStop}%
\bibitem [{\citenamefont {Erpenbeck}\ \emph {et~al.}(2018)\citenamefont
  {Erpenbeck}, \citenamefont {Hertlein}, \citenamefont {Schinabeck},\ and\
  \citenamefont {Thoss}}]{Erpenbeck.18.JCP}%
  \BibitemOpen
  \bibfield  {author} {\bibinfo {author} {\bibfnamefont {A.}~\bibnamefont
  {Erpenbeck}}, \bibinfo {author} {\bibfnamefont {C.}~\bibnamefont {Hertlein}},
  \bibinfo {author} {\bibfnamefont {C.}~\bibnamefont {Schinabeck}}, \ and\
  \bibinfo {author} {\bibfnamefont {M.}~\bibnamefont {Thoss}},\ }\href
  {\doibase 10.1063/1.5041716} {\bibfield  {journal} {\bibinfo  {journal} {J.
  Chem. Phys.}\ }\textbf {\bibinfo {volume} {149}},\ \bibinfo {pages} {064106}
  (\bibinfo {year} {2018})}\BibitemShut {NoStop}%
\bibitem [{\citenamefont {Wilner}\ \emph {et~al.}(2014)\citenamefont {Wilner},
  \citenamefont {Wang}, \citenamefont {Thoss},\ and\ \citenamefont
  {Rabani}}]{Wilner.14.PRB}%
  \BibitemOpen
  \bibfield  {author} {\bibinfo {author} {\bibfnamefont {E.~Y.}\ \bibnamefont
  {Wilner}}, \bibinfo {author} {\bibfnamefont {H.}~\bibnamefont {Wang}},
  \bibinfo {author} {\bibfnamefont {M.}~\bibnamefont {Thoss}}, \ and\ \bibinfo
  {author} {\bibfnamefont {E.}~\bibnamefont {Rabani}},\ }\href {\doibase
  10.1103/PhysRevB.89.205129} {\bibfield  {journal} {\bibinfo  {journal} {Phys.
  Rev. B}\ }\textbf {\bibinfo {volume} {89}},\ \bibinfo {pages} {205129}
  (\bibinfo {year} {2014})}\BibitemShut {NoStop}%
\bibitem [{\citenamefont {Wang}\ and\ \citenamefont
  {Thoss}(2009)}]{Wang.14.JCP}%
  \BibitemOpen
  \bibfield  {author} {\bibinfo {author} {\bibfnamefont {H.}~\bibnamefont
  {Wang}}\ and\ \bibinfo {author} {\bibfnamefont {M.}~\bibnamefont {Thoss}},\
  }\href {\doibase 10.1063/1.3173823} {\bibfield  {journal} {\bibinfo
  {journal} {J. Chem. Phys.}\ }\textbf {\bibinfo {volume} {131}},\ \bibinfo
  {pages} {024114} (\bibinfo {year} {2009})}\BibitemShut {NoStop}%
\bibitem [{\citenamefont {Werner}\ \emph {et~al.}(2009)\citenamefont {Werner},
  \citenamefont {Oka},\ and\ \citenamefont {Millis}}]{Werner.09.PRB}%
  \BibitemOpen
  \bibfield  {author} {\bibinfo {author} {\bibfnamefont {P.}~\bibnamefont
  {Werner}}, \bibinfo {author} {\bibfnamefont {T.}~\bibnamefont {Oka}}, \ and\
  \bibinfo {author} {\bibfnamefont {A.~J.}\ \bibnamefont {Millis}},\ }\href
  {\doibase 10.1103/PhysRevB.79.035320} {\bibfield  {journal} {\bibinfo
  {journal} {Phys. Rev. B}\ }\textbf {\bibinfo {volume} {79}},\ \bibinfo
  {pages} {035320} (\bibinfo {year} {2009})}\BibitemShut {NoStop}%
\bibitem [{\citenamefont {Gull}\ \emph {et~al.}(2011)\citenamefont {Gull},
  \citenamefont {Millis}, \citenamefont {Lichtenstein}, \citenamefont
  {Rubtsov}, \citenamefont {Troyer},\ and\ \citenamefont
  {Werner}}]{Gull.11.RMP}%
  \BibitemOpen
  \bibfield  {author} {\bibinfo {author} {\bibfnamefont {E.}~\bibnamefont
  {Gull}}, \bibinfo {author} {\bibfnamefont {A.~J.}\ \bibnamefont {Millis}},
  \bibinfo {author} {\bibfnamefont {A.~I.}\ \bibnamefont {Lichtenstein}},
  \bibinfo {author} {\bibfnamefont {A.~N.}\ \bibnamefont {Rubtsov}}, \bibinfo
  {author} {\bibfnamefont {M.}~\bibnamefont {Troyer}}, \ and\ \bibinfo {author}
  {\bibfnamefont {P.}~\bibnamefont {Werner}},\ }\href {\doibase
  10.1103/RevModPhys.83.349} {\bibfield  {journal} {\bibinfo  {journal} {Rev.
  Mod. Phys.}\ }\textbf {\bibinfo {volume} {83}},\ \bibinfo {pages} {349}
  (\bibinfo {year} {2011})}\BibitemShut {NoStop}%
\bibitem [{\citenamefont {Cohen}\ \emph {et~al.}(2013)\citenamefont {Cohen},
  \citenamefont {Gull}, \citenamefont {Reichman}, \citenamefont {Millis},\ and\
  \citenamefont {Rabani}}]{Cohen.13.PRB}%
  \BibitemOpen
  \bibfield  {author} {\bibinfo {author} {\bibfnamefont {G.}~\bibnamefont
  {Cohen}}, \bibinfo {author} {\bibfnamefont {E.}~\bibnamefont {Gull}},
  \bibinfo {author} {\bibfnamefont {D.~R.}\ \bibnamefont {Reichman}}, \bibinfo
  {author} {\bibfnamefont {A.~J.}\ \bibnamefont {Millis}}, \ and\ \bibinfo
  {author} {\bibfnamefont {E.}~\bibnamefont {Rabani}},\ }\href {\doibase
  10.1103/PhysRevB.87.195108} {\bibfield  {journal} {\bibinfo  {journal} {Phys.
  Rev. B}\ }\textbf {\bibinfo {volume} {87}},\ \bibinfo {pages} {195108}
  (\bibinfo {year} {2013})}\BibitemShut {NoStop}%
\bibitem [{\citenamefont {M\"uhlbacher}\ and\ \citenamefont
  {Rabani}(2008)}]{Muhlbacher.08.PRL}%
  \BibitemOpen
  \bibfield  {author} {\bibinfo {author} {\bibfnamefont {L.}~\bibnamefont
  {M\"uhlbacher}}\ and\ \bibinfo {author} {\bibfnamefont {E.}~\bibnamefont
  {Rabani}},\ }\href {\doibase 10.1103/PhysRevLett.100.176403} {\bibfield
  {journal} {\bibinfo  {journal} {Phys. Rev. Lett.}\ }\textbf {\bibinfo
  {volume} {100}},\ \bibinfo {pages} {176403} (\bibinfo {year}
  {2008})}\BibitemShut {NoStop}%
\bibitem [{\citenamefont {Weiss}\ \emph {et~al.}(2008)\citenamefont {Weiss},
  \citenamefont {Eckel}, \citenamefont {Thorwart},\ and\ \citenamefont
  {Egger}}]{Weiss.08.PRB}%
  \BibitemOpen
  \bibfield  {author} {\bibinfo {author} {\bibfnamefont {S.}~\bibnamefont
  {Weiss}}, \bibinfo {author} {\bibfnamefont {J.}~\bibnamefont {Eckel}},
  \bibinfo {author} {\bibfnamefont {M.}~\bibnamefont {Thorwart}}, \ and\
  \bibinfo {author} {\bibfnamefont {R.}~\bibnamefont {Egger}},\ }\href
  {\doibase 10.1103/PhysRevB.77.195316} {\bibfield  {journal} {\bibinfo
  {journal} {Phys. Rev. B}\ }\textbf {\bibinfo {volume} {77}},\ \bibinfo
  {pages} {195316} (\bibinfo {year} {2008})}\BibitemShut {NoStop}%
\bibitem [{\citenamefont {Segal}\ \emph {et~al.}(2010)\citenamefont {Segal},
  \citenamefont {Millis},\ and\ \citenamefont {Reichman}}]{Segal.10.PRB}%
  \BibitemOpen
  \bibfield  {author} {\bibinfo {author} {\bibfnamefont {D.}~\bibnamefont
  {Segal}}, \bibinfo {author} {\bibfnamefont {A.~J.}\ \bibnamefont {Millis}}, \
  and\ \bibinfo {author} {\bibfnamefont {D.~R.}\ \bibnamefont {Reichman}},\
  }\href {\doibase 10.1103/PhysRevB.82.205323} {\bibfield  {journal} {\bibinfo
  {journal} {Phys. Rev. B}\ }\textbf {\bibinfo {volume} {82}},\ \bibinfo
  {pages} {205323} (\bibinfo {year} {2010})}\BibitemShut {NoStop}%
\bibitem [{\citenamefont {Kornilovitch}\ \emph {et~al.}(2002)\citenamefont
  {Kornilovitch}, \citenamefont {Bratkovsky},\ and\ \citenamefont
  {Stanley~Williams}}]{Kornilovitch.02.PRB}%
  \BibitemOpen
  \bibfield  {author} {\bibinfo {author} {\bibfnamefont {P.~E.}\ \bibnamefont
  {Kornilovitch}}, \bibinfo {author} {\bibfnamefont {A.~M.}\ \bibnamefont
  {Bratkovsky}}, \ and\ \bibinfo {author} {\bibfnamefont {R.}~\bibnamefont
  {Stanley~Williams}},\ }\href {\doibase 10.1103/PhysRevB.66.165436} {\bibfield
   {journal} {\bibinfo  {journal} {Phys. Rev. B}\ }\textbf {\bibinfo {volume}
  {66}},\ \bibinfo {pages} {165436} (\bibinfo {year} {2002})}\BibitemShut
  {NoStop}%
\bibitem [{\citenamefont {Verdozzi}\ \emph {et~al.}(2006)\citenamefont
  {Verdozzi}, \citenamefont {Stefanucci},\ and\ \citenamefont
  {Almbladh}}]{Verdozzi.06.PRL}%
  \BibitemOpen
  \bibfield  {author} {\bibinfo {author} {\bibfnamefont {C.}~\bibnamefont
  {Verdozzi}}, \bibinfo {author} {\bibfnamefont {G.}~\bibnamefont
  {Stefanucci}}, \ and\ \bibinfo {author} {\bibfnamefont {C.}~\bibnamefont
  {Almbladh}},\ }\href {\doibase 10.1103/PhysRevLett.97.046603} {\bibfield
  {journal} {\bibinfo  {journal} {Phys. Rev. Lett.}\ }\textbf {\bibinfo
  {volume} {97}},\ \bibinfo {pages} {046603} (\bibinfo {year}
  {2006})}\BibitemShut {NoStop}%
\bibitem [{\citenamefont {Arnold}\ \emph {et~al.}(2007)\citenamefont {Arnold},
  \citenamefont {Weigend},\ and\ \citenamefont {Evers}}]{Arnold.07.JCP}%
  \BibitemOpen
  \bibfield  {author} {\bibinfo {author} {\bibfnamefont {A.}~\bibnamefont
  {Arnold}}, \bibinfo {author} {\bibfnamefont {F.}~\bibnamefont {Weigend}}, \
  and\ \bibinfo {author} {\bibfnamefont {F.}~\bibnamefont {Evers}},\ }\href
  {\doibase 10.1063/1.2716664} {\bibfield  {journal} {\bibinfo  {journal} {J.
  Chem. Phys.}\ }\textbf {\bibinfo {volume} {126}},\ \bibinfo {pages} {174101}
  (\bibinfo {year} {2007})}\BibitemShut {NoStop}%
\bibitem [{\citenamefont {Collett}\ and\ \citenamefont
  {Gardiner}(1984)}]{Collett.84.PRA}%
  \BibitemOpen
  \bibfield  {author} {\bibinfo {author} {\bibfnamefont {M.~J.}\ \bibnamefont
  {Collett}}\ and\ \bibinfo {author} {\bibfnamefont {C.~W.}\ \bibnamefont
  {Gardiner}},\ }\href {\doibase 10.1103/PhysRevA.30.1386} {\bibfield
  {journal} {\bibinfo  {journal} {Phys. Rev. A}\ }\textbf {\bibinfo {volume}
  {30}},\ \bibinfo {pages} {1386} (\bibinfo {year} {1984})}\BibitemShut
  {NoStop}%
\bibitem [{\citenamefont {Gardiner}\ and\ \citenamefont
  {Collett}(1985)}]{Gardiner.85.PRA}%
  \BibitemOpen
  \bibfield  {author} {\bibinfo {author} {\bibfnamefont {C.~W.}\ \bibnamefont
  {Gardiner}}\ and\ \bibinfo {author} {\bibfnamefont {M.~J.}\ \bibnamefont
  {Collett}},\ }\href {\doibase 10.1103/PhysRevA.31.3761} {\bibfield  {journal}
  {\bibinfo  {journal} {Phys. Rev. A}\ }\textbf {\bibinfo {volume} {31}},\
  \bibinfo {pages} {3761} (\bibinfo {year} {1985})}\BibitemShut {NoStop}%
\bibitem [{\citenamefont {Search}\ \emph {et~al.}(2002)\citenamefont {Search},
  \citenamefont {P\"otting}, \citenamefont {Zhang},\ and\ \citenamefont
  {Meystre}}]{Search.02.PRA}%
  \BibitemOpen
  \bibfield  {author} {\bibinfo {author} {\bibfnamefont {C.~P.}\ \bibnamefont
  {Search}}, \bibinfo {author} {\bibfnamefont {S.}~\bibnamefont {P\"otting}},
  \bibinfo {author} {\bibfnamefont {W.}~\bibnamefont {Zhang}}, \ and\ \bibinfo
  {author} {\bibfnamefont {P.}~\bibnamefont {Meystre}},\ }\href {\doibase
  10.1103/PhysRevA.66.043616} {\bibfield  {journal} {\bibinfo  {journal} {Phys.
  Rev. A}\ }\textbf {\bibinfo {volume} {66}},\ \bibinfo {pages} {043616}
  (\bibinfo {year} {2002})}\BibitemShut {NoStop}%
\bibitem [{\citenamefont {Gardiner}\ and\ \citenamefont
  {Zoller}(2004)}]{Gardiner.04.NULL}%
  \BibitemOpen
  \bibfield  {author} {\bibinfo {author} {\bibfnamefont {C.~W.}\ \bibnamefont
  {Gardiner}}\ and\ \bibinfo {author} {\bibfnamefont {P.}~\bibnamefont
  {Zoller}},\ }\href@noop {} {\emph {\bibinfo {title} {Quantum Noise: A
  Handbook of Markovian and Non-Markovian Quantum Stochastic Methods with
  Applications to Quantum Optics}}}\ (\bibinfo  {publisher} {Springer,
  Berlin},\ \bibinfo {year} {2004})\BibitemShut {NoStop}%
\bibitem [{\citenamefont {Ciuti}\ and\ \citenamefont
  {Carusotto}(2006)}]{Ciuti.06.PRA}%
  \BibitemOpen
  \bibfield  {author} {\bibinfo {author} {\bibfnamefont {C.}~\bibnamefont
  {Ciuti}}\ and\ \bibinfo {author} {\bibfnamefont {I.}~\bibnamefont
  {Carusotto}},\ }\href {\doibase 10.1103/PhysRevA.74.033811} {\bibfield
  {journal} {\bibinfo  {journal} {Phys. Rev. A}\ }\textbf {\bibinfo {volume}
  {74}},\ \bibinfo {pages} {033811} (\bibinfo {year} {2006})}\BibitemShut
  {NoStop}%
\bibitem [{\citenamefont {Genes}\ \emph {et~al.}(2008)\citenamefont {Genes},
  \citenamefont {Vitali}, \citenamefont {Tombesi}, \citenamefont {Gigan},\ and\
  \citenamefont {Aspelmeyer}}]{Genes.08.PRA}%
  \BibitemOpen
  \bibfield  {author} {\bibinfo {author} {\bibfnamefont {C.}~\bibnamefont
  {Genes}}, \bibinfo {author} {\bibfnamefont {D.}~\bibnamefont {Vitali}},
  \bibinfo {author} {\bibfnamefont {P.}~\bibnamefont {Tombesi}}, \bibinfo
  {author} {\bibfnamefont {S.}~\bibnamefont {Gigan}}, \ and\ \bibinfo {author}
  {\bibfnamefont {M.}~\bibnamefont {Aspelmeyer}},\ }\href {\doibase
  10.1103/PhysRevA.77.033804} {\bibfield  {journal} {\bibinfo  {journal} {Phys.
  Rev. A}\ }\textbf {\bibinfo {volume} {77}},\ \bibinfo {pages} {033804}
  (\bibinfo {year} {2008})}\BibitemShut {NoStop}%
\bibitem [{\citenamefont {Clerk}\ \emph {et~al.}(2010)\citenamefont {Clerk},
  \citenamefont {Devoret}, \citenamefont {Girvin}, \citenamefont {Marquardt},\
  and\ \citenamefont {Schoelkopf}}]{Clerk.10.RMP}%
  \BibitemOpen
  \bibfield  {author} {\bibinfo {author} {\bibfnamefont {A.~A.}\ \bibnamefont
  {Clerk}}, \bibinfo {author} {\bibfnamefont {M.~H.}\ \bibnamefont {Devoret}},
  \bibinfo {author} {\bibfnamefont {S.~M.}\ \bibnamefont {Girvin}}, \bibinfo
  {author} {\bibfnamefont {F.}~\bibnamefont {Marquardt}}, \ and\ \bibinfo
  {author} {\bibfnamefont {R.~J.}\ \bibnamefont {Schoelkopf}},\ }\href
  {\doibase 10.1103/RevModPhys.82.1155} {\bibfield  {journal} {\bibinfo
  {journal} {Rev. Mod. Phys.}\ }\textbf {\bibinfo {volume} {82}},\ \bibinfo
  {pages} {1155} (\bibinfo {year} {2010})}\BibitemShut {NoStop}%
\bibitem [{\citenamefont {Rabl}(2011)}]{Rabl.11.PRL}%
  \BibitemOpen
  \bibfield  {author} {\bibinfo {author} {\bibfnamefont {P.}~\bibnamefont
  {Rabl}},\ }\href {\doibase 10.1103/PhysRevLett.107.063601} {\bibfield
  {journal} {\bibinfo  {journal} {Phys. Rev. Lett.}\ }\textbf {\bibinfo
  {volume} {107}},\ \bibinfo {pages} {063601} (\bibinfo {year}
  {2011})}\BibitemShut {NoStop}%
\bibitem [{\citenamefont {Nunnenkamp}\ \emph {et~al.}(2011)\citenamefont
  {Nunnenkamp}, \citenamefont {B\o{}rkje},\ and\ \citenamefont
  {Girvin}}]{Nunnenkamp.11.PRL}%
  \BibitemOpen
  \bibfield  {author} {\bibinfo {author} {\bibfnamefont {A.}~\bibnamefont
  {Nunnenkamp}}, \bibinfo {author} {\bibfnamefont {K.}~\bibnamefont
  {B\o{}rkje}}, \ and\ \bibinfo {author} {\bibfnamefont {S.~M.}\ \bibnamefont
  {Girvin}},\ }\href {\doibase 10.1103/PhysRevLett.107.063602} {\bibfield
  {journal} {\bibinfo  {journal} {Phys. Rev. Lett.}\ }\textbf {\bibinfo
  {volume} {107}},\ \bibinfo {pages} {063602} (\bibinfo {year}
  {2011})}\BibitemShut {NoStop}%
\bibitem [{\citenamefont {Reitz}\ \emph {et~al.}(2019)\citenamefont {Reitz},
  \citenamefont {Sommer},\ and\ \citenamefont {Genes}}]{Reitz.19.PRL}%
  \BibitemOpen
  \bibfield  {author} {\bibinfo {author} {\bibfnamefont {M.}~\bibnamefont
  {Reitz}}, \bibinfo {author} {\bibfnamefont {C.}~\bibnamefont {Sommer}}, \
  and\ \bibinfo {author} {\bibfnamefont {C.}~\bibnamefont {Genes}},\ }\href
  {\doibase 10.1103/PhysRevLett.122.203602} {\bibfield  {journal} {\bibinfo
  {journal} {Phys. Rev. Lett.}\ }\textbf {\bibinfo {volume} {122}},\ \bibinfo
  {pages} {203602} (\bibinfo {year} {2019})}\BibitemShut {NoStop}%
\bibitem [{\citenamefont {Kiilerich}\ and\ \citenamefont
  {M\o{}lmer}(2019)}]{Kiilerich.19.PRL}%
  \BibitemOpen
  \bibfield  {author} {\bibinfo {author} {\bibfnamefont {A.}~\bibnamefont
  {Kiilerich}}\ and\ \bibinfo {author} {\bibfnamefont {K.}~\bibnamefont
  {M\o{}lmer}},\ }\href {\doibase 10.1103/PhysRevLett.123.123604} {\bibfield
  {journal} {\bibinfo  {journal} {Phys. Rev. Lett.}\ }\textbf {\bibinfo
  {volume} {123}},\ \bibinfo {pages} {123604} (\bibinfo {year}
  {2019})}\BibitemShut {NoStop}%
\bibitem [{\citenamefont {Chikkaraddy}\ \emph {et~al.}(2016)\citenamefont
  {Chikkaraddy}, \citenamefont {de~Nijs}, \citenamefont {Benz}, \citenamefont
  {Barrow}, \citenamefont {Scherman}, \citenamefont {Rosta}, \citenamefont
  {Demetriadou}, \citenamefont {Fox}, \citenamefont {Hess},\ and\ \citenamefont
  {Baumberg}}]{Chikkaraddy.16.N}%
  \BibitemOpen
  \bibfield  {author} {\bibinfo {author} {\bibfnamefont {R.}~\bibnamefont
  {Chikkaraddy}}, \bibinfo {author} {\bibfnamefont {B.}~\bibnamefont
  {de~Nijs}}, \bibinfo {author} {\bibfnamefont {F.}~\bibnamefont {Benz}},
  \bibinfo {author} {\bibfnamefont {S.~J.}\ \bibnamefont {Barrow}}, \bibinfo
  {author} {\bibfnamefont {O.~A.}\ \bibnamefont {Scherman}}, \bibinfo {author}
  {\bibfnamefont {E.}~\bibnamefont {Rosta}}, \bibinfo {author} {\bibfnamefont
  {A.}~\bibnamefont {Demetriadou}}, \bibinfo {author} {\bibfnamefont
  {P.}~\bibnamefont {Fox}}, \bibinfo {author} {\bibfnamefont {O.}~\bibnamefont
  {Hess}}, \ and\ \bibinfo {author} {\bibfnamefont {J.~J.}\ \bibnamefont
  {Baumberg}},\ }\href {https://doi.org/10.1038/nature17974} {\bibfield
  {journal} {\bibinfo  {journal} {Nature}\ }\textbf {\bibinfo {volume} {535}},\
  \bibinfo {pages} {127} (\bibinfo {year} {2016})}\BibitemShut {NoStop}%
\bibitem [{\citenamefont {Liu}\ \emph {et~al.}(2017)\citenamefont {Liu},
  \citenamefont {Zhou}, \citenamefont {Yu}, \citenamefont {Zhang},
  \citenamefont {Wang}, \citenamefont {Liu}, \citenamefont {Wei}, \citenamefont
  {Chen},\ and\ \citenamefont {Wang}}]{Liu.17.PRL}%
  \BibitemOpen
  \bibfield  {author} {\bibinfo {author} {\bibfnamefont {R.}~\bibnamefont
  {Liu}}, \bibinfo {author} {\bibfnamefont {Z.}~\bibnamefont {Zhou}}, \bibinfo
  {author} {\bibfnamefont {Y.}~\bibnamefont {Yu}}, \bibinfo {author}
  {\bibfnamefont {T.}~\bibnamefont {Zhang}}, \bibinfo {author} {\bibfnamefont
  {H.}~\bibnamefont {Wang}}, \bibinfo {author} {\bibfnamefont {G.}~\bibnamefont
  {Liu}}, \bibinfo {author} {\bibfnamefont {Y.}~\bibnamefont {Wei}}, \bibinfo
  {author} {\bibfnamefont {H.}~\bibnamefont {Chen}}, \ and\ \bibinfo {author}
  {\bibfnamefont {X.}~\bibnamefont {Wang}},\ }\href {\doibase
  10.1103/PhysRevLett.118.237401} {\bibfield  {journal} {\bibinfo  {journal}
  {Phys. Rev. Lett.}\ }\textbf {\bibinfo {volume} {118}},\ \bibinfo {pages}
  {237401} (\bibinfo {year} {2017})}\BibitemShut {NoStop}%
\bibitem [{\citenamefont {Orgiu}\ \emph {et~al.}(2015)\citenamefont {Orgiu},
  \citenamefont {George}, \citenamefont {Hutchison}, \citenamefont {Devaux},
  \citenamefont {Dayen}, \citenamefont {Doudin}, \citenamefont {Stellacci},
  \citenamefont {Genet}, \citenamefont {Schachenmayer}, \citenamefont {Genes},
  \citenamefont {Pupillo}, \citenamefont {Samorì},\ and\ \citenamefont
  {Ebbesen}}]{Orgiu.15.NM}%
  \BibitemOpen
  \bibfield  {author} {\bibinfo {author} {\bibfnamefont {E.}~\bibnamefont
  {Orgiu}}, \bibinfo {author} {\bibfnamefont {J.}~\bibnamefont {George}},
  \bibinfo {author} {\bibfnamefont {J.~A.}\ \bibnamefont {Hutchison}}, \bibinfo
  {author} {\bibfnamefont {E.}~\bibnamefont {Devaux}}, \bibinfo {author}
  {\bibfnamefont {J.~F.}\ \bibnamefont {Dayen}}, \bibinfo {author}
  {\bibfnamefont {B.}~\bibnamefont {Doudin}}, \bibinfo {author} {\bibfnamefont
  {F.}~\bibnamefont {Stellacci}}, \bibinfo {author} {\bibfnamefont
  {C.}~\bibnamefont {Genet}}, \bibinfo {author} {\bibfnamefont
  {J.}~\bibnamefont {Schachenmayer}}, \bibinfo {author} {\bibfnamefont
  {C.}~\bibnamefont {Genes}}, \bibinfo {author} {\bibfnamefont
  {G.}~\bibnamefont {Pupillo}}, \bibinfo {author} {\bibfnamefont
  {P.}~\bibnamefont {Samorì}}, \ and\ \bibinfo {author} {\bibfnamefont
  {T.~W.}\ \bibnamefont {Ebbesen}},\ }\href {https://doi.org/10.1038/nmat4392}
  {\bibfield  {journal} {\bibinfo  {journal} {Nat. Mater.}\ }\textbf {\bibinfo
  {volume} {14}},\ \bibinfo {pages} {1123} (\bibinfo {year}
  {2015})}\BibitemShut {NoStop}%
\bibitem [{\citenamefont {Hagenm\"uller}\ \emph {et~al.}(2017)\citenamefont
  {Hagenm\"uller}, \citenamefont {Schachenmayer}, \citenamefont {Sch\"utz},
  \citenamefont {Genes},\ and\ \citenamefont {Pupillo}}]{Hagenmuller.17.PRL}%
  \BibitemOpen
  \bibfield  {author} {\bibinfo {author} {\bibfnamefont {D.}~\bibnamefont
  {Hagenm\"uller}}, \bibinfo {author} {\bibfnamefont {J.}~\bibnamefont
  {Schachenmayer}}, \bibinfo {author} {\bibfnamefont {S.}~\bibnamefont
  {Sch\"utz}}, \bibinfo {author} {\bibfnamefont {C.}~\bibnamefont {Genes}}, \
  and\ \bibinfo {author} {\bibfnamefont {G.}~\bibnamefont {Pupillo}},\ }\href
  {\doibase 10.1103/PhysRevLett.119.223601} {\bibfield  {journal} {\bibinfo
  {journal} {Phys. Rev. Lett.}\ }\textbf {\bibinfo {volume} {119}},\ \bibinfo
  {pages} {223601} (\bibinfo {year} {2017})}\BibitemShut {NoStop}%
\bibitem [{\citenamefont {Hagenm\"uller}\ \emph {et~al.}(2018)\citenamefont
  {Hagenm\"uller}, \citenamefont {Sch\"utz}, \citenamefont {Schachenmayer},
  \citenamefont {Genes},\ and\ \citenamefont {Pupillo}}]{Hagenmuller.18.PRB}%
  \BibitemOpen
  \bibfield  {author} {\bibinfo {author} {\bibfnamefont {D.}~\bibnamefont
  {Hagenm\"uller}}, \bibinfo {author} {\bibfnamefont {S.}~\bibnamefont
  {Sch\"utz}}, \bibinfo {author} {\bibfnamefont {J.}~\bibnamefont
  {Schachenmayer}}, \bibinfo {author} {\bibfnamefont {C.}~\bibnamefont
  {Genes}}, \ and\ \bibinfo {author} {\bibfnamefont {G.}~\bibnamefont
  {Pupillo}},\ }\href {\doibase 10.1103/PhysRevB.97.205303} {\bibfield
  {journal} {\bibinfo  {journal} {Phys. Rev. B}\ }\textbf {\bibinfo {volume}
  {97}},\ \bibinfo {pages} {205303} (\bibinfo {year} {2018})}\BibitemShut
  {NoStop}%
\bibitem [{\citenamefont {Sowa}\ \emph
  {et~al.}(2017{\natexlab{b}})\citenamefont {Sowa}, \citenamefont {Mol},
  \citenamefont {Briggs},\ and\ \citenamefont {Gauger}}]{Sowa.17.PRB}%
  \BibitemOpen
  \bibfield  {author} {\bibinfo {author} {\bibfnamefont {J.~K.}\ \bibnamefont
  {Sowa}}, \bibinfo {author} {\bibfnamefont {J.~A.}\ \bibnamefont {Mol}},
  \bibinfo {author} {\bibfnamefont {G.~A.~D.}\ \bibnamefont {Briggs}}, \ and\
  \bibinfo {author} {\bibfnamefont {E.~M.}\ \bibnamefont {Gauger}},\ }\href
  {\doibase 10.1103/PhysRevB.95.085423} {\bibfield  {journal} {\bibinfo
  {journal} {Phys. Rev. B}\ }\textbf {\bibinfo {volume} {95}},\ \bibinfo
  {pages} {085423} (\bibinfo {year} {2017}{\natexlab{b}})}\BibitemShut
  {NoStop}%
\bibitem [{\citenamefont {Gullans}\ \emph {et~al.}(2015)\citenamefont
  {Gullans}, \citenamefont {Liu}, \citenamefont {Stehlik}, \citenamefont
  {Petta},\ and\ \citenamefont {Taylor}}]{Petta15}%
  \BibitemOpen
  \bibfield  {author} {\bibinfo {author} {\bibfnamefont {M.~J.}\ \bibnamefont
  {Gullans}}, \bibinfo {author} {\bibfnamefont {Y.-Y.}\ \bibnamefont {Liu}},
  \bibinfo {author} {\bibfnamefont {J.}~\bibnamefont {Stehlik}}, \bibinfo
  {author} {\bibfnamefont {J.~R.}\ \bibnamefont {Petta}}, \ and\ \bibinfo
  {author} {\bibfnamefont {J.~M.}\ \bibnamefont {Taylor}},\ }\href {\doibase
  10.1103/PhysRevLett.114.196802} {\bibfield  {journal} {\bibinfo  {journal}
  {Phys. Rev. Lett.}\ }\textbf {\bibinfo {volume} {114}},\ \bibinfo {pages}
  {196802} (\bibinfo {year} {2015})}\BibitemShut {NoStop}%
\bibitem [{\citenamefont {Wingreen}\ \emph {et~al.}(1989)\citenamefont
  {Wingreen}, \citenamefont {Jacobsen},\ and\ \citenamefont
  {Wilkins}}]{Wingreen.89.PRB}%
  \BibitemOpen
  \bibfield  {author} {\bibinfo {author} {\bibfnamefont {N.~S.}\ \bibnamefont
  {Wingreen}}, \bibinfo {author} {\bibfnamefont {K.~W.}\ \bibnamefont
  {Jacobsen}}, \ and\ \bibinfo {author} {\bibfnamefont {J.~W.}\ \bibnamefont
  {Wilkins}},\ }\href {\doibase 10.1103/PhysRevB.40.11834} {\bibfield
  {journal} {\bibinfo  {journal} {Phys. Rev. B}\ }\textbf {\bibinfo {volume}
  {40}},\ \bibinfo {pages} {11834} (\bibinfo {year} {1989})}\BibitemShut
  {NoStop}%
\bibitem [{\citenamefont {Topp}\ \emph {et~al.}(2015)\citenamefont {Topp},
  \citenamefont {Brandes},\ and\ \citenamefont {Schaller}}]{Topp.15.EPL}%
  \BibitemOpen
  \bibfield  {author} {\bibinfo {author} {\bibfnamefont {G.~E.}\ \bibnamefont
  {Topp}}, \bibinfo {author} {\bibfnamefont {T.}~\bibnamefont {Brandes}}, \
  and\ \bibinfo {author} {\bibfnamefont {G.}~\bibnamefont {Schaller}},\ }\href
  {\doibase 10.1209/0295-5075/110/67003} {\bibfield  {journal} {\bibinfo
  {journal} {Europhys. Lett.}\ }\textbf {\bibinfo {volume} {110}},\ \bibinfo
  {pages} {67003} (\bibinfo {year} {2015})}\BibitemShut {NoStop}%
\bibitem [{\citenamefont {Jussiau}\ \emph {et~al.}(2019)\citenamefont
  {Jussiau}, \citenamefont {Hasegawa},\ and\ \citenamefont
  {Whitney}}]{Jussiau.19.PRB}%
  \BibitemOpen
  \bibfield  {author} {\bibinfo {author} {\bibfnamefont {E.}~\bibnamefont
  {Jussiau}}, \bibinfo {author} {\bibfnamefont {M.}~\bibnamefont {Hasegawa}}, \
  and\ \bibinfo {author} {\bibfnamefont {R.~S.}\ \bibnamefont {Whitney}},\
  }\href {\doibase 10.1103/PhysRevB.100.115411} {\bibfield  {journal} {\bibinfo
   {journal} {Phys. Rev. B}\ }\textbf {\bibinfo {volume} {100}},\ \bibinfo
  {pages} {115411} (\bibinfo {year} {2019})}\BibitemShut {NoStop}%
\bibitem [{\citenamefont {Yang}\ \emph {et~al.}(2015)\citenamefont {Yang},
  \citenamefont {Lin},\ and\ \citenamefont {Zhang}}]{Yang.15.PRB}%
  \BibitemOpen
  \bibfield  {author} {\bibinfo {author} {\bibfnamefont {P.}~\bibnamefont
  {Yang}}, \bibinfo {author} {\bibfnamefont {C.}~\bibnamefont {Lin}}, \ and\
  \bibinfo {author} {\bibfnamefont {W.}~\bibnamefont {Zhang}},\ }\href
  {\doibase 10.1103/PhysRevB.92.165403} {\bibfield  {journal} {\bibinfo
  {journal} {Phys. Rev. B}\ }\textbf {\bibinfo {volume} {92}},\ \bibinfo
  {pages} {165403} (\bibinfo {year} {2015})}\BibitemShut {NoStop}%
\bibitem [{\citenamefont {Sowa}\ \emph {et~al.}(2018)\citenamefont {Sowa},
  \citenamefont {Mol}, \citenamefont {Briggs},\ and\ \citenamefont
  {Gauger}}]{Sowa.18.JCP}%
  \BibitemOpen
  \bibfield  {author} {\bibinfo {author} {\bibfnamefont {J.~K.}\ \bibnamefont
  {Sowa}}, \bibinfo {author} {\bibfnamefont {J.~A.}\ \bibnamefont {Mol}},
  \bibinfo {author} {\bibfnamefont {G.~A.~D.}\ \bibnamefont {Briggs}}, \ and\
  \bibinfo {author} {\bibfnamefont {E.~M.}\ \bibnamefont {Gauger}},\ }\href
  {\doibase 10.1063/1.5049537} {\bibfield  {journal} {\bibinfo  {journal} {J.
  Chem. Phys.}\ }\textbf {\bibinfo {volume} {149}},\ \bibinfo {pages} {154112}
  (\bibinfo {year} {2018})}\BibitemShut {NoStop}%
\bibitem [{\citenamefont {Jauho}\ \emph {et~al.}(1994)\citenamefont {Jauho},
  \citenamefont {Wingreen},\ and\ \citenamefont {Meir}}]{Jauho.94.PRB}%
  \BibitemOpen
  \bibfield  {author} {\bibinfo {author} {\bibfnamefont {A.-P.}\ \bibnamefont
  {Jauho}}, \bibinfo {author} {\bibfnamefont {N.~S.}\ \bibnamefont {Wingreen}},
  \ and\ \bibinfo {author} {\bibfnamefont {Y.}~\bibnamefont {Meir}},\ }\href
  {\doibase 10.1103/PhysRevB.50.5528} {\bibfield  {journal} {\bibinfo
  {journal} {Phys. Rev. B}\ }\textbf {\bibinfo {volume} {50}},\ \bibinfo
  {pages} {5528} (\bibinfo {year} {1994})}\BibitemShut {NoStop}%
\bibitem [{\citenamefont {Datta}(1995)}]{Datta.95.NULL}%
  \BibitemOpen
  \bibfield  {author} {\bibinfo {author} {\bibfnamefont {S.}~\bibnamefont
  {Datta}},\ }\href@noop {} {\emph {\bibinfo {title} {Electronic Transport in
  Mesoscopic Systems}}}\ (\bibinfo  {publisher} {Cambridge University Press,
  New York},\ \bibinfo {year} {1995})\BibitemShut {NoStop}%
\bibitem [{\citenamefont {Breuer}\ and\ \citenamefont
  {Petruccione}(2002)}]{Breuer.02.NULL}%
  \BibitemOpen
  \bibfield  {author} {\bibinfo {author} {\bibfnamefont {H.-P.}\ \bibnamefont
  {Breuer}}\ and\ \bibinfo {author} {\bibfnamefont {F.}~\bibnamefont
  {Petruccione}},\ }\href@noop {} {\emph {\bibinfo {title} {The Theory of Open
  Quantum Systems}}}\ (\bibinfo  {publisher} {Oxford University Press, New
  York},\ \bibinfo {year} {2002})\BibitemShut {NoStop}%
\bibitem [{\citenamefont {Gurvitz}\ and\ \citenamefont
  {Prager}(1996)}]{Gurvitz.96.PRB}%
  \BibitemOpen
  \bibfield  {author} {\bibinfo {author} {\bibfnamefont {S.~A.}\ \bibnamefont
  {Gurvitz}}\ and\ \bibinfo {author} {\bibfnamefont {Y.~S.}\ \bibnamefont
  {Prager}},\ }\href {\doibase 10.1103/PhysRevB.53.15932} {\bibfield  {journal}
  {\bibinfo  {journal} {Phys. Rev. B}\ }\textbf {\bibinfo {volume} {53}},\
  \bibinfo {pages} {15932} (\bibinfo {year} {1996})}\BibitemShut {NoStop}%
\bibitem [{\citenamefont {Schinabeck}(2019)}]{Schinabeck2019}%
  \BibitemOpen
  \bibfield  {author} {\bibinfo {author} {\bibfnamefont {C.}~\bibnamefont
  {Schinabeck}},\ }\emph {\bibinfo {title} {Hierarchical quantum master
  equation approaches to nonequilibrium charge transport through
  single-molecule junctions}},\ \href@noop {} {\bibinfo {type}
  {doctoralthesis}},\ \bibinfo  {school} {Friedrich-Alexander-Universit{\"a}t
  Erlangen-N{\"u}rnberg (FAU)} (\bibinfo {year} {2019})\BibitemShut {NoStop}%
\bibitem [{\citenamefont {Sch{\"a}fer}\ \emph {et~al.}(2019)\citenamefont
  {Sch{\"a}fer}, \citenamefont {Ruggenthaler}, \citenamefont {Appel},\ and\
  \citenamefont {Rubio}}]{Schafer.19.PNAS}%
  \BibitemOpen
  \bibfield  {author} {\bibinfo {author} {\bibfnamefont {C.}~\bibnamefont
  {Sch{\"a}fer}}, \bibinfo {author} {\bibfnamefont {M.}~\bibnamefont
  {Ruggenthaler}}, \bibinfo {author} {\bibfnamefont {H.}~\bibnamefont {Appel}},
  \ and\ \bibinfo {author} {\bibfnamefont {A.}~\bibnamefont {Rubio}},\ }\href
  {\doibase 10.1073/pnas.1814178116} {\bibfield  {journal} {\bibinfo  {journal}
  {Proc. Natl. Acad. Sci. U.S.A.}\ }\textbf {\bibinfo {volume} {116}},\
  \bibinfo {pages} {4883} (\bibinfo {year} {2019})}\BibitemShut {NoStop}%
\bibitem [{\citenamefont {del Pino}\ \emph {et~al.}(2018)\citenamefont {del
  Pino}, \citenamefont {Schr\"oder}, \citenamefont {Chin}, \citenamefont
  {Feist},\ and\ \citenamefont {Garcia-Vidal}}]{Del.18.PRL}%
  \BibitemOpen
  \bibfield  {author} {\bibinfo {author} {\bibfnamefont {J.}~\bibnamefont {del
  Pino}}, \bibinfo {author} {\bibfnamefont {F.~A. Y.~N.}\ \bibnamefont
  {Schr\"oder}}, \bibinfo {author} {\bibfnamefont {A.~W.}\ \bibnamefont
  {Chin}}, \bibinfo {author} {\bibfnamefont {J.}~\bibnamefont {Feist}}, \ and\
  \bibinfo {author} {\bibfnamefont {F.~J.}\ \bibnamefont {Garcia-Vidal}},\
  }\href {\doibase 10.1103/PhysRevLett.121.227401} {\bibfield  {journal}
  {\bibinfo  {journal} {Phys. Rev. Lett.}\ }\textbf {\bibinfo {volume} {121}},\
  \bibinfo {pages} {227401} (\bibinfo {year} {2018})}\BibitemShut {NoStop}%
\bibitem [{\citenamefont {Maguire}\ \emph {et~al.}(2019)\citenamefont
  {Maguire}, \citenamefont {Iles-Smith},\ and\ \citenamefont
  {Nazir}}]{Maguire.19.PRL}%
  \BibitemOpen
  \bibfield  {author} {\bibinfo {author} {\bibfnamefont {H.}~\bibnamefont
  {Maguire}}, \bibinfo {author} {\bibfnamefont {J.}~\bibnamefont {Iles-Smith}},
  \ and\ \bibinfo {author} {\bibfnamefont {A.}~\bibnamefont {Nazir}},\ }\href
  {\doibase 10.1103/PhysRevLett.123.093601} {\bibfield  {journal} {\bibinfo
  {journal} {Phys. Rev. Lett.}\ }\textbf {\bibinfo {volume} {123}},\ \bibinfo
  {pages} {093601} (\bibinfo {year} {2019})}\BibitemShut {NoStop}%
\bibitem [{\citenamefont {Sanvitto}\ and\ \citenamefont
  {K\'ena-Cohen}(2016)}]{Sanvitto.16.NM}%
  \BibitemOpen
  \bibfield  {author} {\bibinfo {author} {\bibfnamefont {D.}~\bibnamefont
  {Sanvitto}}\ and\ \bibinfo {author} {\bibfnamefont {S.}~\bibnamefont
  {K\'ena-Cohen}},\ }\href {https://doi.org/10.1038/nmat4668} {\bibfield
  {journal} {\bibinfo  {journal} {Nat. Mater.}\ }\textbf {\bibinfo {volume}
  {15}},\ \bibinfo {pages} {1061} (\bibinfo {year} {2016})}\BibitemShut
  {NoStop}%
\bibitem [{\citenamefont {Ebbesen}(2016)}]{Ebbesen.16.ACR}%
  \BibitemOpen
  \bibfield  {author} {\bibinfo {author} {\bibfnamefont {T.~W.}\ \bibnamefont
  {Ebbesen}},\ }\href {\doibase 10.1021/acs.accounts.6b00295} {\bibfield
  {journal} {\bibinfo  {journal} {Acc. Chem. Res.}\ }\textbf {\bibinfo {volume}
  {49}},\ \bibinfo {pages} {2403} (\bibinfo {year} {2016})}\BibitemShut
  {NoStop}%
\bibitem [{\citenamefont {Galego}\ \emph {et~al.}(2015)\citenamefont {Galego},
  \citenamefont {Garcia-Vidal},\ and\ \citenamefont {Feist}}]{Galego.15.PRX}%
  \BibitemOpen
  \bibfield  {author} {\bibinfo {author} {\bibfnamefont {J.}~\bibnamefont
  {Galego}}, \bibinfo {author} {\bibfnamefont {F.~J.}\ \bibnamefont
  {Garcia-Vidal}}, \ and\ \bibinfo {author} {\bibfnamefont {J.}~\bibnamefont
  {Feist}},\ }\href {\doibase 10.1103/PhysRevX.5.041022} {\bibfield  {journal}
  {\bibinfo  {journal} {Phys. Rev. X}\ }\textbf {\bibinfo {volume} {5}},\
  \bibinfo {pages} {041022} (\bibinfo {year} {2015})}\BibitemShut {NoStop}%
\bibitem [{\citenamefont {Herrera}\ and\ \citenamefont
  {Spano}(2016)}]{Herrera.16.PRL}%
  \BibitemOpen
  \bibfield  {author} {\bibinfo {author} {\bibfnamefont {F.}~\bibnamefont
  {Herrera}}\ and\ \bibinfo {author} {\bibfnamefont {F.~C.}\ \bibnamefont
  {Spano}},\ }\href {\doibase 10.1103/PhysRevLett.116.238301} {\bibfield
  {journal} {\bibinfo  {journal} {Phys. Rev. Lett.}\ }\textbf {\bibinfo
  {volume} {116}},\ \bibinfo {pages} {238301} (\bibinfo {year}
  {2016})}\BibitemShut {NoStop}%
\bibitem [{\citenamefont {\'Cwik}\ \emph {et~al.}(2014)\citenamefont {\'Cwik},
  \citenamefont {Reja}, \citenamefont {Littlewood},\ and\ \citenamefont
  {Keeling}}]{Cwik.14.EPL}%
  \BibitemOpen
  \bibfield  {author} {\bibinfo {author} {\bibfnamefont {J.~A.}\ \bibnamefont
  {\'Cwik}}, \bibinfo {author} {\bibfnamefont {S.}~\bibnamefont {Reja}},
  \bibinfo {author} {\bibfnamefont {P.~B.}\ \bibnamefont {Littlewood}}, \ and\
  \bibinfo {author} {\bibfnamefont {J.}~\bibnamefont {Keeling}},\ }\href
  {http://stacks.iop.org/0295-5075/105/i=4/a=47009} {\bibfield  {journal}
  {\bibinfo  {journal} {Europhys. Lett.}\ }\textbf {\bibinfo {volume} {105}},\
  \bibinfo {pages} {47009} (\bibinfo {year} {2014})}\BibitemShut {NoStop}%
\bibitem [{\citenamefont {Herrera}\ and\ \citenamefont
  {Spano}(2017)}]{Herrera.17.PRL}%
  \BibitemOpen
  \bibfield  {author} {\bibinfo {author} {\bibfnamefont {F.}~\bibnamefont
  {Herrera}}\ and\ \bibinfo {author} {\bibfnamefont {F.~C.}\ \bibnamefont
  {Spano}},\ }\href {\doibase 10.1103/PhysRevLett.118.223601} {\bibfield
  {journal} {\bibinfo  {journal} {Phys. Rev. Lett.}\ }\textbf {\bibinfo
  {volume} {118}},\ \bibinfo {pages} {223601} (\bibinfo {year}
  {2017})}\BibitemShut {NoStop}%
\bibitem [{\citenamefont {Herrera}\ and\ \citenamefont
  {Spano}(2018)}]{Herrera.18.ACSP}%
  \BibitemOpen
  \bibfield  {author} {\bibinfo {author} {\bibfnamefont {F.}~\bibnamefont
  {Herrera}}\ and\ \bibinfo {author} {\bibfnamefont {F.~C.}\ \bibnamefont
  {Spano}},\ }\href {\doibase 10.1021/acsphotonics.7b00728} {\bibfield
  {journal} {\bibinfo  {journal} {ACS Photonics}\ }\textbf {\bibinfo {volume}
  {5}},\ \bibinfo {pages} {65} (\bibinfo {year} {2018})}\BibitemShut {NoStop}%
\bibitem [{\citenamefont {Zeb}\ \emph {et~al.}(2018)\citenamefont {Zeb},
  \citenamefont {Kirton},\ and\ \citenamefont {Keeling}}]{Zeb.18.ACSP}%
  \BibitemOpen
  \bibfield  {author} {\bibinfo {author} {\bibfnamefont {M.~A.}\ \bibnamefont
  {Zeb}}, \bibinfo {author} {\bibfnamefont {P.~G.}\ \bibnamefont {Kirton}}, \
  and\ \bibinfo {author} {\bibfnamefont {J.}~\bibnamefont {Keeling}},\ }\href
  {\doibase 10.1021/acsphotonics.7b00916} {\bibfield  {journal} {\bibinfo
  {journal} {ACS Photonics}\ }\textbf {\bibinfo {volume} {5}},\ \bibinfo
  {pages} {249} (\bibinfo {year} {2018})}\BibitemShut {NoStop}%
\bibitem [{\citenamefont {H\"artle}\ \emph
  {et~al.}(2013{\natexlab{b}})\citenamefont {H\"artle}, \citenamefont
  {Butzin},\ and\ \citenamefont {Thoss}}]{Hartle.13.PRB}%
  \BibitemOpen
  \bibfield  {author} {\bibinfo {author} {\bibfnamefont {R.}~\bibnamefont
  {H\"artle}}, \bibinfo {author} {\bibfnamefont {M.}~\bibnamefont {Butzin}}, \
  and\ \bibinfo {author} {\bibfnamefont {M.}~\bibnamefont {Thoss}},\ }\href
  {\doibase 10.1103/PhysRevB.87.085422} {\bibfield  {journal} {\bibinfo
  {journal} {Phys. Rev. B}\ }\textbf {\bibinfo {volume} {87}},\ \bibinfo
  {pages} {085422} (\bibinfo {year} {2013}{\natexlab{b}})}\BibitemShut
  {NoStop}%
\bibitem [{\citenamefont {Simine}\ \emph {et~al.}(2015)\citenamefont {Simine},
  \citenamefont {Chen},\ and\ \citenamefont {Segal}}]{Simine.15.JPCC}%
  \BibitemOpen
  \bibfield  {author} {\bibinfo {author} {\bibfnamefont {L.}~\bibnamefont
  {Simine}}, \bibinfo {author} {\bibfnamefont {W.}~\bibnamefont {Chen}}, \ and\
  \bibinfo {author} {\bibfnamefont {D.}~\bibnamefont {Segal}},\ }\href
  {\doibase 10.1021/jp512648f} {\bibfield  {journal} {\bibinfo  {journal} {J.
  Phys. Chem. C}\ }\textbf {\bibinfo {volume} {119}},\ \bibinfo {pages} {12097}
  (\bibinfo {year} {2015})}\BibitemShut {NoStop}%
\bibitem [{\citenamefont {Agarwalla}\ and\ \citenamefont
  {Segal}(2018)}]{Bijay.18.PRB}%
  \BibitemOpen
  \bibfield  {author} {\bibinfo {author} {\bibfnamefont {B.}~\bibnamefont
  {Agarwalla}}\ and\ \bibinfo {author} {\bibfnamefont {D.}~\bibnamefont
  {Segal}},\ }\href {\doibase 10.1103/PhysRevB.98.155438} {\bibfield  {journal}
  {\bibinfo  {journal} {Phys. Rev. B}\ }\textbf {\bibinfo {volume} {98}},\
  \bibinfo {pages} {155438} (\bibinfo {year} {2018})}\BibitemShut {NoStop}%
\bibitem [{\citenamefont {Gurvitz}(1998)}]{Gurvitz.98.PRB}%
  \BibitemOpen
  \bibfield  {author} {\bibinfo {author} {\bibfnamefont {S.~A.}\ \bibnamefont
  {Gurvitz}},\ }\href {\doibase 10.1103/PhysRevB.57.6602} {\bibfield  {journal}
  {\bibinfo  {journal} {Phys. Rev. B}\ }\textbf {\bibinfo {volume} {57}},\
  \bibinfo {pages} {6602} (\bibinfo {year} {1998})}\BibitemShut {NoStop}%
\bibitem [{\citenamefont {Schaller}\ \emph {et~al.}(2009)\citenamefont
  {Schaller}, \citenamefont {Kie\ss{}lich},\ and\ \citenamefont
  {Brandes}}]{Schaller.09.PRB}%
  \BibitemOpen
  \bibfield  {author} {\bibinfo {author} {\bibfnamefont {G.}~\bibnamefont
  {Schaller}}, \bibinfo {author} {\bibfnamefont {G.}~\bibnamefont
  {Kie\ss{}lich}}, \ and\ \bibinfo {author} {\bibfnamefont {T.}~\bibnamefont
  {Brandes}},\ }\href {\doibase 10.1103/PhysRevB.80.245107} {\bibfield
  {journal} {\bibinfo  {journal} {Phys. Rev. B}\ }\textbf {\bibinfo {volume}
  {80}},\ \bibinfo {pages} {245107} (\bibinfo {year} {2009})}\BibitemShut
  {NoStop}%
\bibitem [{\citenamefont {Perrin}\ \emph {et~al.}(2015)\citenamefont {Perrin},
  \citenamefont {Galan}, \citenamefont {Eelkema}, \citenamefont {Grozema},
  \citenamefont {Thijssen},\ and\ \citenamefont {van~der Zant}}]{Zantdiode}%
  \BibitemOpen
  \bibfield  {author} {\bibinfo {author} {\bibfnamefont {M.~L.}\ \bibnamefont
  {Perrin}}, \bibinfo {author} {\bibfnamefont {E.}~\bibnamefont {Galan}},
  \bibinfo {author} {\bibfnamefont {R.}~\bibnamefont {Eelkema}}, \bibinfo
  {author} {\bibfnamefont {F.}~\bibnamefont {Grozema}}, \bibinfo {author}
  {\bibfnamefont {J.~M.}\ \bibnamefont {Thijssen}}, \ and\ \bibinfo {author}
  {\bibfnamefont {H.~S.~J.}\ \bibnamefont {van~der Zant}},\ }\href
  {https://doi.org/10.1021/jp512803s} {\bibfield  {journal} {\bibinfo
  {journal} {J. Phys. Chem. C}\ }\textbf {\bibinfo {volume} {119}},\ \bibinfo
  {pages} {5697} (\bibinfo {year} {2015})}\BibitemShut {NoStop}%
\bibitem [{\citenamefont {Kilgour}\ and\ \citenamefont
  {Segal}(2015{\natexlab{a}})}]{Kilgourdiode}%
  \BibitemOpen
  \bibfield  {author} {\bibinfo {author} {\bibfnamefont {M.}~\bibnamefont
  {Kilgour}}\ and\ \bibinfo {author} {\bibfnamefont {D.}~\bibnamefont
  {Segal}},\ }\href {https://doi.org/10.1021/acs.jpcc.5b08818} {\bibfield
  {journal} {\bibinfo  {journal} {J. Phys. Chem. C}\ }\textbf {\bibinfo
  {volume} {119}},\ \bibinfo {pages} {25291} (\bibinfo {year}
  {2015}{\natexlab{a}})}\BibitemShut {NoStop}%
\bibitem [{\citenamefont {Selzer}\ \emph {et~al.}(2005)\citenamefont {Selzer},
  \citenamefont {Cai}, \citenamefont {Cabassi}, \citenamefont {Yao},
  \citenamefont {Tour}, \citenamefont {Mayer},\ and\ \citenamefont
  {Allara}}]{Selzer.05.NL}%
  \BibitemOpen
  \bibfield  {author} {\bibinfo {author} {\bibfnamefont {Y.}~\bibnamefont
  {Selzer}}, \bibinfo {author} {\bibfnamefont {L.}~\bibnamefont {Cai}},
  \bibinfo {author} {\bibfnamefont {M.~A.}\ \bibnamefont {Cabassi}}, \bibinfo
  {author} {\bibfnamefont {Y.}~\bibnamefont {Yao}}, \bibinfo {author}
  {\bibfnamefont {J.~M.}\ \bibnamefont {Tour}}, \bibinfo {author}
  {\bibfnamefont {T.~S.}\ \bibnamefont {Mayer}}, \ and\ \bibinfo {author}
  {\bibfnamefont {D.~L.}\ \bibnamefont {Allara}},\ }\href {\doibase
  10.1021/nl048372j} {\bibfield  {journal} {\bibinfo  {journal} {Nano Lett.}\
  }\textbf {\bibinfo {volume} {5}},\ \bibinfo {pages} {61} (\bibinfo {year}
  {2005})}\BibitemShut {NoStop}%
\bibitem [{\citenamefont {Lee}\ \emph {et~al.}(2012)\citenamefont {Lee},
  \citenamefont {Yamada}, \citenamefont {Tanaka}, \citenamefont {Chang},
  \citenamefont {Asai},\ and\ \citenamefont {Tada}}]{Lee.12.ACSN}%
  \BibitemOpen
  \bibfield  {author} {\bibinfo {author} {\bibfnamefont {S.}~\bibnamefont
  {Lee}}, \bibinfo {author} {\bibfnamefont {R.}~\bibnamefont {Yamada}},
  \bibinfo {author} {\bibfnamefont {S.}~\bibnamefont {Tanaka}}, \bibinfo
  {author} {\bibfnamefont {G.}~\bibnamefont {Chang}}, \bibinfo {author}
  {\bibfnamefont {Y.}~\bibnamefont {Asai}}, \ and\ \bibinfo {author}
  {\bibfnamefont {H.}~\bibnamefont {Tada}},\ }\href {\doibase
  10.1021/nn3006976} {\bibfield  {journal} {\bibinfo  {journal} {ACS Nano}\
  }\textbf {\bibinfo {volume} {6}},\ \bibinfo {pages} {5078} (\bibinfo {year}
  {2012})}\BibitemShut {NoStop}%
\bibitem [{\citenamefont {Kilgour}\ and\ \citenamefont
  {Segal}(2015{\natexlab{b}})}]{Kilgour.15.JCP}%
  \BibitemOpen
  \bibfield  {author} {\bibinfo {author} {\bibfnamefont {M.}~\bibnamefont
  {Kilgour}}\ and\ \bibinfo {author} {\bibfnamefont {D.}~\bibnamefont
  {Segal}},\ }\href {\doibase 10.1063/1.4926395} {\bibfield  {journal}
  {\bibinfo  {journal} {J. Chem. Phys.}\ }\textbf {\bibinfo {volume} {143}},\
  \bibinfo {pages} {024111} (\bibinfo {year} {2015}{\natexlab{b}})}\BibitemShut
  {NoStop}%
\bibitem [{\citenamefont {Garrigues}\ \emph
  {et~al.}(2016{\natexlab{a}})\citenamefont {Garrigues}, \citenamefont {Yuan},
  \citenamefont {Wang}, \citenamefont {Singh}, \citenamefont {del Barco},\ and\
  \citenamefont {Nijhuis}}]{Garrigues.16.DT}%
  \BibitemOpen
  \bibfield  {author} {\bibinfo {author} {\bibfnamefont {A.~R.}\ \bibnamefont
  {Garrigues}}, \bibinfo {author} {\bibfnamefont {L.}~\bibnamefont {Yuan}},
  \bibinfo {author} {\bibfnamefont {L.}~\bibnamefont {Wang}}, \bibinfo {author}
  {\bibfnamefont {S.}~\bibnamefont {Singh}}, \bibinfo {author} {\bibfnamefont
  {E.}~\bibnamefont {del Barco}}, \ and\ \bibinfo {author} {\bibfnamefont
  {C.~A.}\ \bibnamefont {Nijhuis}},\ }\href {\doibase 10.1039/C6DT03204D}
  {\bibfield  {journal} {\bibinfo  {journal} {Dalton Trans.}\ }\textbf
  {\bibinfo {volume} {45}},\ \bibinfo {pages} {17153} (\bibinfo {year}
  {2016}{\natexlab{a}})}\BibitemShut {NoStop}%
\bibitem [{\citenamefont {Garrigues}\ \emph
  {et~al.}(2016{\natexlab{b}})\citenamefont {Garrigues}, \citenamefont {Yuan},
  \citenamefont {Wang}, \citenamefont {Mucciolo}, \citenamefont {Thompon},
  \citenamefont {del Barco},\ and\ \citenamefont {Nijhuis}}]{Garrigues.16.SR}%
  \BibitemOpen
  \bibfield  {author} {\bibinfo {author} {\bibfnamefont {A.~R.}\ \bibnamefont
  {Garrigues}}, \bibinfo {author} {\bibfnamefont {L.}~\bibnamefont {Yuan}},
  \bibinfo {author} {\bibfnamefont {L.}~\bibnamefont {Wang}}, \bibinfo {author}
  {\bibfnamefont {E.~R.}\ \bibnamefont {Mucciolo}}, \bibinfo {author}
  {\bibfnamefont {D.}~\bibnamefont {Thompon}}, \bibinfo {author} {\bibfnamefont
  {E.}~\bibnamefont {del Barco}}, \ and\ \bibinfo {author} {\bibfnamefont
  {C.~A.}\ \bibnamefont {Nijhuis}},\ }\href {https://doi.org/10.1038/srep26517}
  {\bibfield  {journal} {\bibinfo  {journal} {Sci. Rep.}\ }\textbf {\bibinfo
  {volume} {6}},\ \bibinfo {pages} {26517} (\bibinfo {year}
  {2016}{\natexlab{b}})}\BibitemShut {NoStop}%
\bibitem [{\citenamefont {Kim}\ and\ \citenamefont {Segal}(2017)}]{Kim.17.JCP}%
  \BibitemOpen
  \bibfield  {author} {\bibinfo {author} {\bibfnamefont {H.}~\bibnamefont
  {Kim}}\ and\ \bibinfo {author} {\bibfnamefont {D.}~\bibnamefont {Segal}},\
  }\href {\doibase 10.1063/1.4981022} {\bibfield  {journal} {\bibinfo
  {journal} {J. Chem. Phys.}\ }\textbf {\bibinfo {volume} {146}},\ \bibinfo
  {pages} {164702} (\bibinfo {year} {2017})}\BibitemShut {NoStop}%
\bibitem [{\citenamefont {Engelkes}\ \emph {et~al.}(2004)\citenamefont
  {Engelkes}, \citenamefont {Beebe},\ and\ \citenamefont
  {Frisbie}}]{Engelkes.04.JACS}%
  \BibitemOpen
  \bibfield  {author} {\bibinfo {author} {\bibfnamefont {V.~B.}\ \bibnamefont
  {Engelkes}}, \bibinfo {author} {\bibfnamefont {J.~M.}\ \bibnamefont {Beebe}},
  \ and\ \bibinfo {author} {\bibfnamefont {C.~D.}\ \bibnamefont {Frisbie}},\
  }\href {\doibase 10.1021/ja046274u} {\bibfield  {journal} {\bibinfo
  {journal} {J. Am. Chem. Soc.}\ }\textbf {\bibinfo {volume} {126}},\ \bibinfo
  {pages} {14287} (\bibinfo {year} {2004})}\BibitemShut {NoStop}%
\bibitem [{\citenamefont {Li}\ \emph {et~al.}(2016)\citenamefont {Li},
  \citenamefont {Xiang}, \citenamefont {Palma}, \citenamefont {Asai},\ and\
  \citenamefont {Tao}}]{Tao16}%
  \BibitemOpen
  \bibfield  {author} {\bibinfo {author} {\bibfnamefont {Y.}~\bibnamefont
  {Li}}, \bibinfo {author} {\bibfnamefont {L.}~\bibnamefont {Xiang}}, \bibinfo
  {author} {\bibfnamefont {J.~L.}\ \bibnamefont {Palma}}, \bibinfo {author}
  {\bibfnamefont {Y.}~\bibnamefont {Asai}}, \ and\ \bibinfo {author}
  {\bibfnamefont {N.}~\bibnamefont {Tao}},\ }\href
  {https://doi.org/10.1038/ncomms11294} {\bibfield  {journal} {\bibinfo
  {journal} {Nat. Commun.}\ }\textbf {\bibinfo {volume} {7}},\ \bibinfo {pages}
  {11294} (\bibinfo {year} {2016})}\BibitemShut {NoStop}%
\bibitem [{\citenamefont {Salomon}\ \emph {et~al.}(2003)\citenamefont
  {Salomon}, \citenamefont {Cahen}, \citenamefont {Lindsay}, \citenamefont
  {Tomfohr}, \citenamefont {Engelkes},\ and\ \citenamefont
  {Frisbie}}]{Salomon.03.AM}%
  \BibitemOpen
  \bibfield  {author} {\bibinfo {author} {\bibfnamefont {A.}~\bibnamefont
  {Salomon}}, \bibinfo {author} {\bibfnamefont {D.}~\bibnamefont {Cahen}},
  \bibinfo {author} {\bibfnamefont {S.}~\bibnamefont {Lindsay}}, \bibinfo
  {author} {\bibfnamefont {J.}~\bibnamefont {Tomfohr}}, \bibinfo {author}
  {\bibfnamefont {V.~B.}\ \bibnamefont {Engelkes}}, \ and\ \bibinfo {author}
  {\bibfnamefont {C.~D.}\ \bibnamefont {Frisbie}},\ }\href {\doibase
  10.1002/adma.200306091} {\bibfield  {journal} {\bibinfo  {journal} {Adv.
  Mater.}\ }\textbf {\bibinfo {volume} {15}},\ \bibinfo {pages} {1881}
  (\bibinfo {year} {2003})}\BibitemShut {NoStop}%
\bibitem [{\citenamefont {Tal}\ \emph {et~al.}(2008)\citenamefont {Tal},
  \citenamefont {Krieger}, \citenamefont {Leerink},\ and\ \citenamefont {van
  Ruitenbeek}}]{Tal08}%
  \BibitemOpen
  \bibfield  {author} {\bibinfo {author} {\bibfnamefont {O.}~\bibnamefont
  {Tal}}, \bibinfo {author} {\bibfnamefont {M.}~\bibnamefont {Krieger}},
  \bibinfo {author} {\bibfnamefont {B.}~\bibnamefont {Leerink}}, \ and\
  \bibinfo {author} {\bibfnamefont {J.~M.}\ \bibnamefont {van Ruitenbeek}},\
  }\href {\doibase 10.1103/PhysRevLett.100.196804} {\bibfield  {journal}
  {\bibinfo  {journal} {Phys. Rev. Lett.}\ }\textbf {\bibinfo {volume} {100}},\
  \bibinfo {pages} {196804} (\bibinfo {year} {2008})}\BibitemShut {NoStop}%
\bibitem [{\citenamefont {Kumar}\ \emph {et~al.}(2012)\citenamefont {Kumar},
  \citenamefont {Avriller}, \citenamefont {Yeyati},\ and\ \citenamefont {van
  Ruitenbeek}}]{Kumar12}%
  \BibitemOpen
  \bibfield  {author} {\bibinfo {author} {\bibfnamefont {M.}~\bibnamefont
  {Kumar}}, \bibinfo {author} {\bibfnamefont {R.}~\bibnamefont {Avriller}},
  \bibinfo {author} {\bibfnamefont {A.}~\bibnamefont {Yeyati}}, \ and\ \bibinfo
  {author} {\bibfnamefont {J.~M.}\ \bibnamefont {van Ruitenbeek}},\ }\href
  {\doibase 10.1103/PhysRevLett.108.146602} {\bibfield  {journal} {\bibinfo
  {journal} {Phys. Rev. Lett.}\ }\textbf {\bibinfo {volume} {108}},\ \bibinfo
  {pages} {146602} (\bibinfo {year} {2012})}\BibitemShut {NoStop}%
\bibitem [{\citenamefont {Avriller}\ and\ \citenamefont
  {Levy~Yeyati}(2009)}]{Yeyati09}%
  \BibitemOpen
  \bibfield  {author} {\bibinfo {author} {\bibfnamefont {R.}~\bibnamefont
  {Avriller}}\ and\ \bibinfo {author} {\bibfnamefont {A.}~\bibnamefont
  {Levy~Yeyati}},\ }\href {\doibase 10.1103/PhysRevB.80.041309} {\bibfield
  {journal} {\bibinfo  {journal} {Phys. Rev. B}\ }\textbf {\bibinfo {volume}
  {80}},\ \bibinfo {pages} {041309} (\bibinfo {year} {2009})}\BibitemShut
  {NoStop}%
\bibitem [{\citenamefont {Haupt}\ \emph {et~al.}(2009)\citenamefont {Haupt},
  \citenamefont {Novotn\'y},\ and\ \citenamefont {Belzig}}]{Haupt09}%
  \BibitemOpen
  \bibfield  {author} {\bibinfo {author} {\bibfnamefont {F.}~\bibnamefont
  {Haupt}}, \bibinfo {author} {\bibfnamefont {T.}~\bibnamefont {Novotn\'y}}, \
  and\ \bibinfo {author} {\bibfnamefont {W.}~\bibnamefont {Belzig}},\ }\href
  {\doibase 10.1103/PhysRevLett.103.136601} {\bibfield  {journal} {\bibinfo
  {journal} {Phys. Rev. Lett.}\ }\textbf {\bibinfo {volume} {103}},\ \bibinfo
  {pages} {136601} (\bibinfo {year} {2009})}\BibitemShut {NoStop}%
\bibitem [{\citenamefont {Haupt}\ \emph {et~al.}(2010)\citenamefont {Haupt},
  \citenamefont {Novotn\'y},\ and\ \citenamefont {Belzig}}]{Haupt10}%
  \BibitemOpen
  \bibfield  {author} {\bibinfo {author} {\bibfnamefont {F.}~\bibnamefont
  {Haupt}}, \bibinfo {author} {\bibfnamefont {T.}~\bibnamefont {Novotn\'y}}, \
  and\ \bibinfo {author} {\bibfnamefont {W.}~\bibnamefont {Belzig}},\ }\href
  {\doibase 10.1103/PhysRevB.82.165441} {\bibfield  {journal} {\bibinfo
  {journal} {Phys. Rev. B}\ }\textbf {\bibinfo {volume} {82}},\ \bibinfo
  {pages} {165441} (\bibinfo {year} {2010})}\BibitemShut {NoStop}%
\bibitem [{\citenamefont {Novotn\'y}\ \emph {et~al.}(2011)\citenamefont
  {Novotn\'y}, \citenamefont {Haupt},\ and\ \citenamefont {Belzig}}]{Haupt11}%
  \BibitemOpen
  \bibfield  {author} {\bibinfo {author} {\bibfnamefont {T.}~\bibnamefont
  {Novotn\'y}}, \bibinfo {author} {\bibfnamefont {F.}~\bibnamefont {Haupt}}, \
  and\ \bibinfo {author} {\bibfnamefont {W.}~\bibnamefont {Belzig}},\ }\href
  {\doibase 10.1103/PhysRevB.84.113107} {\bibfield  {journal} {\bibinfo
  {journal} {Phys. Rev. B}\ }\textbf {\bibinfo {volume} {84}},\ \bibinfo
  {pages} {113107} (\bibinfo {year} {2011})}\BibitemShut {NoStop}%
\bibitem [{\citenamefont {Mahan}(2000)}]{Mahan.00.NULL}%
  \BibitemOpen
  \bibfield  {author} {\bibinfo {author} {\bibfnamefont {G.~D.}\ \bibnamefont
  {Mahan}},\ }\href@noop {} {\emph {\bibinfo {title} {Many-Particle Physics}}}\
  (\bibinfo  {publisher} {Plenum, New York},\ \bibinfo {year}
  {2000})\BibitemShut {NoStop}%
\end{thebibliography}
%

\end{document}